\documentclass[a4paper,fleqn,usenatbib,twocolumn,useAMS]{mnras}

\newif\ifAMStwofonts

\usepackage{graphicx}	
\usepackage{amsmath}	
\usepackage{amssymb}	
\usepackage{epsfig}
\usepackage{color}

\usepackage{pdflscape}	

\def\pmb#1{\mbox{\boldmath$#1$}}
\def\gtsim {>\kern-1.2em\lower1.1ex\hbox{$\sim$}}
\def\ltsim {<\kern-1.2em\lower1.1ex\hbox{$\sim$}}
\def\gtsim {>\kern-1.2em\lower1.1ex\hbox{$\sim$}}
\def\ltsim {<\kern-1.2em\lower1.1ex\hbox{$\sim$}}

\def\be{\begin{equation}}
\def\ee{\end{equation}}

\def\pmbmt#1{\pmb{\sf #1}}
\def\rmi{{\rm i}}

\begin{document}
\title[Overstable Convective Modes in Early Type Stars]{
Overstable Convective Modes in Rotating Early Type Stars}

\author[U. Lee]{
Umin Lee$^{1}$\thanks{E-mail: lee@astr.tohoku.ac.jp}
\\
$^{1}$Astronomical Institute, Tohoku University, Sendai, Miyagi 980-8578, Japan\\
}

\date{Accepted XXX. Received YYY; in original form ZZZ}
\pubyear{2015}

\maketitle

\begin{abstract}
We calculate overstable convective (OsC) modes of $2M_\odot$, $4M_\odot$, and $20M_\odot$ main sequence stars.
To compute non-adiabatic OsC modes in the core, we assume $(\nabla\cdot\pmb{F}_C)^\prime=0$ as a prescription for the approximation called frozen-in convection in pulsating stars where $\pmb{F}_C$ is the convective energy flux and the prime $^\prime$ 
indicates Eulerian perturbation.
We find that the general properties of the OsC modes 
are roughly the same as those obtained by \citet{LeeSaio20} who assumed $\delta (\nabla\cdot\pmb{F}_C)=0$, except that no OsC modes behave like inertial modes when they tend toward complete stabilization with increasing rotation frequency where
$\delta$ indicates the Lagrangian perturbation.
As the rotation frequency of the stars increases, the OsC modes are stabilized to resonantly excite $g$-modes in the envelope when the core rotates slightly faster than the envelope.
The frequency of the OsC modes that excite envelope $g$-modes is approximately given by $\sigma\sim |m\Omega_c|$ in the inertial frame and hence $\sigma_{m=-2}\approx2\sigma_{m=-1}$
where $m$ is the azimuthal wavenumber of the modes and $\Omega_c$ is the rotation frequency of the core.
We find that the modal properties of OsC modes do not strongly depend on the mass of the stars.

We discuss angular momentum transport by OsC modes in resonance with envelope $g$-modes in the main sequence stars.
We suggest that angular momentum transfer takes place from the core to the envelope and
that the OsC modes may help the stars rotate uniformly and keep the rotation frequency of the core low
during their evolution as main sequence stars. 

\end{abstract}



\section{introduction}

Low frequency photometric variations
have been detected in many rotating early type stars such as A-type stars
(e.g., \citealt{Balona13, Balona17}) and B-type stars
(e.g., \citealt{DegrooteAckeSamadietal2011,Balona16,Balonaetal19,BalonaOzuyar20}).
Their frequencies are consistent with rotation frequency of the stars and hence
the variabilities are called rotational modulation.
The origin of rotational modulation in early type stars is not necessarily
well understood.
We usually assume that rotational modulation is produced by inhomogeneous
brightness distribution on the surface of rotating stars and that the origin of
the inhomogeneity is attributed to the existence of global magnetic fields
at the stellar surface.
However, early type stars do not posses a thick surface convection zone and hence dynamo mechanism
is not necessarily an efficient mechanism for generating global surface magnetic fields.
It has been suggested that subsurface convection zones in early type stars
can generate surface magnetic fields in hot massive stars (\citealt{CantielloBraithwaite11})
and in A- and late B-type stars (\citealt{CantielloBraithwaite19}). 
It is important to note that although \citet{CantielloBraithwaite19} have also predicted regions  
in the H-R diagram where subsurface convection is unlikely to
produce significant magnetic fields, 
rotational modulation is observed to be present in such regions (\citealt{BalonaOzuyar20}).
It may be desirable to finds a generating mechanism for rotational modulation that does not need
surface spots and magnetic fields.

Rotational modulations have also been identified in pulsating variables such as
$\gamma$ Dor stars (e.g., \citealt{VanReethMombargMathisetal2018}) and slowly pulsating B (SPB) stars and $\beta$ Cephei stars (see Table 1 of \cite{BalonaOzuyar20}).
Assuming that the rotational modulations are produced by spots on the surface of rotating stars
and using observed low frequency $g$-modes to derive the rotation rate
in the near-core region for $\gamma$ Dor stars, 
\citet{VanReethMombargMathisetal2018} suggested that almost rigid rotation prevails in the envelope of the stars.

Convective modes, also called $g^{-}$-modes, in early type stars 
are confined in the convective core and
are unstable in the sense that the amplitudes grow exponentially with time.
Assuming uniform rotation, \citet{LeeSaio86} have numerically shown for a $10M_\odot$ main sequence star that
low $m$ convective modes in the core become overstable when the star rotates and that
as the rotation speed increases, the overstable convective (OsC) modes 
are stabilized to be oscillatory in time and to
resonantly excite low frequency $g$-modes in the envelope where $m$ is the azimuthal wavenumber of the modes.
Recently, \citet{LeeSaio20} computed OsC modes of $2M_\odot$ main sequence stars
with some improvements to the previous study (\citealt{LeeSaio86}) and showed that low $m$ OsC modes in the core 
can resonantly excite prograde sectoral $g$-modes in the envelope
if the core rotates slightly faster than the envelope.
It may be interesting to note that no effective excitation of envelope $g$-modes by OsC modes 
takes place for uniform rotation.
If the excited $g$-modes have significant amplitudes at the stellar surface, they will be observed as
low frequency oscillations.
\citet{LeeSaio20} thus proposed that the OsC modes are 
responsible for rotational modulations observed in rotating A-type main sequence stars (e.g. \citealt{Balona13}).
In fact, the oscillation frequency $\sigma$ of the excited $g$-modes in the inertial frame is in a good approximation given 
by $\sigma\approx |m\Omega_c|$ with $\Omega_c$ being the rotation frequency of the core, and
the low frequency oscillations of $\sigma\approx |m\Omega_c|$ for small $|m|$ will be recognized as rotational modulations.
\citet{LeeSaio20} also suggested that 
if $m=-1$ and $m=-2$ OsC modes simultaneously excite $g$-modes in the envelope,
photometric variations
with the frequencies $\sigma\approx \Omega_c$ and $\approx 2\Omega_c$ will be observed as 
rotational modulation with a low frequency and its first harmonic (\citealt{Balona13}).

Since we are to compute convective modes in the core of rotating stars, 
some comments may be needed on
how to treat turbulent convective fluid motions in pulsating stars, 
which has been a difficult problem to solve, particularly when we are interested in
non-adiabatic analyses to discuss the pulsational stability of oscillation modes.
To describe interactions between pulsations and convective fluid motions in non-rotating stars, 
we may use the theory of time-dependent convection  
(e.g., \citealt{Dupretetal05}), which has been successfully applied to explain the drivng mechanism
for $g$-modes in $\gamma$ Dor stars.
Note that the theory of time-dependent convection has also been applied to rotating stars by \cite{BouabidDupretSalmonetal2013},
who used the traditional approximation of rotation (TAR) (e.g., \citealt{LeeSaio97}).
However, it is difficult to apply the theory to rotating stars without TAR and
we usually employ a simplifying approximation, called frozen-in convection, 
for turbulent convection in rotating and pulsating stars.
There are several prescriptions for the approximation.
For example, the frozen-in convection in pulsating stars may be prescribed by
$\delta(\nabla\cdot\pmb{F}_C)=0$ (e.g., \citealt{LeeSaio20}) or by $(\nabla\cdot\pmb{F}_C)^\prime=0$
(e.g., \citealt{LeeBaraffe95}) where $\delta$ and the prime $^\prime$ indicate Lagrangian and Eulerian perturbation, respectively.
See also \citet{Unnoetal1989} for other possible prescriptions.
We need to examine whether or not the different prescriptions $\delta(\nabla\cdot\pmb{F}_C)=0$ and $(\nabla\cdot\pmb{F}_C)^\prime=0$
for the frozen-in convection lead to significant differences in the properties of OsC modes.

Slowly pulsating B stars (\citealt{Waelkens91}) are known to show slow photometric variabilities due to $g$-mode pulsations 
excited by the iron opacity bump mechanism (e.g., \citealt{GautschySaio1993,Dziembowskietal93}).
Since many low frequency $g$-modes are excited in a SPB star,
precise observational determination of the frequencies provide us with a good information concerning
the internal structure of the stars (e.g., \citealt{DegrooteAertsBaglinetal2010,DegrooteAertsMicheletal2012,
PapicsTMoravvejiAerts2014,PapicsTkachenkoAertsetal2015,PapicsTkachenkoVanReethetal2017}).
If the stars rapidly rotate, even if the low frequency modes are reasonably well described under TAR,
some complexities due to rapid rotation may arise in
the frequency spectra of low frequency $g$-modes.
Note that in this paper we use the word \lq\lq rapid rotation" somewhat loosely to
suggest that the rotation velocity is greater than about half the breakup rotation velocity.
For example, in $\gamma$ Dor stars, many low frequency $g$-mods in the envelope are excited by the mechanism
called convection blocking in the subsurface convection zone (e.g., \citealt{Guziketal2000}, \citealt{Dupretetal05}, 
but see \citealt{Kahramanetal20}).
For rapidly rotating $\gamma$ Dor stars, period spacings $\Delta P_n=P_{n+1}-P_n$ of observed $g$-modes have been used to derive the rotation rate in the radiative regions in the envelope close to the convective core of the stars (e.g., \citealt{BouabidDupretSalmonetal2013,VanReethTkachenkoAerts2016}) where $P_n$ is the oscillation period of $g$-mode and $n$ denotes its radial order.
\citet{Quazzanietal20} suggested for $\gamma$ Dor stars that 
the period spacings may be disturbed to describe a deep dip when plotted as a function of $P_n$ if the $g$-modes are in resonance with an inertial mode in the convective core.
Such a resonance can provide useful information concerning the convective core itself (\citealt{Saioetal21}).
It is likely that similar resonance phenomena between $g$-modes and an inertial mode take place in rotating SPB stars, in which $g$-modes are excited by the opacity bump mechanism.
We also expect that resonances may occur between
opacity driven $g$-modes and OsC modes in rapidly rotating SPB stars and
it is one of our interests to see whether or not this really happens.

Angular momentum transport by non-radial oscillations in rotating stars has been discussed by many authors, including  \citet{Ando83}, 
\citet{LeeSaio93}, \citet{TalonKumarZahn02}, \citet{RogersGlatzmaier06}, \citet{TownsendGoldsteinZweibel18}, and \citet{Neineretal20}.
See a recent review by \citet{AertsMathisRogers19} for angular momentum transport in rotating stars.
As \citet{Ando83} discussed, prograde waves extract angular momentum from rotating fluid to decelerate its rotation 
where the waves are excited, and deposit angular momentum to accelerate the rotation where the waves are damped.
For example, \citet{TalonKumarZahn02} and \citet{RogersGlatzmaier06} carried out numerical simulations to follow the evolution of 
internal rotation for solar models, 
taking account of angular momentum transport by gravity waves, which are assumed to be excited by turbulent fluid motions
in the convective envelope and to suffer radiative and viscous dissipations in the radiative core.
They found that the radiative dissipation tends to strengthen and the viscous dissipation to smooth out differential rotation in the radiative core
and that the competition between the two effects may lead to quasi-periodic changes in the internal rotation.
\citet{TownsendGoldsteinZweibel18} also numerically followed the evolution of internal rotation in SPB stars, 
taking account of angular momentum transport by a $g$-mode excited by the opacity bump mechanism.
Assuming finite oscillation amplitudes consistent with observations for the $g$-mode,
they found that the rotation speed in the surface layers is significantly decelerated
as a result of angular momentum redistribution in the envelope.
More recently, \citet{Neineretal20} discussed angular momentum transport by low frequency $g$-modes
stochastically excited in the core of massive main sequence stars.
They suggested that the surface layers are significantly accelerated if the stars rotate rapidly.

Since OsC modes that resonantly excite envelope $g$-modes have amplitudes both in the core and in the envelope of
rotating stars (e.g., \citealt{LeeSaio20}), they are expected to play an important role in angular momentum transport between the core and the envelope of the stars.
Prograde OsC modes are excited by convective instability and $\epsilon$ mechanism in the core.
On the other hand,
prograde $g$-modes driven by the OsC modes suffer from radiative dissipation in the envelope.
We guess that OsC modes in resonance with envelope $g$-modes can be a carrier of angular momentum
from the core to the envelope in rotating early type main sequence stars.

In this paper, we carry out non-adiabatic calculations of low $m$ OsC modes for $2M_\odot$, $4M_\odot$, and $20M_\odot$
main sequence stars, assuming that the core rotates slightly faster than the envelope.
We calculate OsC modes in $2M_\odot$ main sequence stars to compare the results obtained by using the two different
prescriptions for the approximation of frozen-in convection.
$4M_\odot$ main sequence models correspond to SPB stars, in which we expect OsC modes
to coexist with low frequency $g$-modes excited by the opacity bump mechanism.
Massive main sequence stars have a large convective core and rather thick subsurface convection zones
due to the iron opacity bump in the envelope.
It is therefore interesting to see whether or not OsC modes in $20M_\odot$ main sequence stars behave differently from those of $2M_\odot$ stars.
A brief account of method of calculation is given in \S 2 and numerical results for the low $m$ OsC modes are
presented in \S 3.
We discuss angular momentum transport by the OsC modes in resonance with envelope $g$-modes in \S 4.
We also discuss OsC modes with negative energy of oscillation in \S 5 and
we conclude in \S 6.
In the Appendix, we discuss low frequency modes in the convective core 
assuming the two different prescriptions for the approximation of frozen-in convection.

\section{Method of Calculation}

\begin{figure*}
\resizebox{0.66\columnwidth}{!}{
\includegraphics{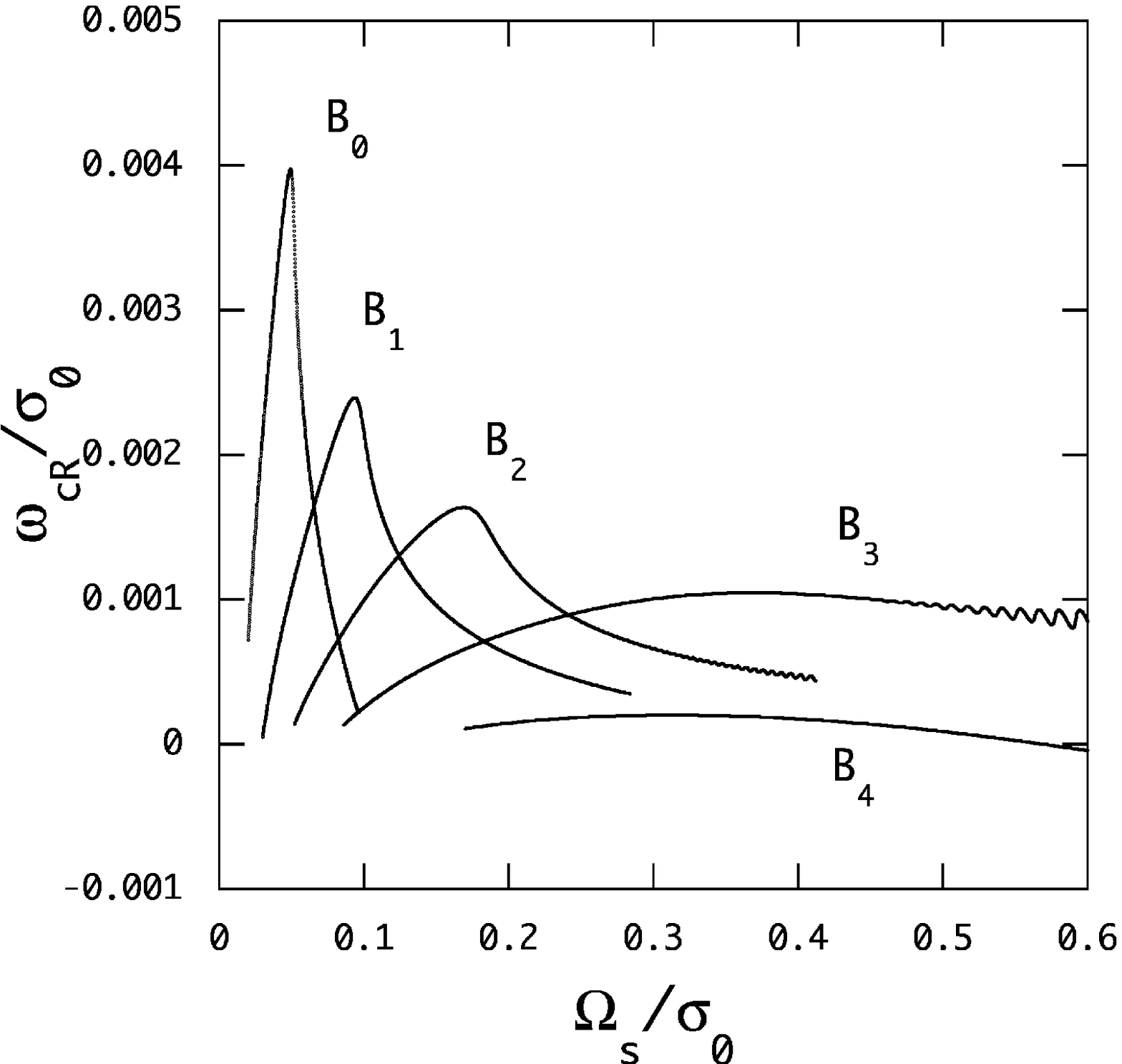}}
\resizebox{0.66\columnwidth}{!}{
\includegraphics{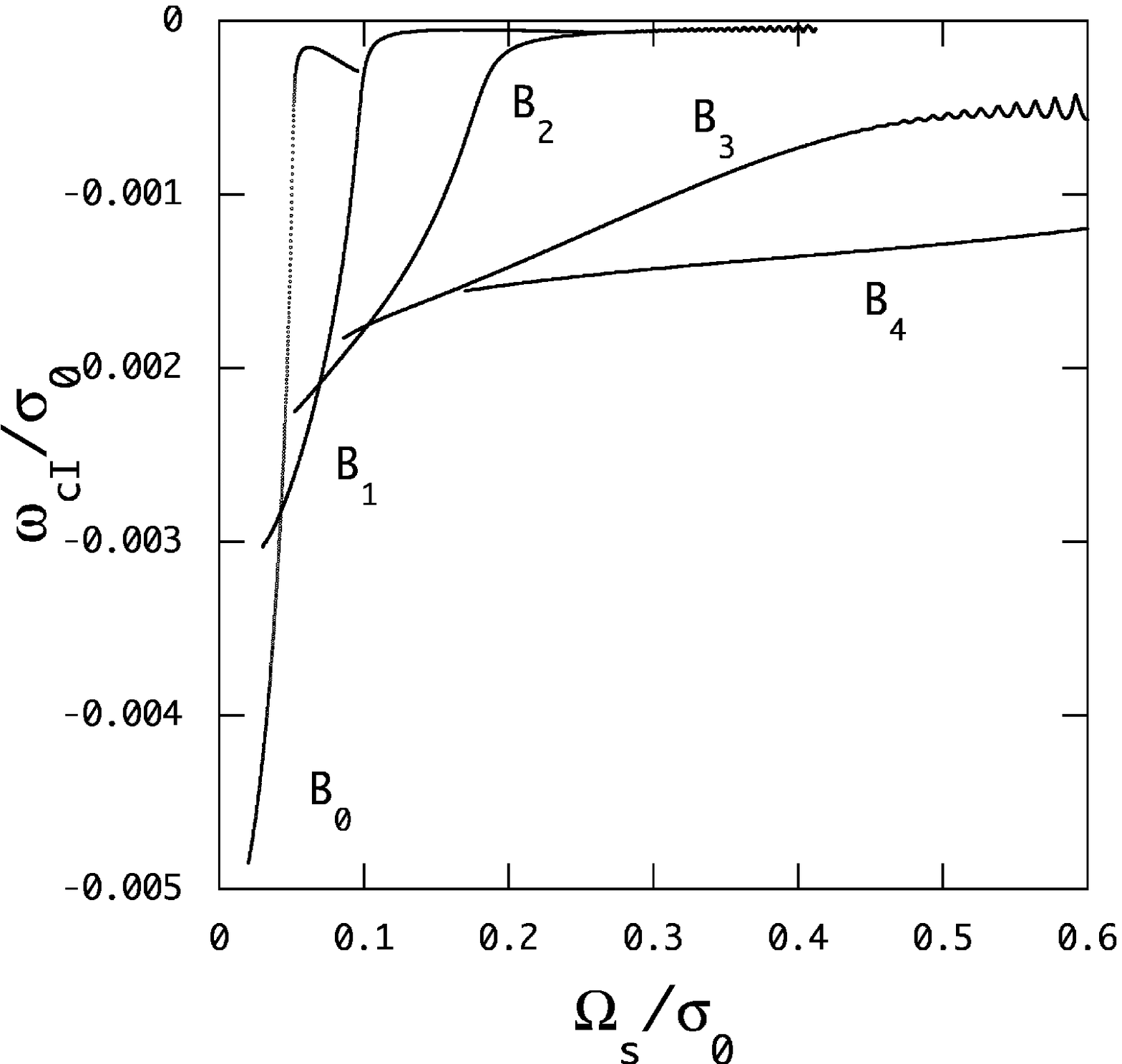}}
\resizebox{0.66\columnwidth}{!}{
\includegraphics{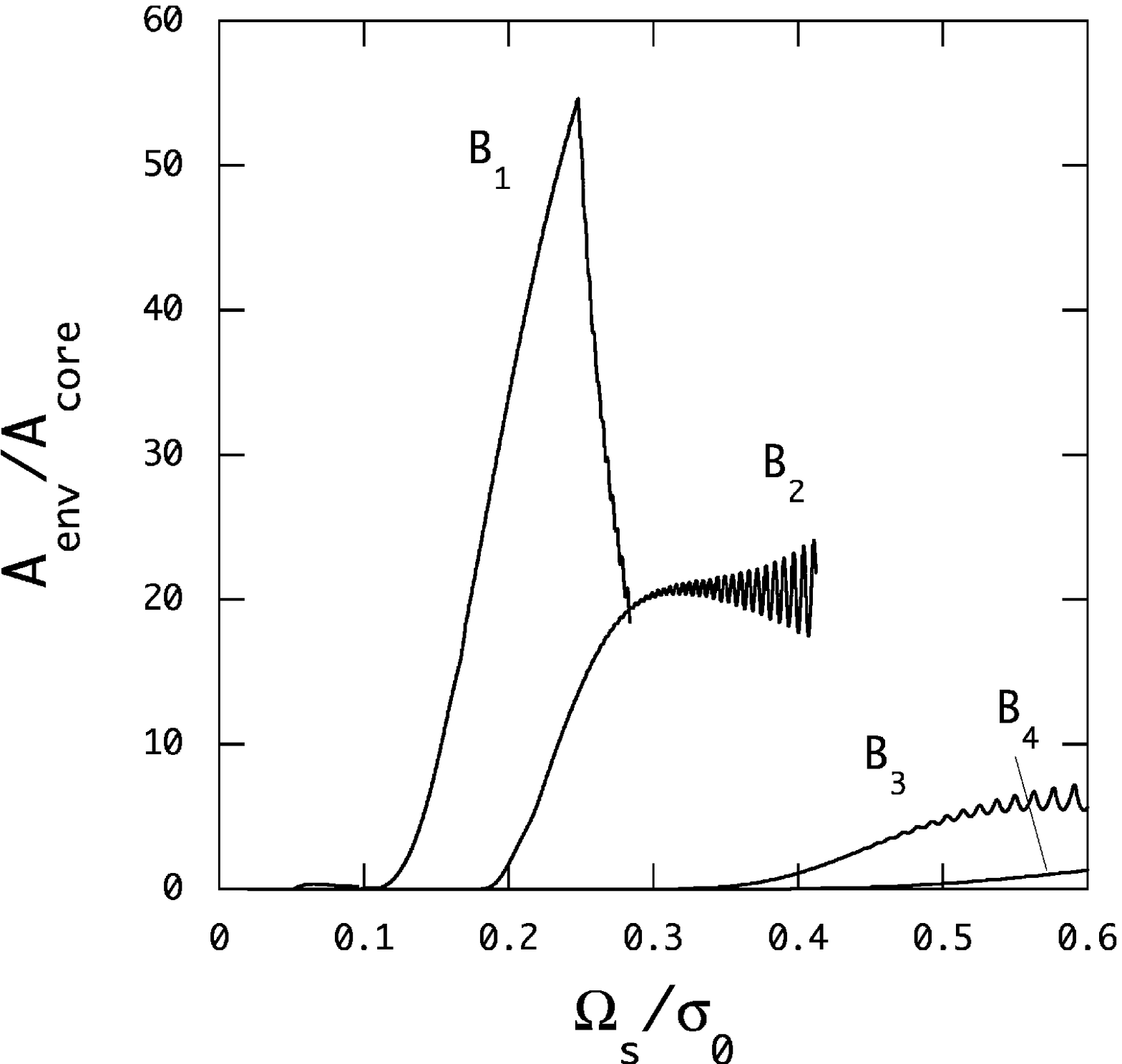}}
\resizebox{0.66\columnwidth}{!}{
\includegraphics{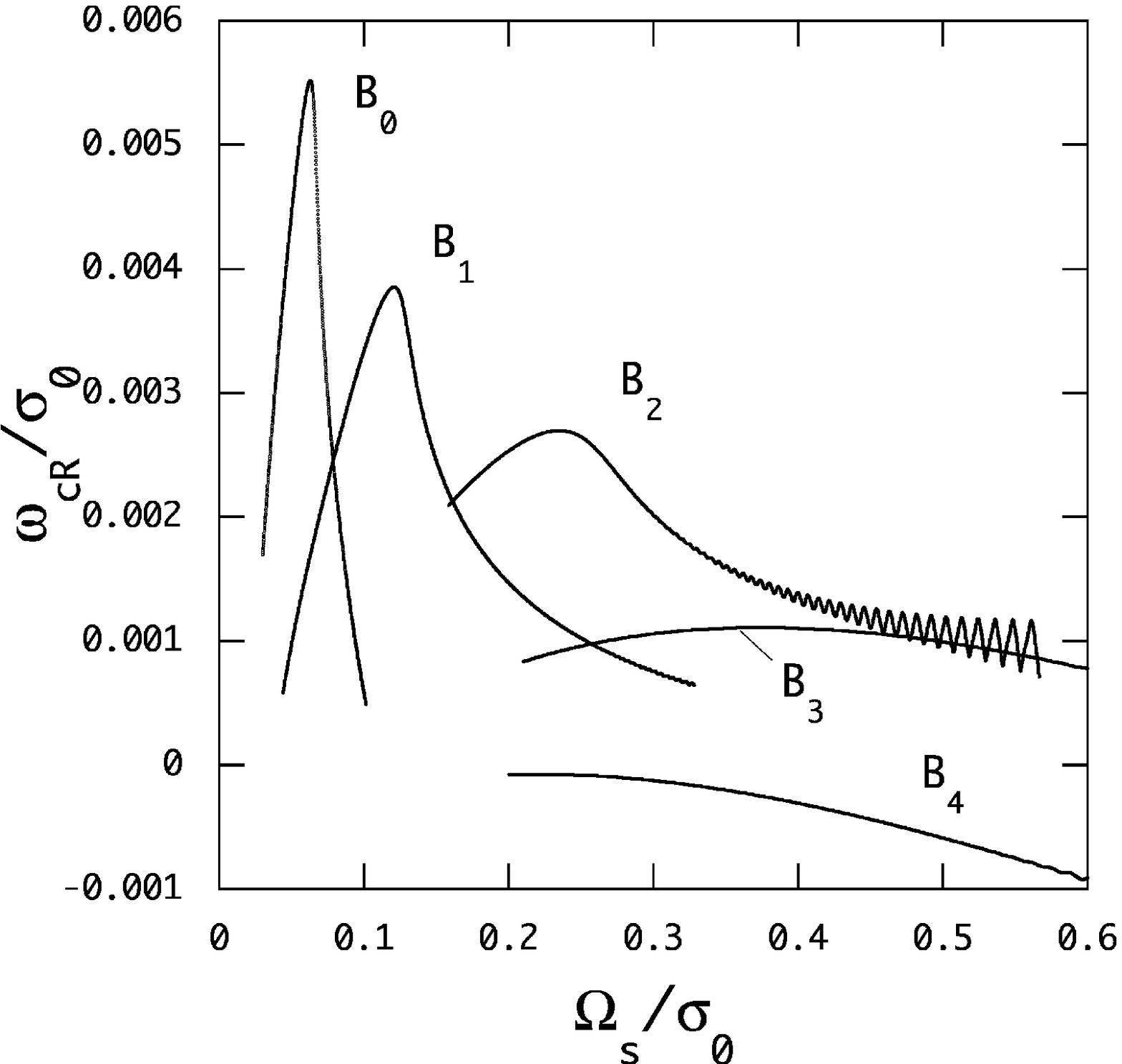}}
\resizebox{0.66\columnwidth}{!}{
\includegraphics{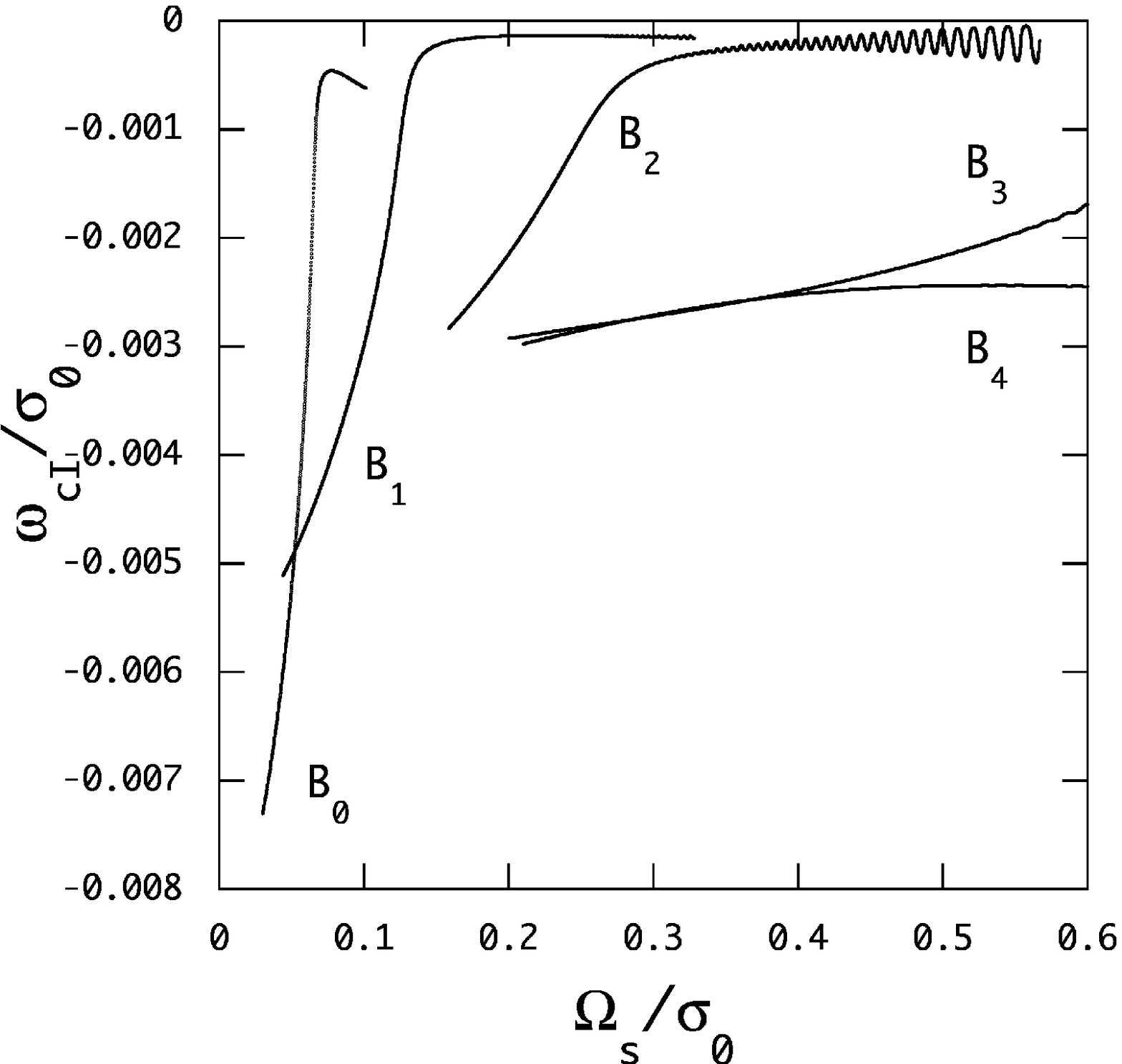}}
\resizebox{0.66\columnwidth}{!}{
\includegraphics{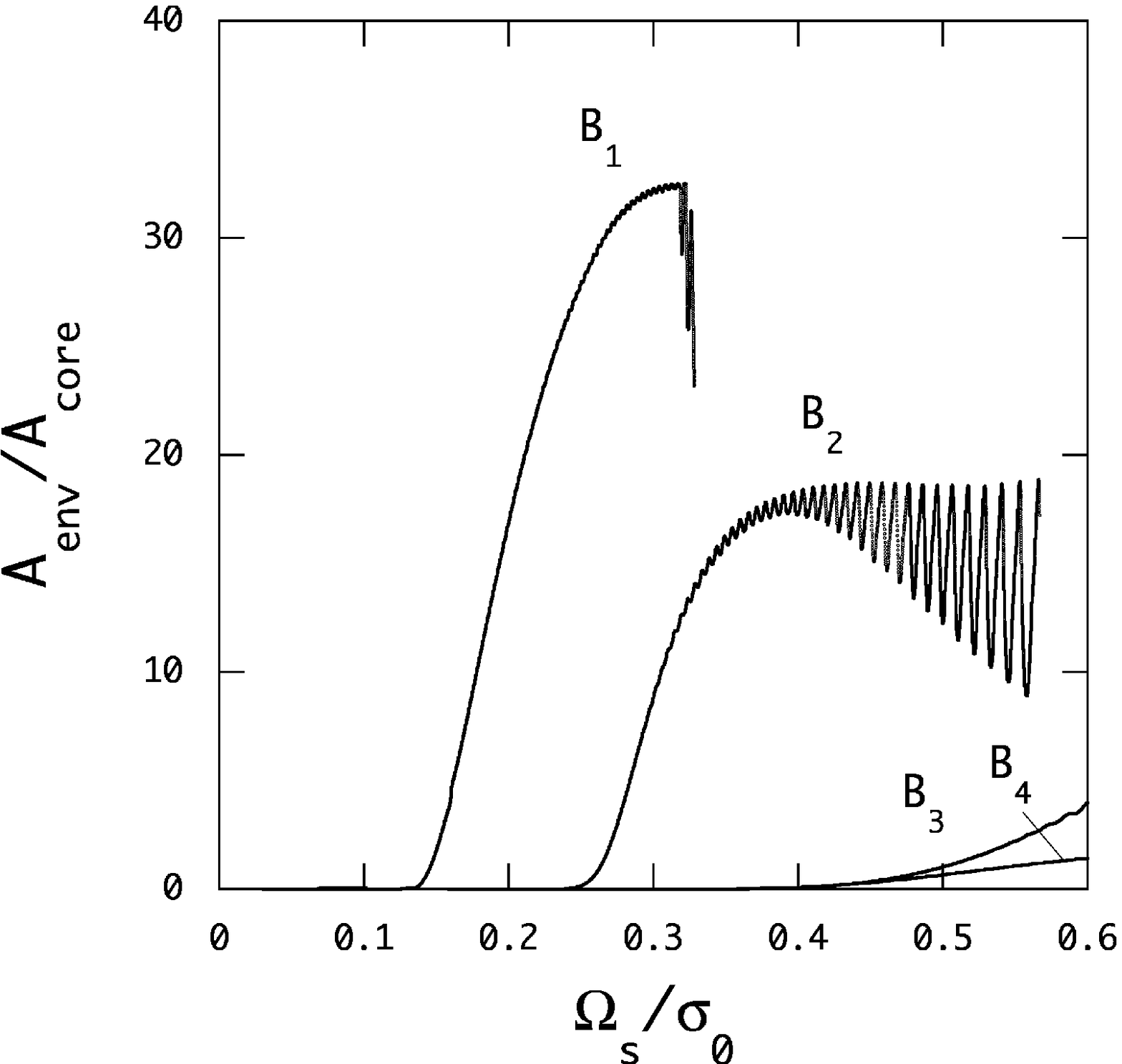}}
\caption{Complex $\overline\omega_c$ and the ratio $A_{\rm env}/A_{\rm core}$ of $m=-1$ (upper panels) and $m=-2$ (lower panels)
OsC modes of the $2M_\odot$ ZAMS model versus $\overline\Omega_s$ for $b=1.1$
where $A_{\rm env}$ is the maximum amplitude of $xH_l$ in the envelope and $A_{\rm core}$ is that of 
$xS_l$ in the core.}
\label{fig:omega_m2md1b1p1}
\resizebox{0.66\columnwidth}{!}{
\includegraphics{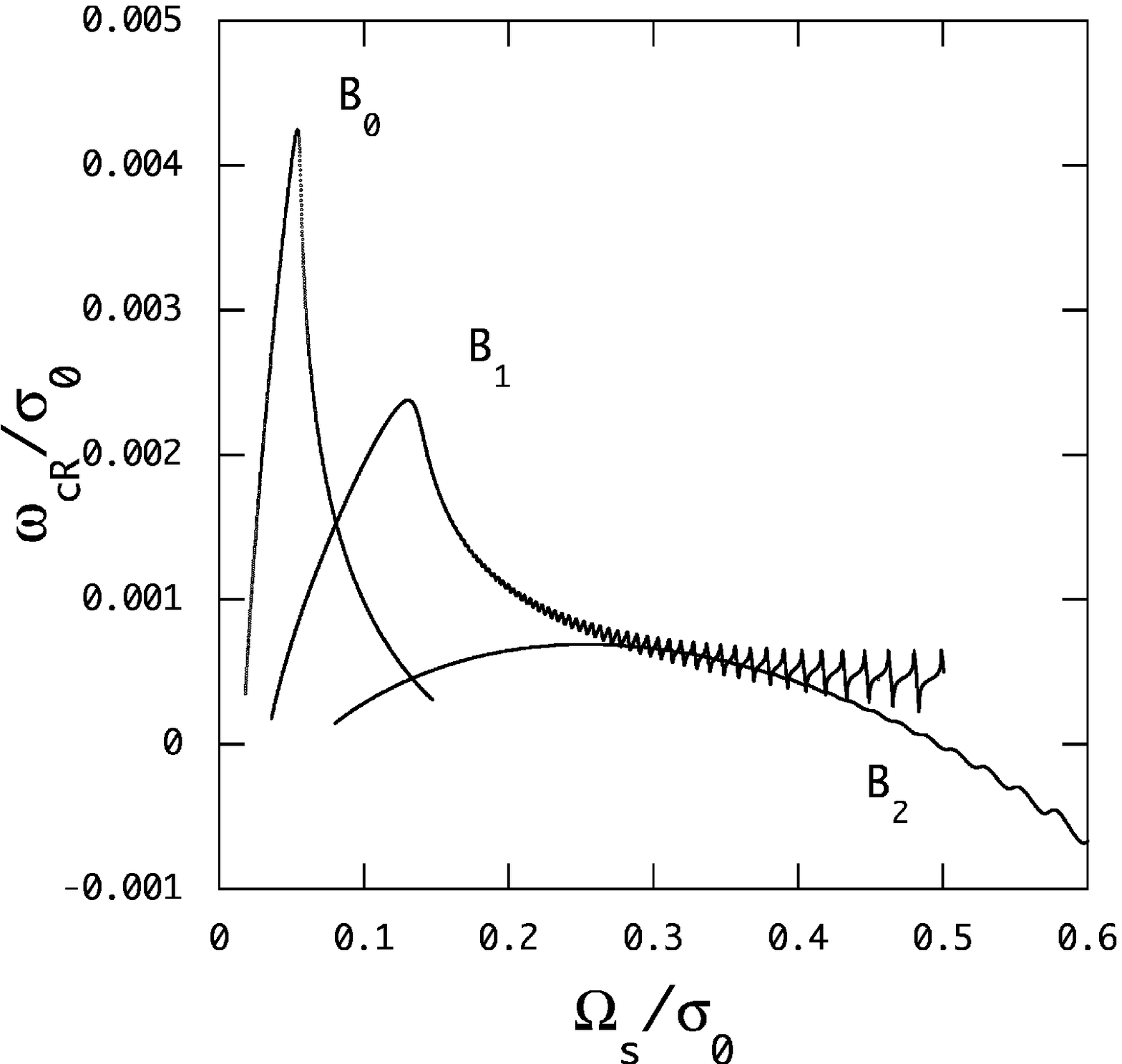}}
\resizebox{0.66\columnwidth}{!}{
\includegraphics{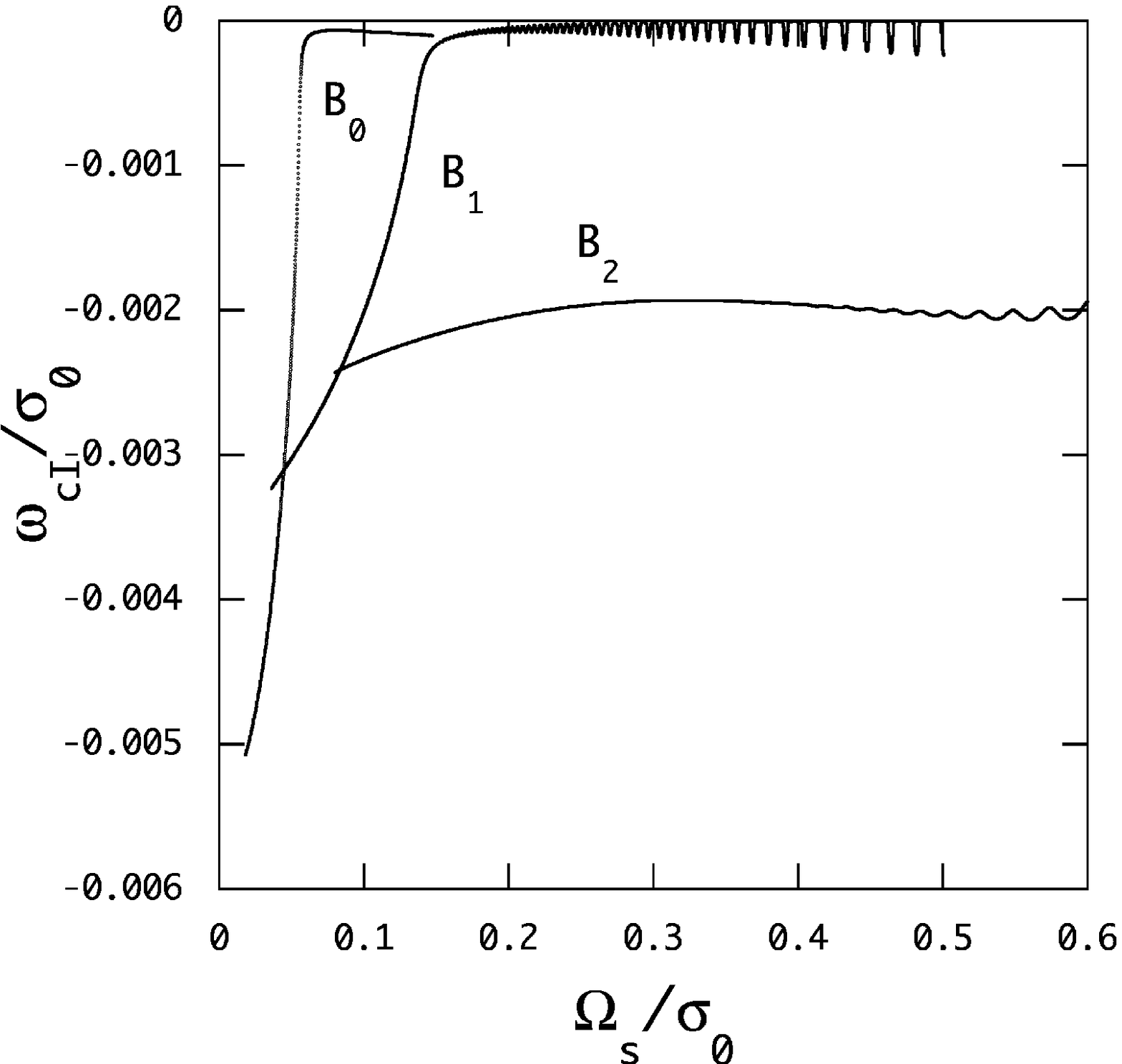}}
\resizebox{0.66\columnwidth}{!}{
\includegraphics{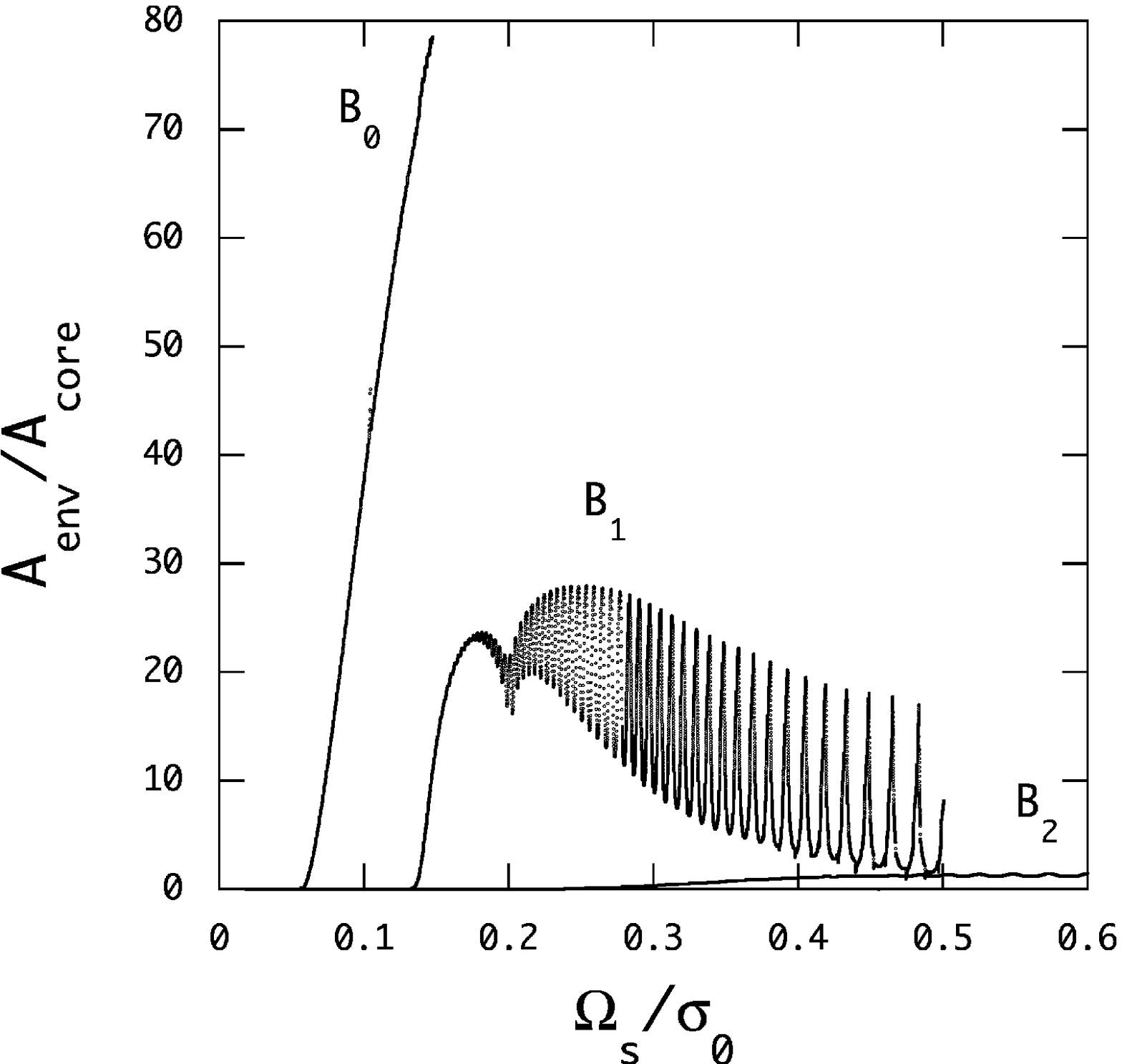}}
\resizebox{0.66\columnwidth}{!}{
\includegraphics{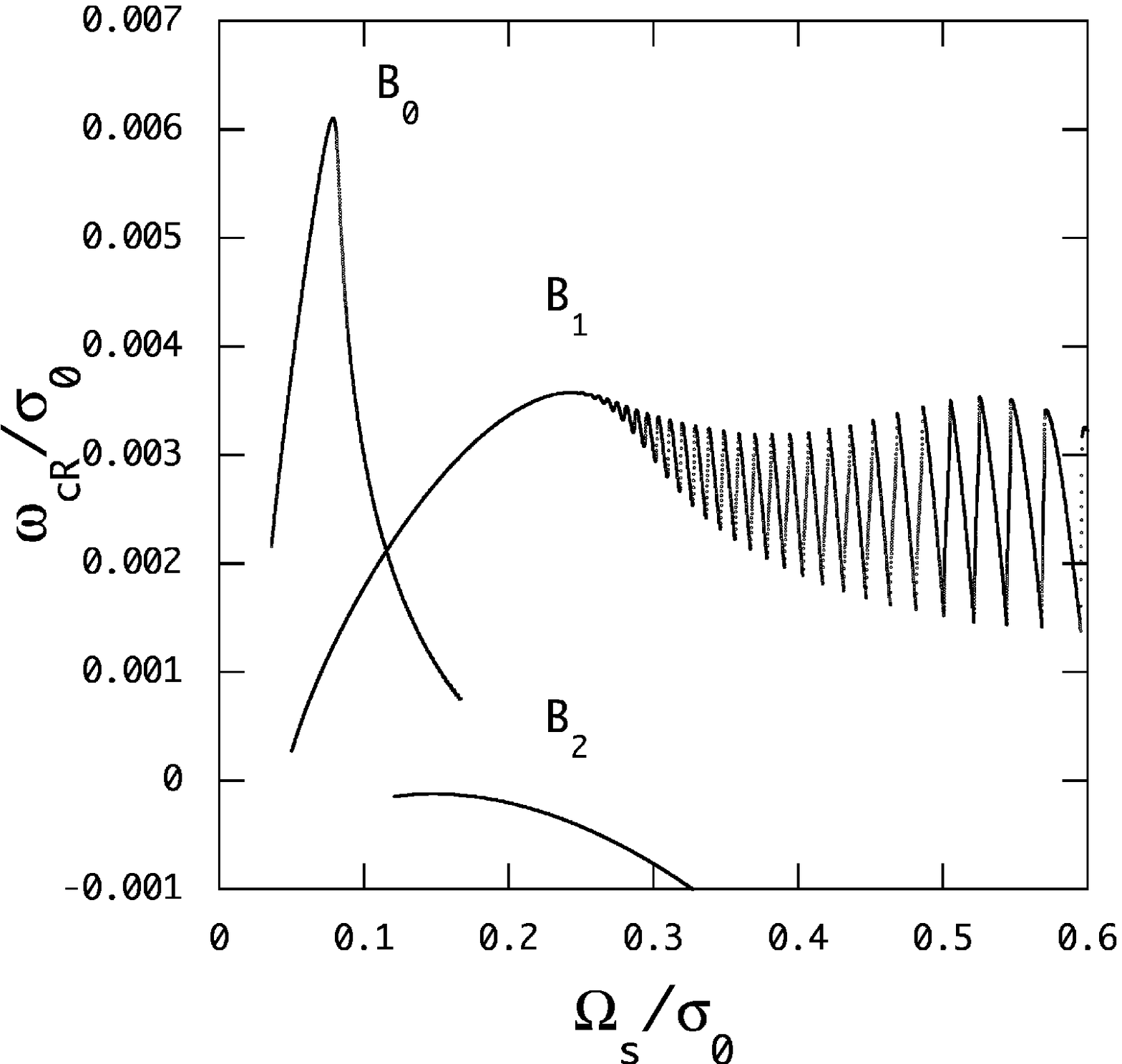}}
\resizebox{0.66\columnwidth}{!}{
\includegraphics{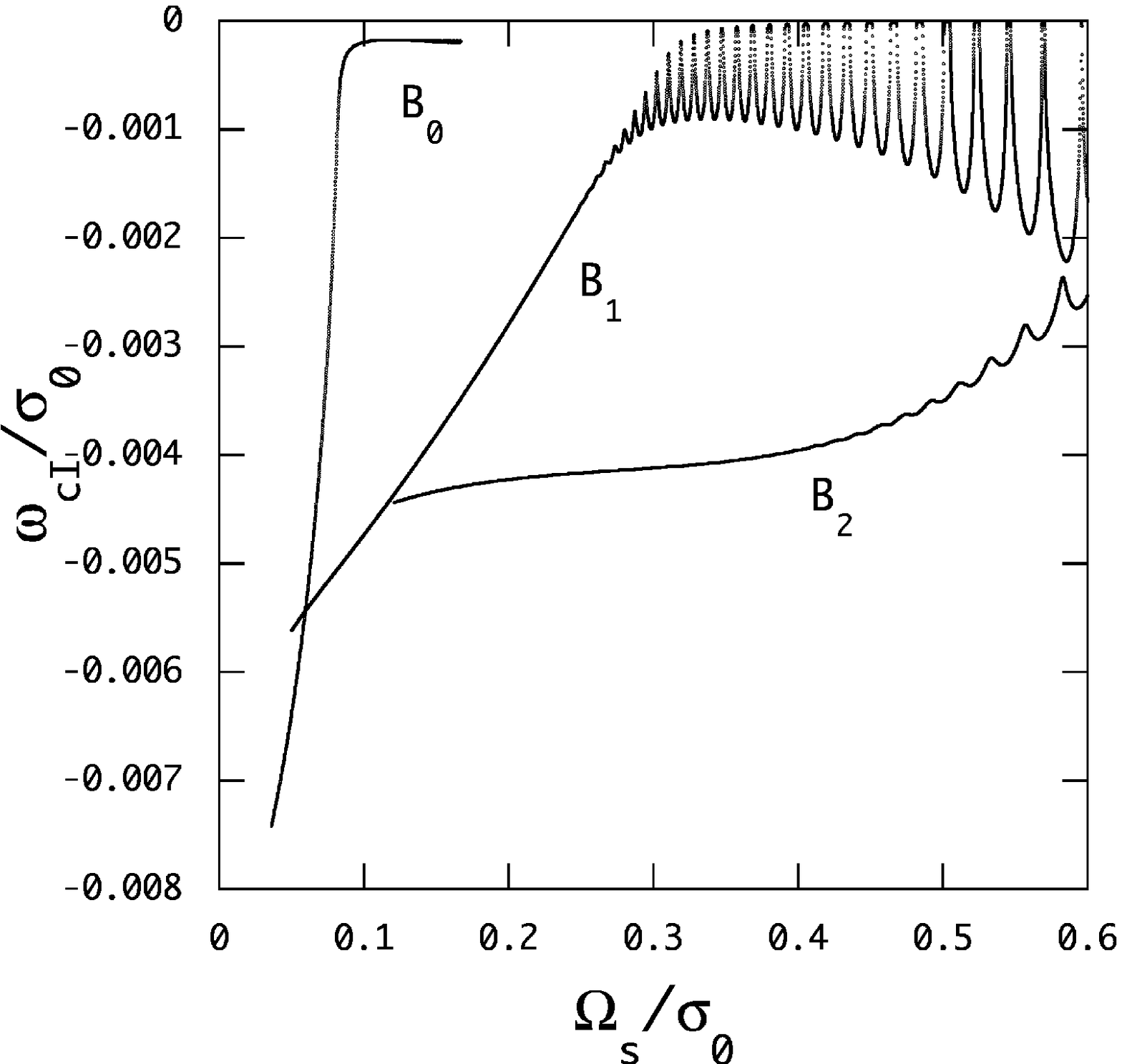}}
\resizebox{0.66\columnwidth}{!}{
\includegraphics{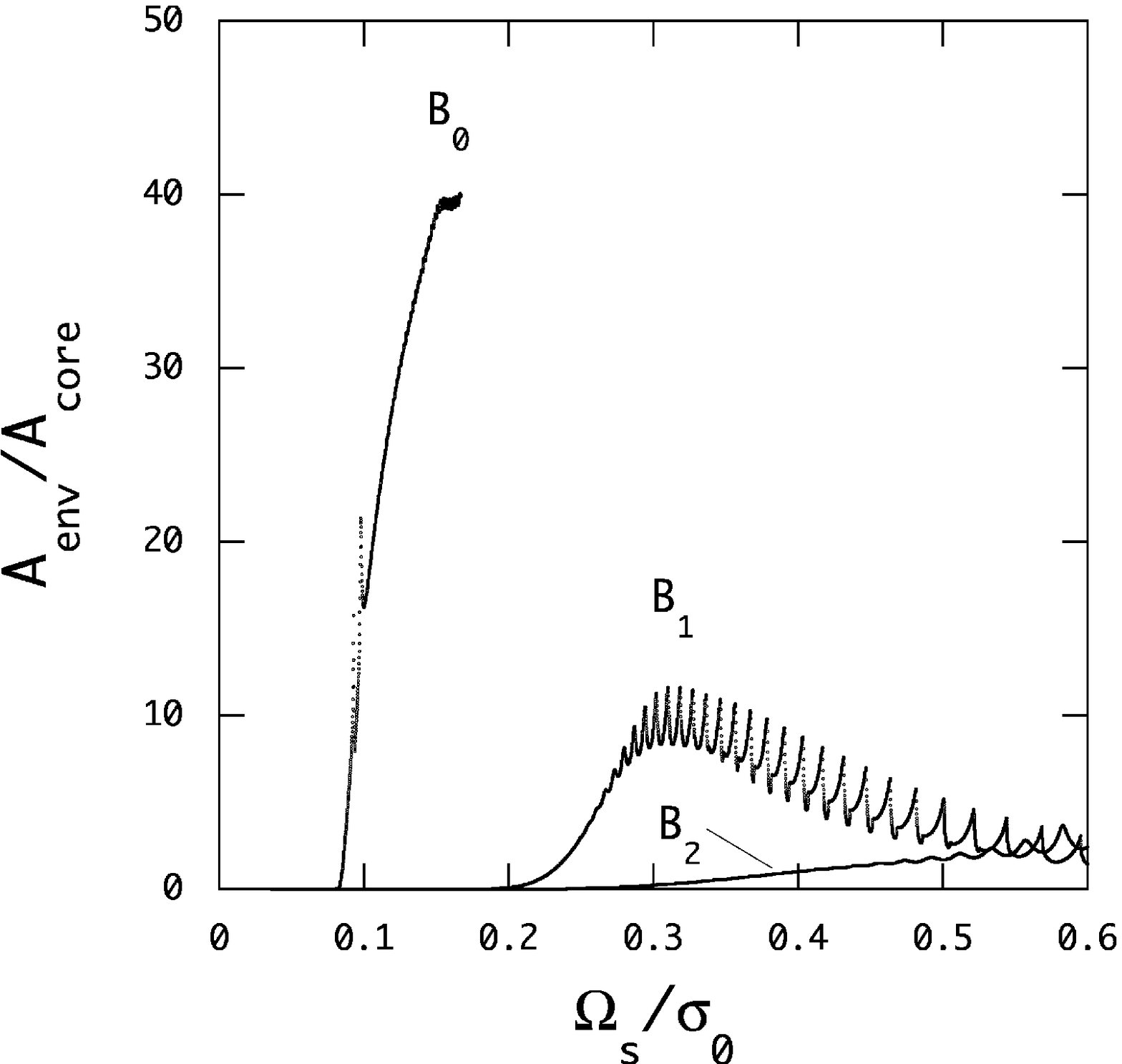}}
\caption{Same as Fig.\ref{fig:omega_m2md1b1p1} but for $b=1.2$}
\label{fig:omega_m2md1b1p2}
\end{figure*}

We compute non-adiabatic low frequency oscillations in $2M_\odot$, $4M_\odot$, and $20M_\odot$ main sequence stars, 
taking account of the effects of differential rotation on the oscillation modes.
The back ground models used for mode calculations are computed by using a stellar evolution code, originally written by 
\citet{Paczynski1970}, with OPAL opacity \citep{IglesiasRogers96} for the initial composition $X=0.7$ and $Y=0.28$.
The models have a convective core and an envelope, which consists of radiative layers 
and geometrically thin subsurface convection zones.
The method of calculation of non-adiabatic oscillations of differentially rotating stars
is the same as that used by \citet{LeeSaio93}, except that for the approximation called 
frozen-in convection we employ a prescription given by $(\nabla\cdot\pmb{F}_{\rm C})^\prime=0$,
instead of $\delta(\nabla\cdot\pmb{F}_{\rm C})=0$.
In general, the difference in the prescriptions for frozen-in convection should have no significant influences on
the stability results of $p$- and $g$-modes.
But, we found that the modal properties of low frequency modes in the convective core 
depends on the prescriptions. 
In fact, we found that for the prescription $\delta(\nabla\cdot\pmb{F}_{\rm C})=0$,
there appears low frequency modes which are destabilized by nuclear energy generation and are confined in the core.
We may call them core modes.
We found that the core modes exist even when the super-adiabatic temperature gradient $\nabla-\nabla_{ad}$ vanishes in the core
and that they do not have adiabatic counterparts where $\nabla={d\ln T/d\ln p}$ and $\nabla_{ad}=(\partial\ln T/\partial\ln p)_{ad}$ with $T$ and $p$ being the temperature and the pressure, respectively.
We therefore have decided in this paper to employ the
prescription $(\nabla\cdot\pmb{F}_{\rm C})^\prime=0$ so that we can discuss non-adiabatic OsC modes free from
the core modes.
Note that OsC modes exist only when $\nabla-\nabla_{ad}>0$ and non-adiabatic OsC modes have adiabatic counterparts.
See the Appendix A for further discussions.

To represent oscillation modes in a rotating star, we use series expansion for the perturbations.
The displacement vector $\pmb{\xi}(r,\theta,\phi,t)$ may be represented by
\be
{\xi_r\over r}=\sum_{j=1}^{j_{\rm max}}S_{l_j}Y_{l_j}^me^{\rmi\sigma t},
\ee
\be
{\xi_\theta\over r}=\sum_{j=1}^{j_{\rm max}}\left(H_{l_j}{\partial\over\partial\theta}Y_{l_j}^m
+T_{l'_j}{1\over\sin\theta}{\partial\over\partial\phi}Y_{l'_j}^m\right)e^{\rmi\sigma t},
\ee
\be
{\xi_\phi\over r}=\sum_{j=1}^{j_{\rm max}}\left(H_{l_j}{1\over\sin\theta}{\partial\over\partial\phi}Y_{l_j}^m-T_{l'_j}{\partial\over\partial\theta}Y_{l'_j}^m\right)e^{\rmi\sigma t},
\label{eq:xiphiexpand}
\ee
and the Eulerian pressure perturbation $p'(r,\theta,\phi,t)$ by
\be
p'=\sum_{j=1}^{j_{\rm max}}p'_{l_j}Y_{l_j}^me^{\rmi\sigma t},
\ee
where $Y_l^m$ is the spherical harmonic function $Y_l^m(\theta,\phi)$, and $S_l$, $H_l$, $T_{l^\prime}$, and $p_l$ are the expansion coefficients which depend only on $r$, and
$l_j=|m|+2(j-1)$ and $l'_j=l_j+1$ for even modes and $l_j=|m|+2j-1$ and $l'_j=l_j-1$ for odd modes
with $j=1,~2,~\cdots,~j_{\rm max}$ (see, e.g., \citealt{LeeSaio86}).
The parameter $j_{\rm max}$ gives the length of expansions.
Substituting these expansions into linearized basic equations, we obtain a set of linear ordinary differential equations
for the expansion coefficients  
\citep[e.g.,][]{LeeSaio86}.
The set of differential equations for non-adiabatic oscillations in differentially rotating stars
for the prescription $(\nabla\cdot\pmb{F}_{\rm C})^\prime=0$ are given in the Appendix B. 
In this study we employ the Cowling approximation (\citealt{Cowling41}), neglecting the Euler perturbation of the gravitational potential. 
We also ignore the terms associated with centrifugal force, which is justified because most of the kinetic energy
of the low frequency modes is confined into deep interior.
For the series expansions, we use $j_{\rm max}=10$ to 15, with which the frequencies and eigenfunctions become insensitive to $j_{\rm max}$.

\begin{figure*}
\resizebox{0.66\columnwidth}{!}{
\includegraphics{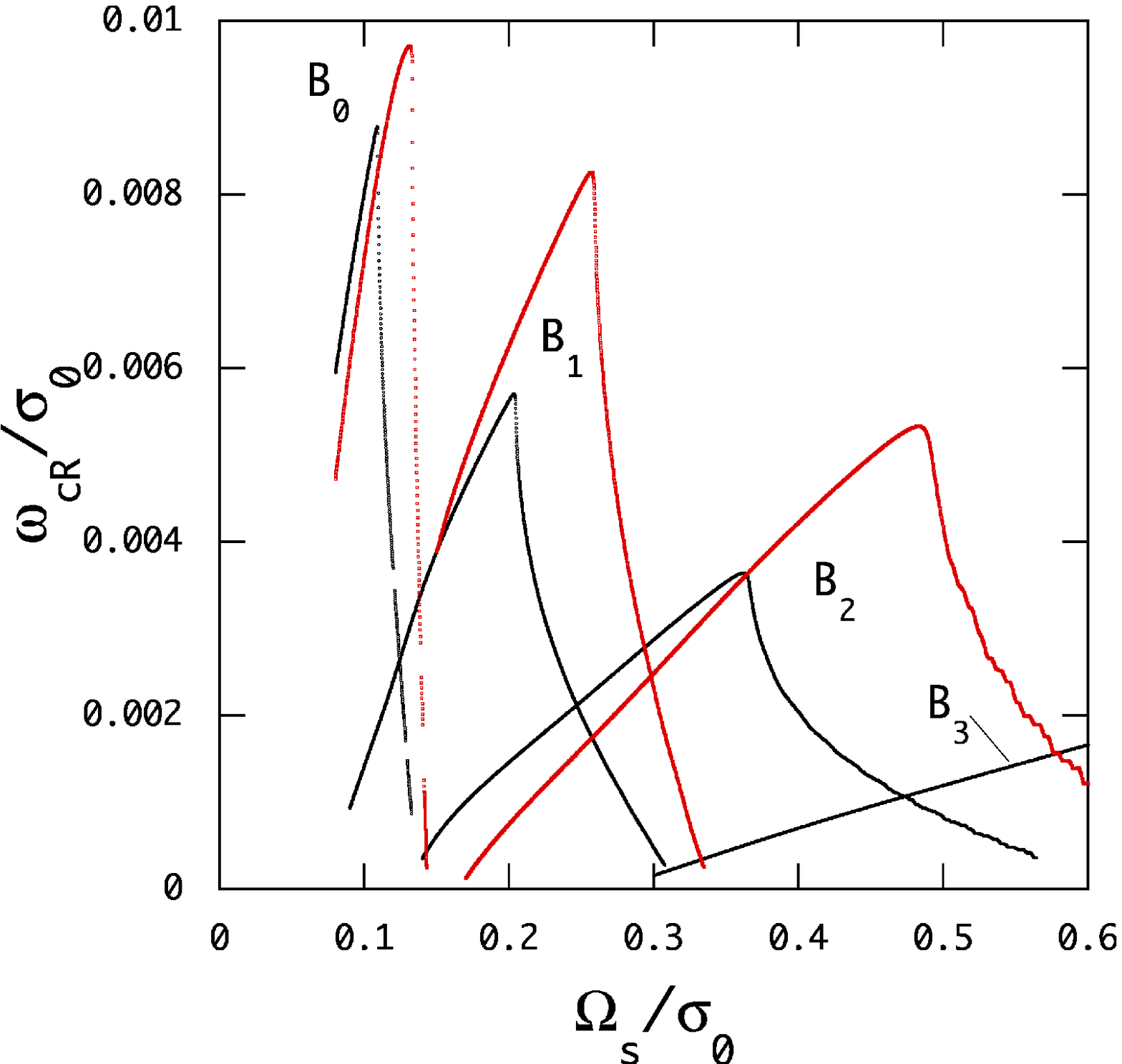}}
\resizebox{0.66\columnwidth}{!}{
\includegraphics{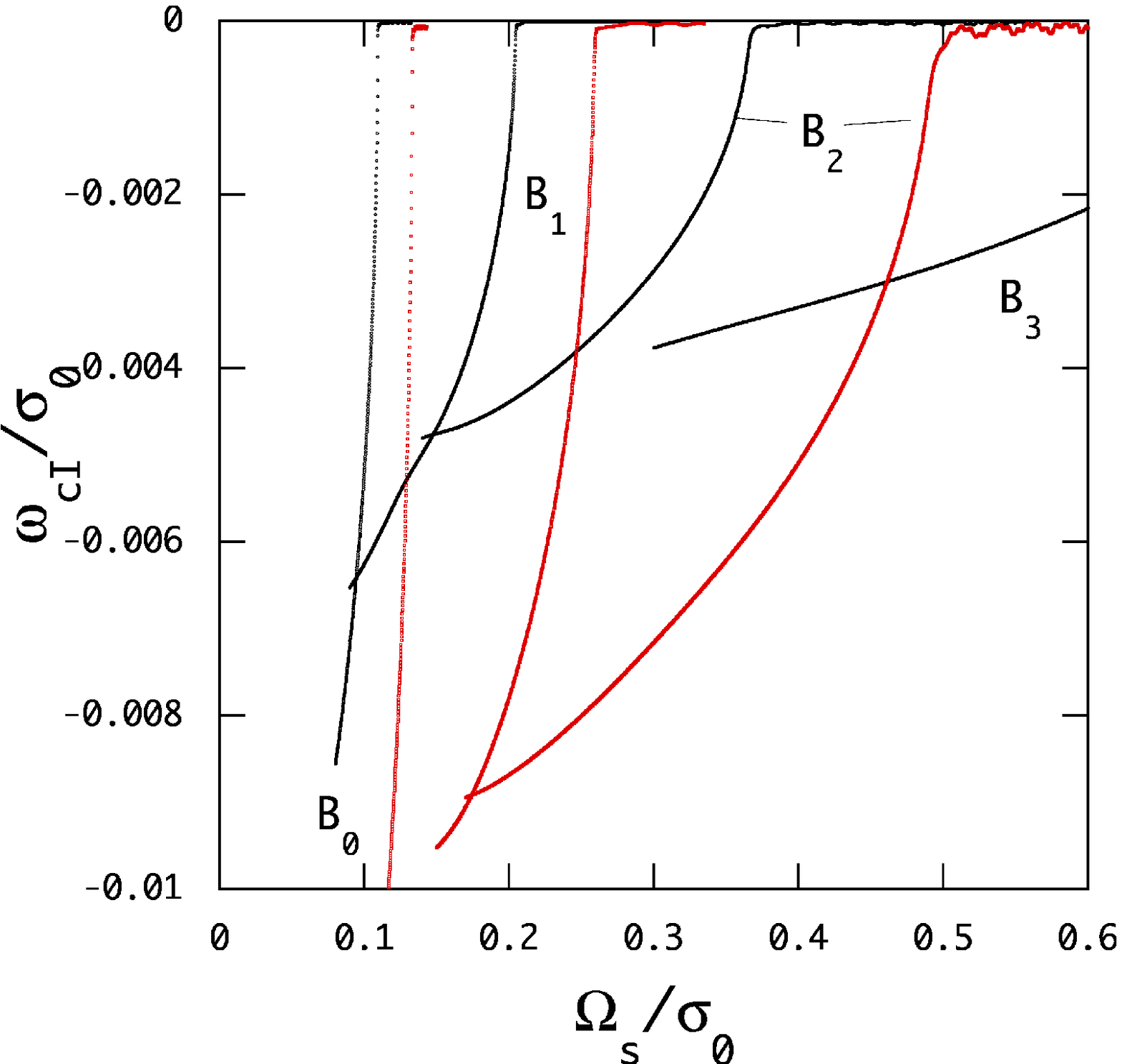}}
\resizebox{0.66\columnwidth}{!}{
\includegraphics{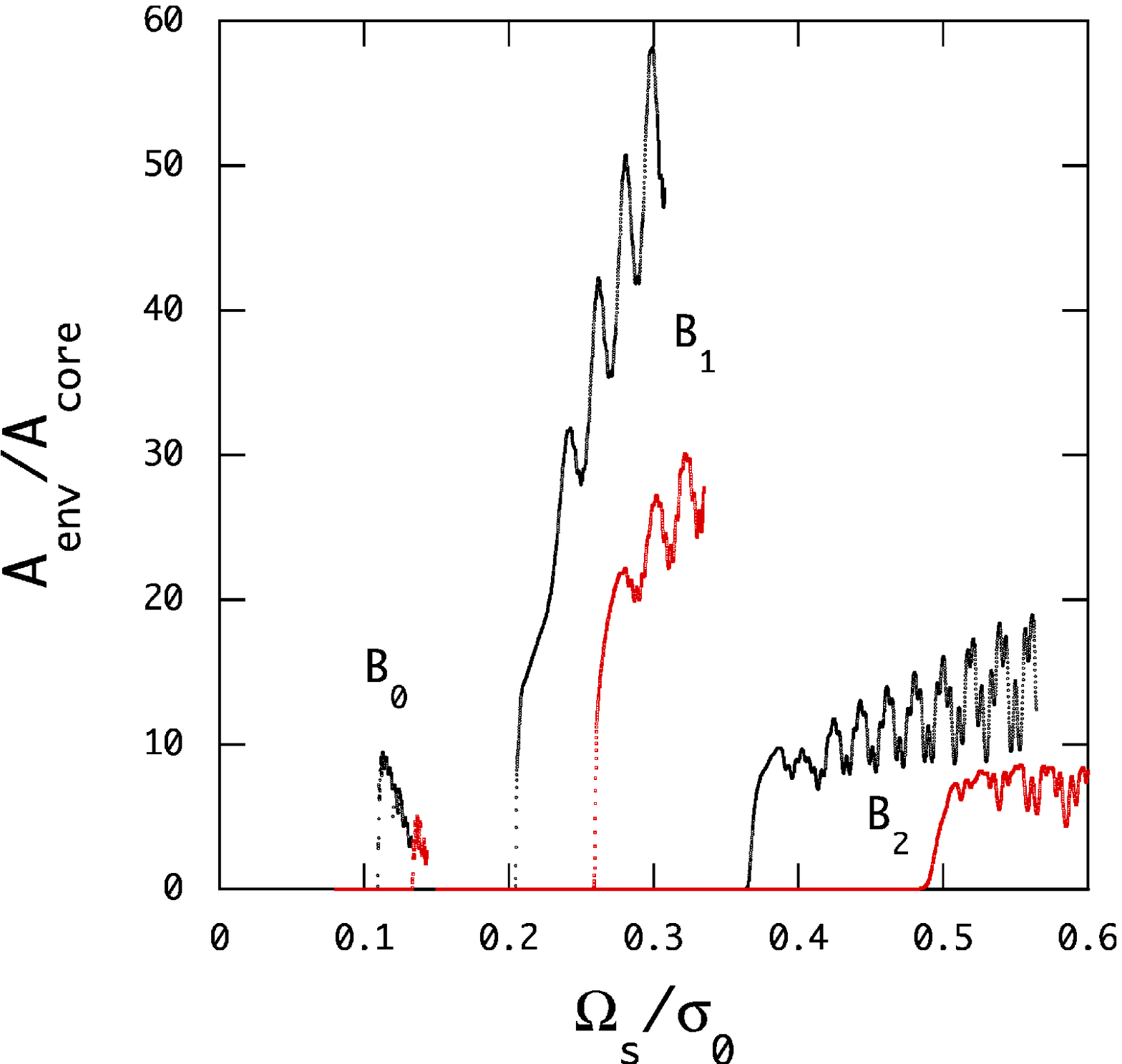}}
\resizebox{0.66\columnwidth}{!}{
\includegraphics{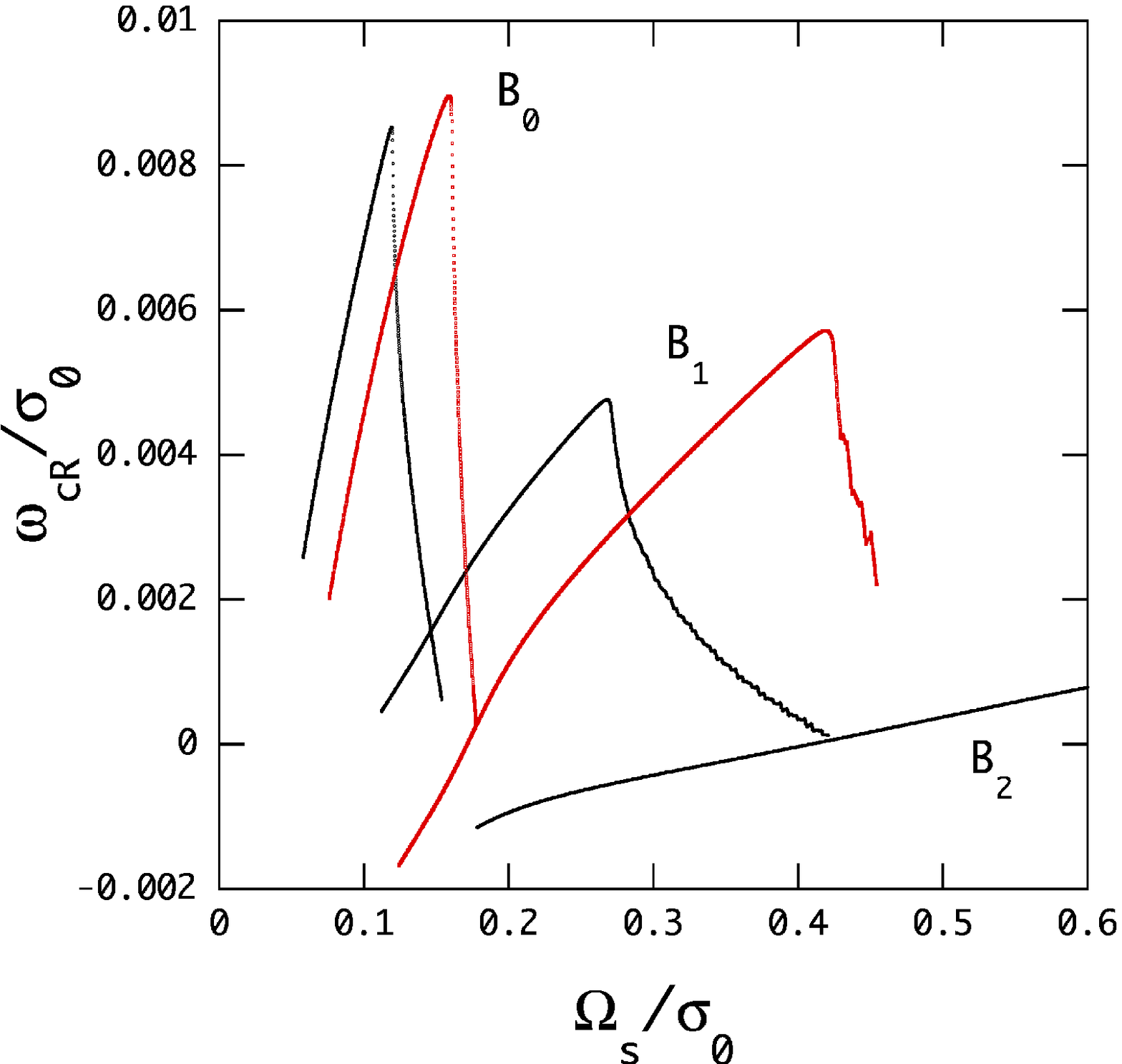}}
\resizebox{0.66\columnwidth}{!}{
\includegraphics{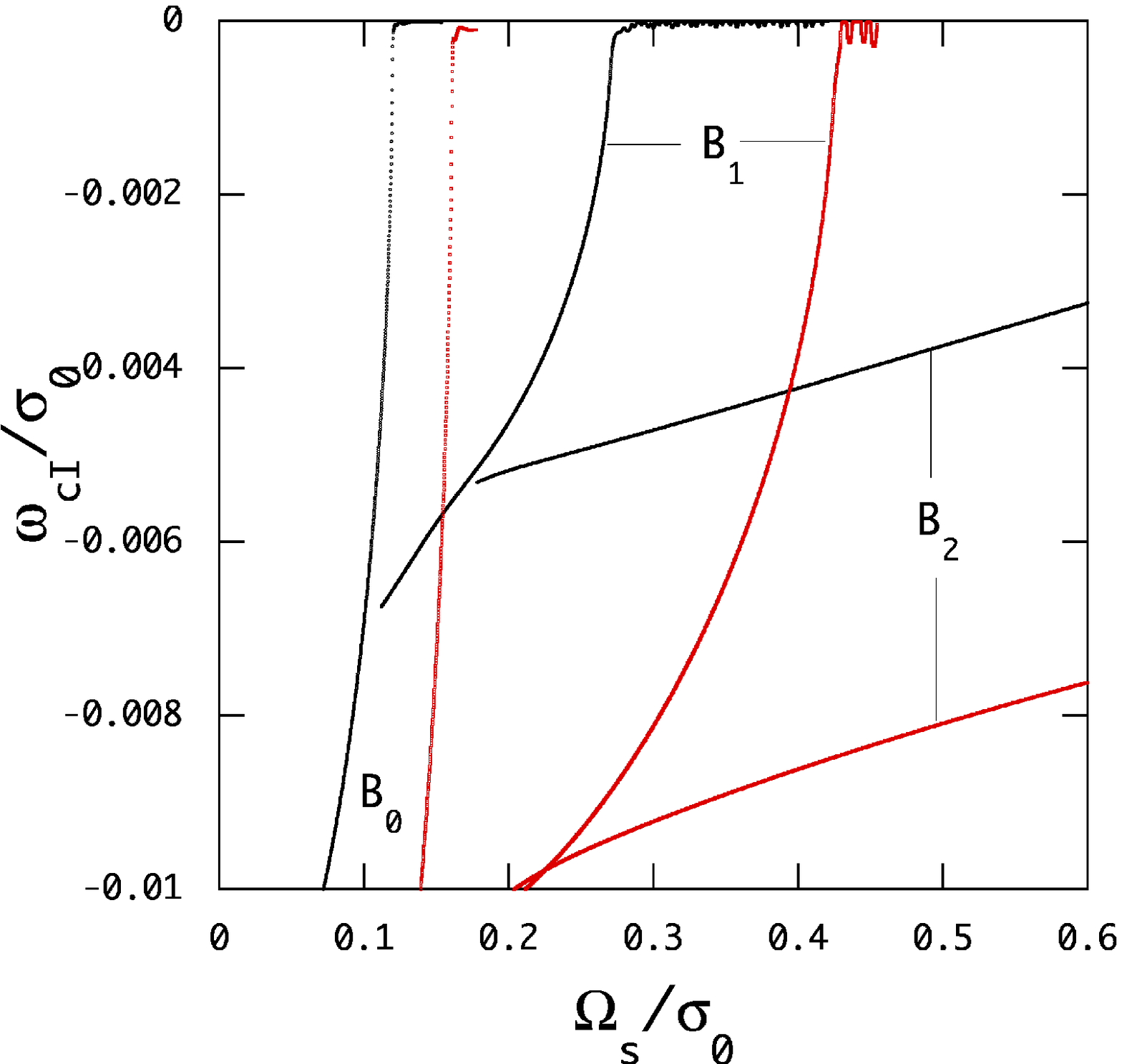}}
\resizebox{0.66\columnwidth}{!}{
\includegraphics{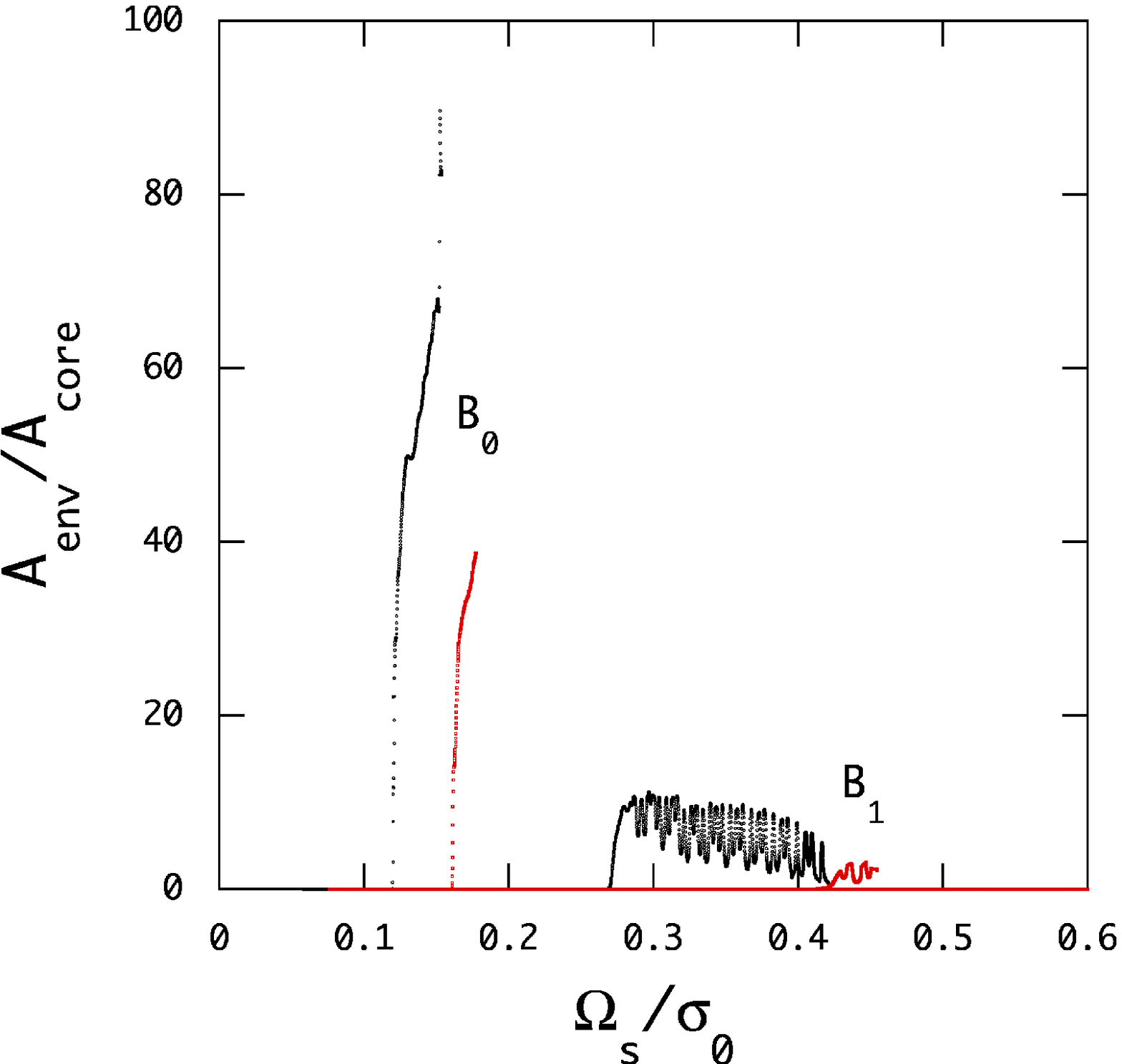}}
\caption{Complex $\overline\omega_c$ and the ratio $A_{\rm env}/A_{\rm core}$ of $m=-1$ (black dots) and $m=-2$ (red dots)
OsC modes of the $2M_\odot$ model with $X_c=0.2$ are plotted versus $\overline\Omega_s$ for $b=1.1$ (upper panels) and for $b=1.2$ (lower panels).}
\label{fig:omega_m2md154}
\end{figure*}

For differentially rotating stars, we assume a rotation law given by (e.g., \citealt{Lee88})
\be
\Omega(r)=\Omega_s\left[1+{b-1\over 1+e^{a(x-x_c)}}\right],
\label{eq:difrot}
\ee
where $x=r/R$, and $x_c$ denotes the outer boundary of the convective core, $\Omega_s$ is the rotation speed at the stellar surface, and $a$ and $b$ are parameters. 
Uniform rotation is given by $b=1$.
The condition $b>1$ implies that the core rotates faster than the envelope.
In this paper we use $a=100$, for which $\Omega(r)$ stays $\approx b\Omega_s$ for $x<x_c$
but decreases steeply to $\Omega_s$ around $x_c$.

For oscillation modes in differentially rotating stars, we use the symbol $\sigma$ to represent the angular frequency
(eigenfrequency) of oscillation in the inertial frame.
Although the inertial frame frequency $\sigma$ does not depend on $r$, the frequency $\omega=\sigma+m\Omega(r)$ in
a local co-rotating frame depends on $r$ (but ${\rm Im}(\omega)={\rm Im}(\sigma)$).
If we let $\omega_c$ denote an oscillation frequency  
in the co-rotating frame of the core, 
the frequency $\omega_s$ in the co-rotating frame of the envelope is given by
\be
\omega_s=\omega_c-m(\Omega_c-\Omega_s)\approx \omega_c-m\Omega_s(b-1),
\label{eq:omega_env}
\ee
where $\omega_c=\sigma+m\Omega_c$ with $\Omega_c=\Omega(0)$ and $b\approx \Omega_c/\Omega_s$.
If a prograde convective mode has a frequency $\omega_c>0$ for $m<0$ in the core,
the frequency $\omega_c$ should be shifted to  
$\omega_s$ in the envelope.
Then the $g$-modes in resonance with $\omega_s$ in the envelope should have a
radial order much lower than that of a $g$-mode having the frequency $\omega_c$ in the envelope,
which is one of the important effects of differential rotation on the modal properties of low frequency modes.

In this paper, we let $\overline\omega$ and $\overline\Omega_s$ denote
dimensionless frequencies defined as $\overline\omega=\omega/\sigma_0$ and 
$\overline\Omega_s=\Omega_s/\sigma_0$ where $\sigma_0=\sqrt{GM/R^3}$ with $M$ and $R$ being
the mass and radius of the star and $G$ the gravitational constant.
We also let $\omega_{\rm R}$ and $\omega_{\rm I}$ denote
the real and imaginary part of the complex frequency $\omega=\omega_{\rm R}+\rmi\omega_{\rm I}$, respectively, and
we note that unstable modes have negative $\omega_{\rm I}$.

\section{OsC Modes in Differentially Rotating Main Sequence Stars}

For modal analysis, we use main sequence models with $X_c=0.7$ (ZAMS model) and $X_c=0.2$ (evolved model), where $X_c$ is the mass fraction of hydrogen at the stellar centre.
In the convective core, we assume a finite value for the superadiabatic temperature gradient $\nabla-\nabla_{ad}=10^{-5}$ as in \citet{LeeSaio20}.
It is difficult to correctly estimate the value of $\nabla-\nabla_{ad}$ in the core of rotating stars 
(e.g., \citealt{Stevenson79}).
We guess that $\nabla-\nabla_{ad}$ for rotating stars could be much larger than that estimated for non-rotating stars.

\subsection{$2M_\odot$ models}

To compare with the OsC modes obtained by \citet{LeeSaio20} who assumed
$\delta(\nabla\cdot\pmb{F}_C)=0$ for the convective energy flux,
we have computed low $m$ OsC modes in $2M_\odot$ main sequence models assuming $(\nabla\cdot\pmb{F}_C)^\prime=0$.
Since OsC modes in uniformly rotating stars do not effectively excite $g$-modes as shown by
\citet{LeeSaio20},
we consider OsC modes in weakly differentially rotating stars given by $b=1.1$ or $b=1.2$.
The complex eigenfrequency
$\overline\omega_c$ and the ratio $A_{\rm env}/A_{\rm core}$ of $m=-1$ and $m=-2$ OsC modes in the ZAMS model are
plotted as a function of $\overline{\Omega}_s$
in Fig.\ref{fig:omega_m2md1b1p1} for $b=1.1$ and in Fig.\ref{fig:omega_m2md1b1p2} for $b=1.2$, where
the OsC modes are labeled $B_n$ with $n$ being the number of radial nodes of $S_{l_1}$ in the convective core (see \citealt{Lee19}).
As $\overline\Omega_s$ increases from $\overline\Omega_s\sim 0$, $\overline\omega_{c{\rm R}}$ of a $B_n$-mode
increases to reach a maximum and then decreases, describing a peaked curve $\overline\omega_{c{\rm R}}(\overline\Omega_s)$.
As the radial order $n$ of $B_n$-modes increases, the height and width of the peak becomes lower and broader and
the peak itself shifts to higher $\overline\Omega_s$.
On the other hand, the imaginary part $|\overline\omega_{c{\rm I}}|$, in general, decreases as $\overline\Omega_s$ increases, 
indicating that the OsC modes are stabilized by rotation.
Effective stabilization by rotation, however,
occurs only for low radial order $B_n$-modes and as $n$ increases, stabilization effects becomes weaker, 
that is, $|\overline\omega_{c{\rm I}}|$ depends on $\overline\Omega_s$ only weakly and tends to stay large.
When $|\overline\omega_{c{\rm I}}|$ of OsC modes in the core becomes vanishingly small as a result of rotational stabilization, there occurs resonant excitation of envelope $g$-modes by OsC modes and we have the ratio $A_{\rm env}/A_{\rm core}\gtrsim 1$.
Even if $|\overline\omega_{c{\rm I}}|$ of OsC modes is not vanishingly small, however,
resonant excitation of $g$-modes can occur when the Doppler shifted frequency $\overline\omega_{s{\rm R}}$ of the modes
is large enough to be coupled with low radial order $g$-modes in the envelope (\citealt{LeeSaio20}).
Note that resonances between the OsC mode and envelope $g$-modes 
manifest themselves as quasi-periodic fluctuations of $\overline\omega_c$ and $A_{\rm env}/A_{\rm core}$ as a function of $\overline\Omega_s$ unless the radial orders of $g$-modes are extremely high.
We also find that resonant excitation of $g$-modes takes place even if OsC modes have a co-rotation point 
defined by $\overline\omega_{\rm R}(r)=0$, the existence of which is suggested by $\overline\omega_{c{\rm R}}<0$.
For example, for the $m=-2$ and $b=1.1$ $B_4$-mode, we obtain $A_{\rm env}/A_{\rm core}\gtrsim 1$
when $\overline\Omega_s\gtrsim 0.4$, for which $\overline\omega_{c{\rm R}}<0$.
Note that we cannot properly compute OsC modes when $\overline\omega_{c}\approx 0$, 
which is the reason why we had to stop computing some of OsC modes beyond certain values of $\overline\Omega_s$.

We carry out similar computations of low $m$ OsC modes for the evolved model with $X_c=0.2$ and the results for $b=1.1$
are shown in Fig.\ref{fig:omega_m2md154}.
Because the radius of the evolved model is larger than that of the ZAMS model, 
the stabilizing effect of core rotation $\Omega_c\approx b\Omega_s=b\overline\Omega_s\sigma_0$ on the OsC modes
for the evolved model is weaker than for the ZAMS model for a given value of $\overline\Omega_s$.
Note that $\sigma_0$ for the former is smaller than for the latter.
For $\overline\Omega_s\lesssim 0.6$, the low radial order $B_n$-modes in the figure are almost all well stabilized to
have vanishingly small $|\overline\omega_{c{\rm I}}|$
and show resonant fluctuations of $\overline\omega_c$
and $A_{\rm env}/A_{\rm core}$ as a function of $\overline\Omega_s$.
Note that the $m=-1$ $B_3$-mode is not strongly stabilized by rotation and does not excite envelope $g$-modes for $\overline\Omega_s\lesssim 0.6$.
It is also interesting to note that the $B_0$-modes can excite $g$-modes even for $b=1.1$, which does not 
occur in the ZAMS model.

\begin{figure}
\resizebox{0.9\columnwidth}{!}{
\includegraphics{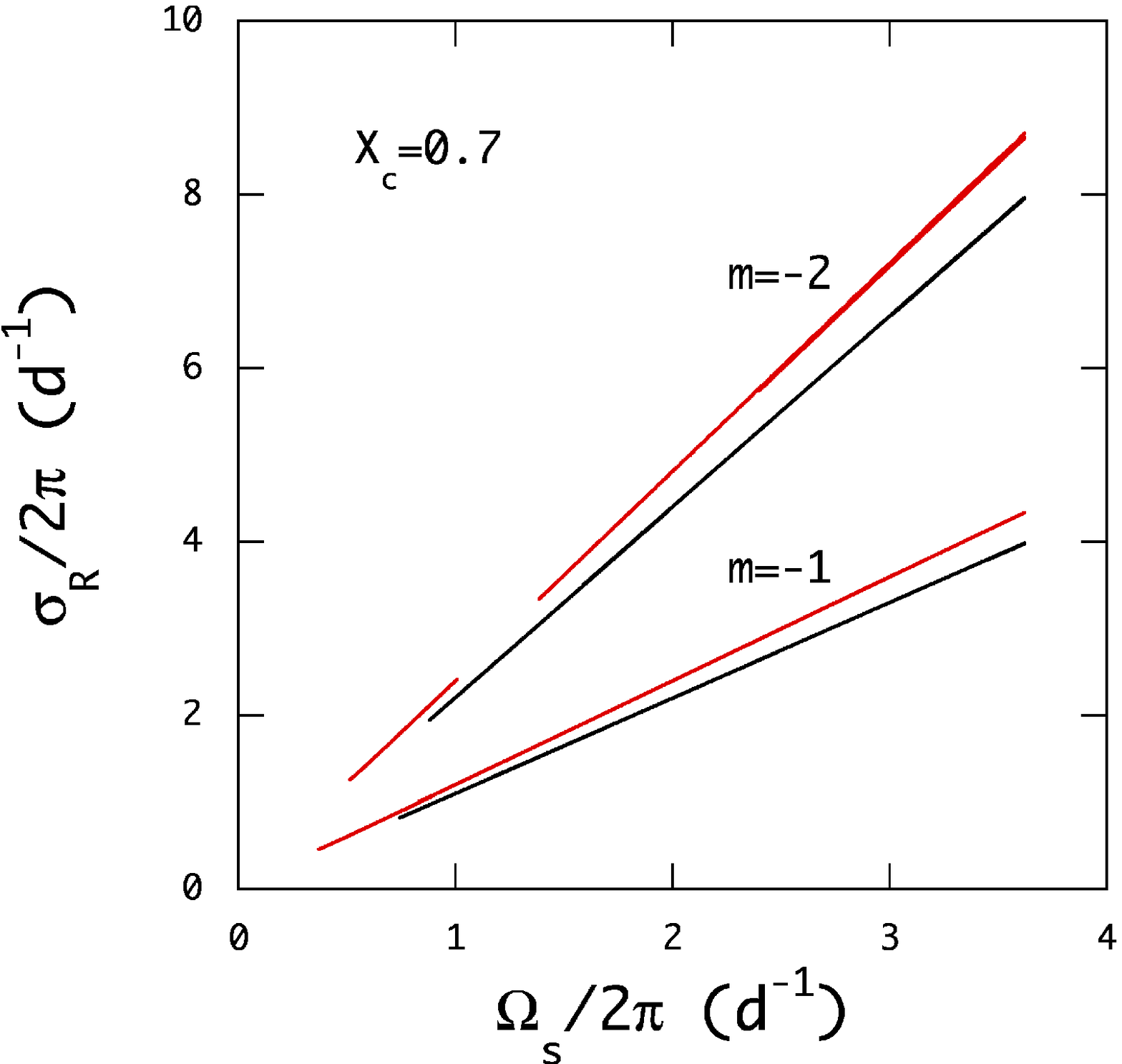}}
\resizebox{0.9\columnwidth}{!}{
\includegraphics{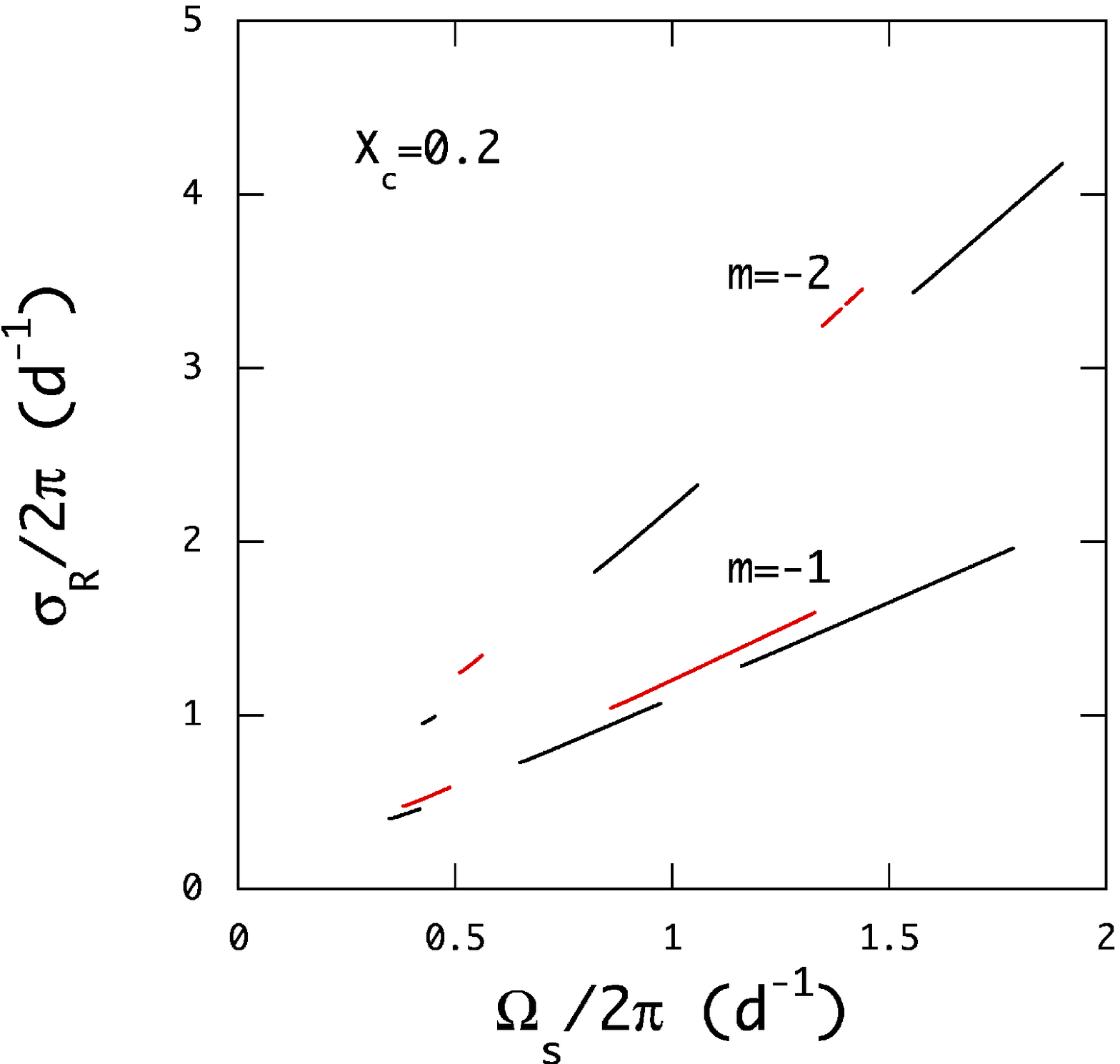}}
\caption{Oscillation frequency $\sigma/2\pi$ of the OsC modes with $A_{\rm env}/A_{\rm core}\ge1$
versus rotation frequency $\Omega_s/2\pi$ for the $2M_\odot$ ZAMS model ($X_c=0.7$) and the evolved model ($X_c=0.2$)
for $b=1.1$ (black dots) and $b=1.2$ (red dots), where the OsC modes for $\overline\Omega_s\le 0.6$ are
plotted.
}
\label{fig:sig-ome_m2}
\end{figure}

In Fig.\ref{fig:sig-ome_m2}, the inertial frame frequency
$\sigma_{\rm R}/2\pi$ of the OsC modes that excite envelope $g$-modes
so that $A_{\rm env}/A_{\rm core}\ge1$ is plotted against $\Omega_s/2\pi$
for the $2M_\odot$ ZAMS model ($X_c=0.7$) and evolved model ($X_c=0.2$).
The figure shows that the frequency $\sigma_{\rm R}$ of the OsC modes is approximately proportional to $|m\Omega_s|$, 
and this comes from the fact
that $\sigma_{\rm R}=\omega_{c{\rm R}}+m\Omega_c\approx m\Omega_c\approx mb\Omega_s$ 
since $|\omega_{c{\rm R}}|\ll|m\Omega_c|$ for the OsC modes. 
We thus obtain $\sigma_{m=-2}\approx 2\sigma_{m=-1}$ for the OsC modes.
The frequency $\sigma_{\rm R}$ of the OsC modes at a given $\overline\Omega_s$ for $b=1.2$ is slightly higher than that for $b=1.1$.
For the ZAMS model, we find that in wide ranges of $\Omega_s/2\pi$ both $m=-1$ and $m=-2$ OsC modes simultaneously excite $g$-modes although there exists a break of $\Omega_s/2\pi$ in which no effective $g$-mode excitation occurs
for the $m=-2$ OsC modes for $b=1.2$.
Since the $B_0$-modes excite envelope $g$-modes for $b=1.2$,
the lower limit to $\Omega_s/2\pi$ for the OsC modes to have $A_{\rm env}/A_{\rm core}\gtrsim 1$
extends to smaller values, compared to that for $b=1.1$.
For the evolved model, on the other hand, simultaneous excitation of $m=-1$ and $m=-2$ $g$-modes occurs only limited intervals
of $\Omega_s/2\pi$.
The difficulty in $g$-mode excitation by OsC modes in the evolved model may be caused by 
the $\mu$-gradient zone outside the convective core.

The general properties of OsC modes obtained by assuming $(\nabla\cdot\pmb{F}_C)^\prime=0$ are quite similar to
those of OsC modes calculated by \citet{LeeSaio20} assuming $\delta(\nabla\cdot\pmb{F}_C)=0$, except that 
for $(\nabla\cdot\pmb{F}_C)^\prime=0$ we find no OsC modes that follow the relation $\overline\omega_{c{\rm R}}\propto\overline\Omega_c\approx b\overline\Omega_s$ when they tend towards complete stabilization with increasing $\overline\Omega_s$.
Note that the ratio $\omega_{{\rm R}}/\Omega$ is approximately constant for inertial modes.

\subsection{$4M_\odot$ models}

\begin{figure}
\resizebox{0.9\columnwidth}{!}{
\includegraphics{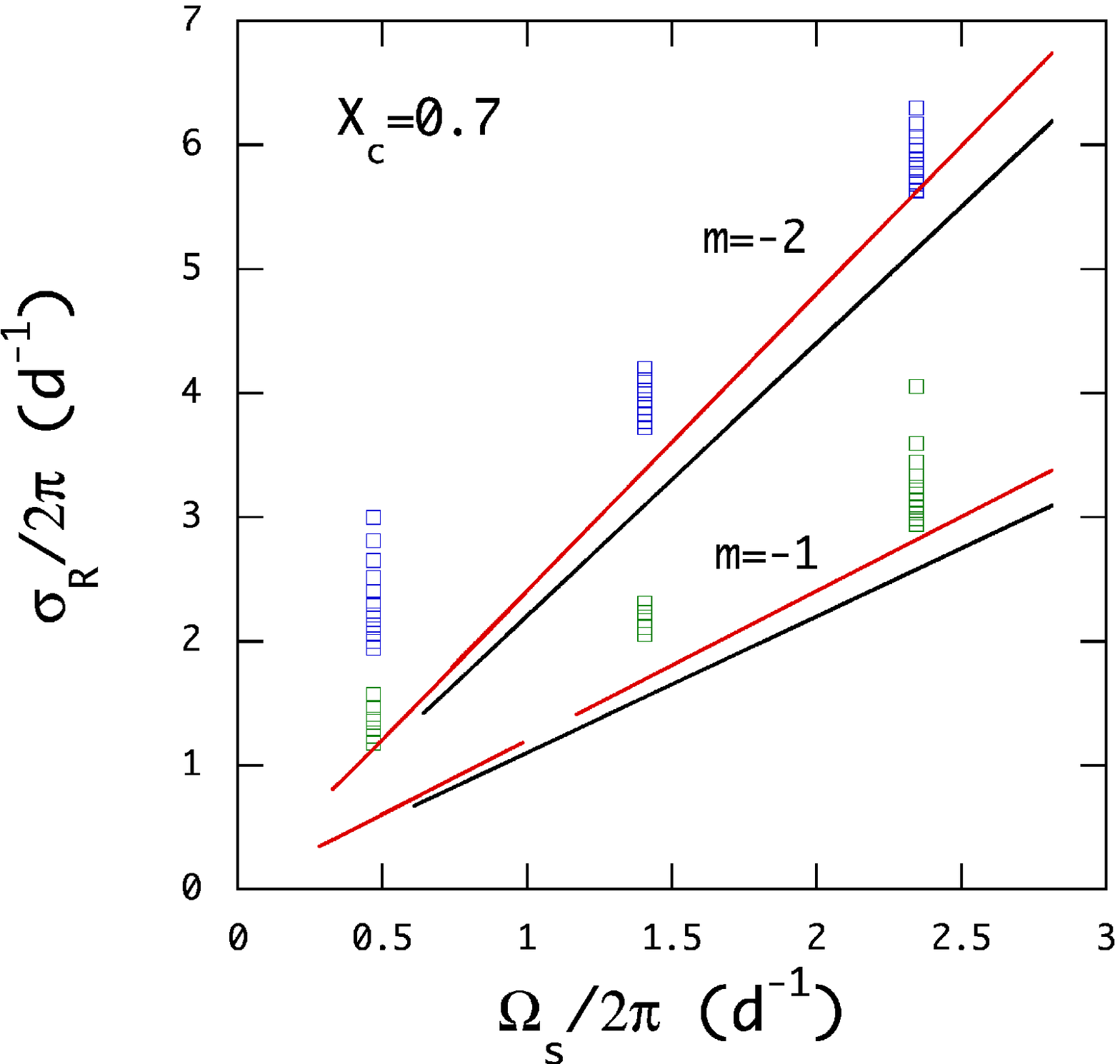}}
\resizebox{0.9\columnwidth}{!}{
\includegraphics{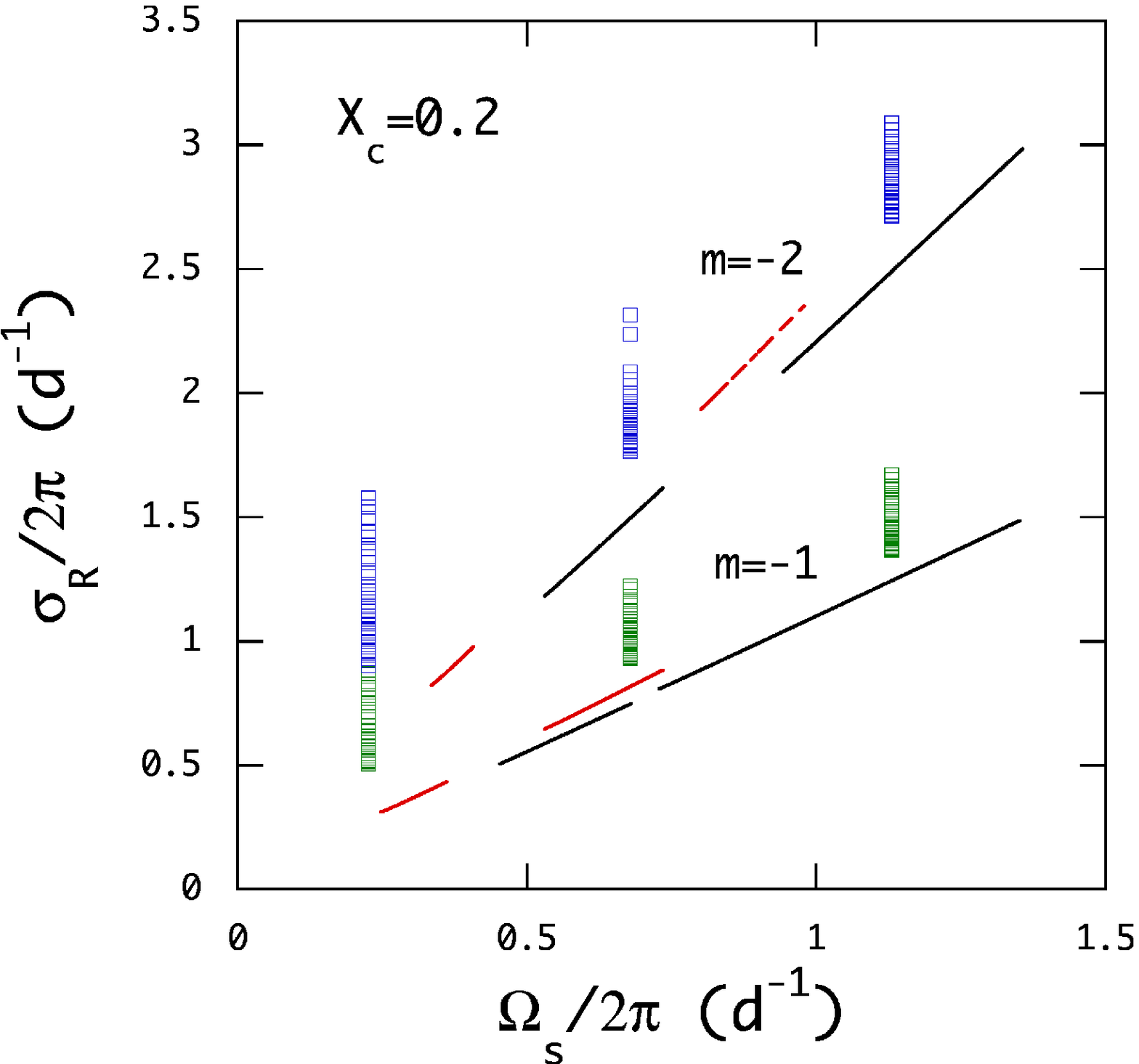}}
\caption{Same as Fig.\ref{fig:sig-ome_m2} but for $4M_\odot$ main sequence models.
The vertically aligned open squares indicate $m=-1$ (green) and $m=-2$ (blue) prograde sectoral $g$-modes excited by the opacity bump mechanism for $\overline\Omega_s=0.1$, 0.3, and 0.5 for $b=1.2$.
}
\label{fig:sigmar}
\end{figure}

\begin{figure}
\resizebox{0.9\columnwidth}{!}{
\includegraphics{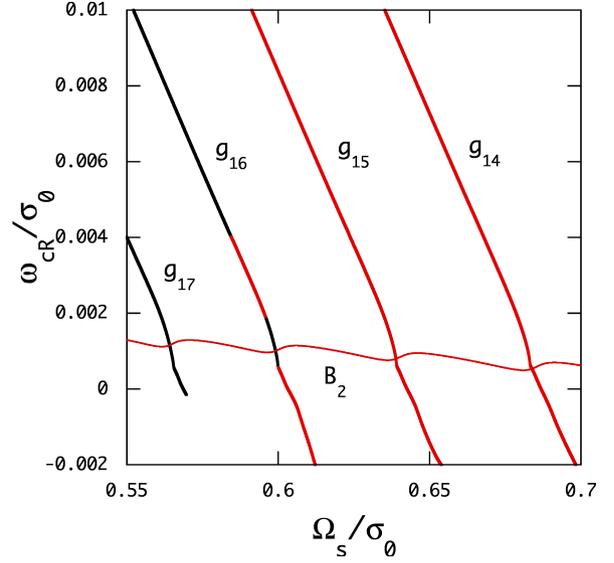}}
\caption{$\overline\omega_{c{\rm R}}$ as a function of $\overline\Omega_s$ for the $m=-1$ $B_2$-mode in the core and $g$-modes
in the envelope of the $4M_\odot$ ZAMS model for $b=1.2$ where the black lines and red lines indicate stable and unstable , respectively.
}
\label{fig:osc-gmode}
\end{figure}

\begin{figure}
\resizebox{0.9\columnwidth}{!}{
\includegraphics{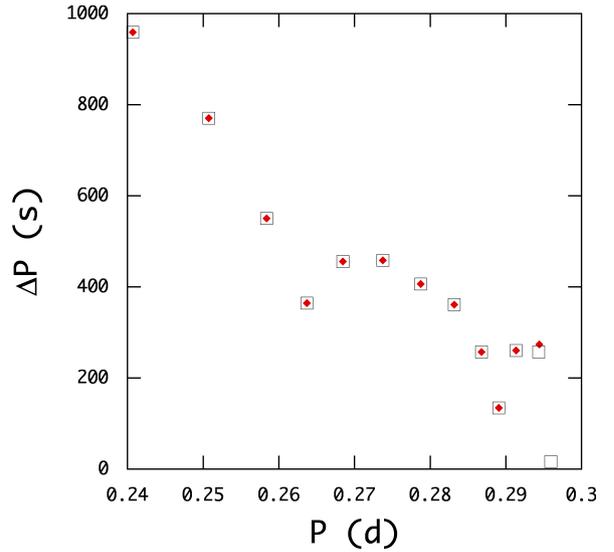}}
\caption{Period spacing $\Delta P$ versus the period $P$ of the opacity driven $g$-modes and OsC $B_2$-mode
of the $4M_\odot$ ZAMS star
at $\overline\Omega_s=0.6$ for $m=-1$ and $b=1.2$ where $\Delta P=P_{n+1}-P_n$ and $P=(P_{n+1}+P_n)/2$
with $P_n$ being the period of the low frequency modes.
The open squares are for $\Delta P$ with the OsC mode included and the filled red diamonds for $\Delta P$
with the OsC mode excluded.
}
\label{fig:pdp}
\end{figure}

\begin{figure}
\resizebox{0.9\columnwidth}{!}{
\includegraphics{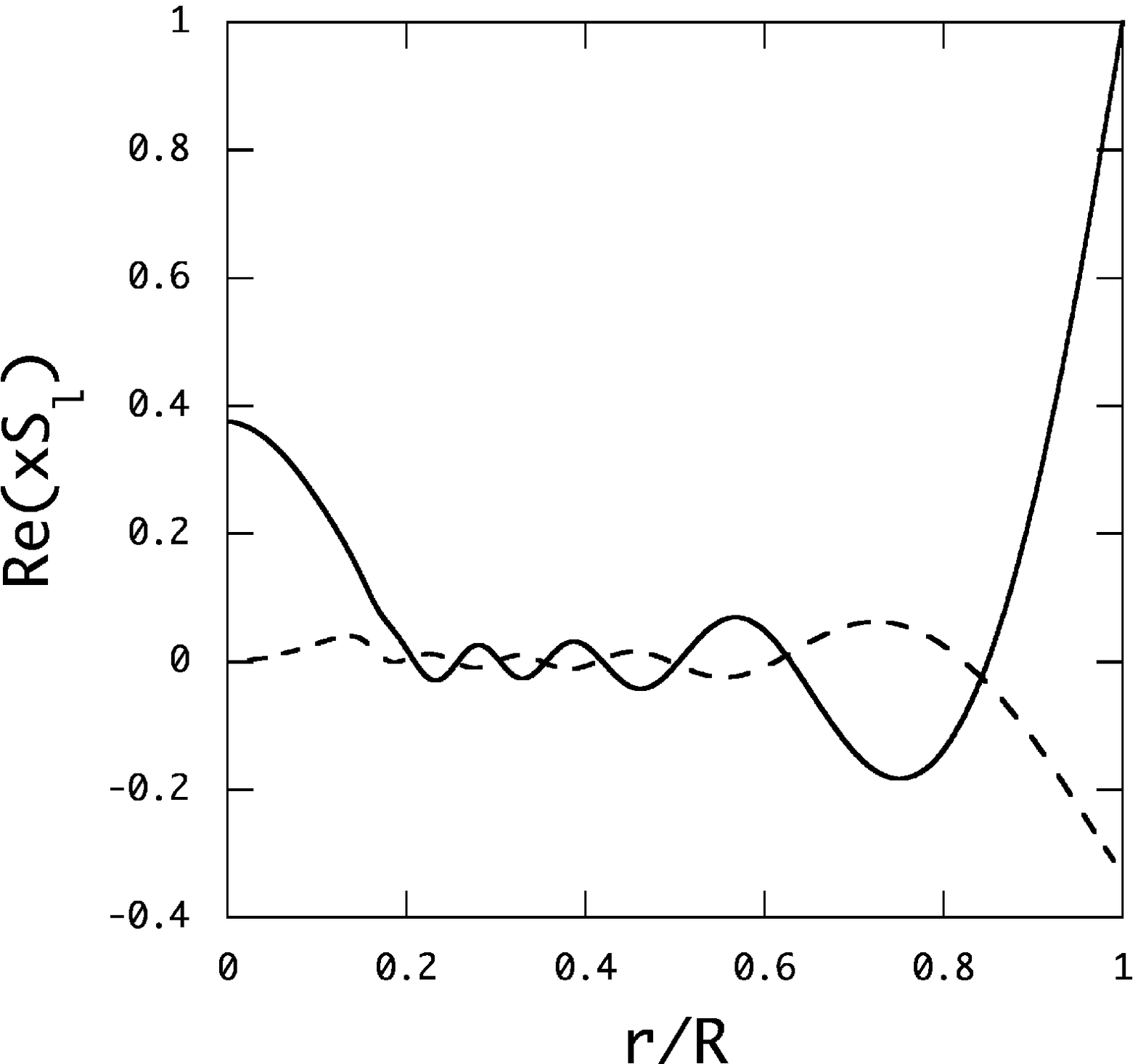}}
\caption{${\rm Re}(xS_l)$ versus $x=r/R$ for the $m=-1$ $g_8$-mode of the $4M_\odot$ ZAMS model at $\overline\Omega_s=0.6$
for $b=1.2$ where $\sigma_{\rm R}/2\pi=3.919~({\rm d}^{-1})$ and the solid line and dashed line are for $l=1$ and 3, respectively.
}
\label{fig:b2-inertial}
\end{figure}

For slowly pulsating B (SPB) stars, we compute OsC modes of $4M_\odot$ main sequence stars for $X_c=0.7$ (ZAMS model) and
$X_c=0.2$ (evolved model). 
The behavior of low $m$ OsC modes as a function of $\overline\Omega_s$ for the $4M_\odot$ models 
is quite similar to that found for the $2M_\odot$ main sequence models.
This is shown by Fig. \ref{fig:sigmar} in which
$\sigma_{\rm R}/2\pi$ of the OsC modes having $A_{\rm env}/A_{\rm core}\ge 1$
are plotted against $\Omega_s/2\pi$.
Note that the vertically aligned open squares in the figure indicate prograde sectoral $g$-modes excited by the opacity bump mechanism for $\overline\Omega_s=0.1$, 0.3, and 0.5.
Again for the OsC modes in SPB stars, we obtain $\sigma_{\rm R} \propto m\Omega_s$ 
so that $\sigma_{m=-2}\approx 2\sigma_{m=-1}$.
The OsC modes with $A_{\rm env}/A_{\rm core}\ge1$ for $b=1.2$ extend to lower rotation frequencies $\Omega_s/2\pi$ than those for $b=1.1$, and this extension occurs both for the ZAMS model and for the evolved model.

\begin{figure*}
\resizebox{0.66\columnwidth}{!}{
\includegraphics{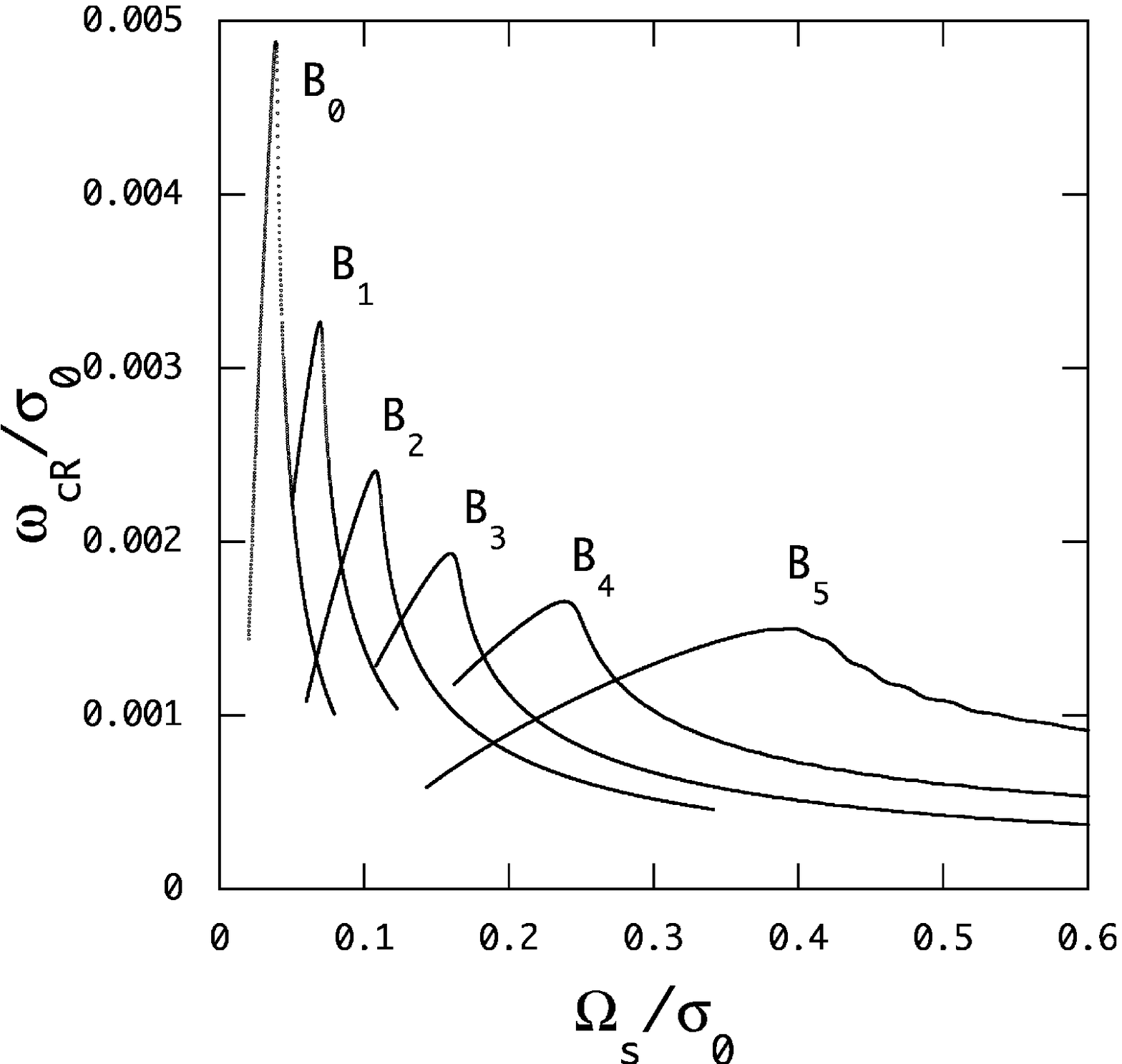}}
\resizebox{0.66\columnwidth}{!}{
\includegraphics{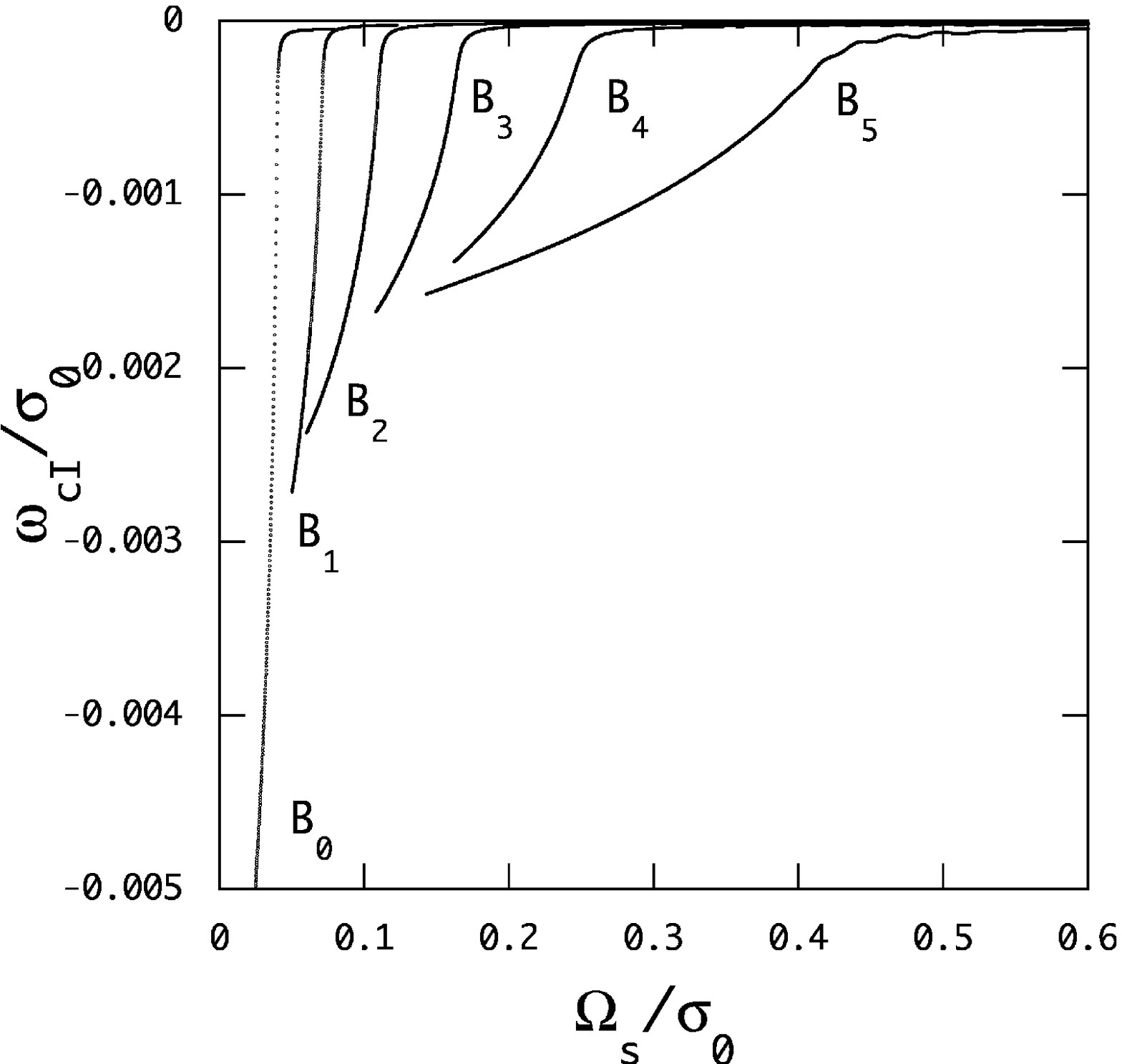}}
\resizebox{0.66\columnwidth}{!}{
\includegraphics{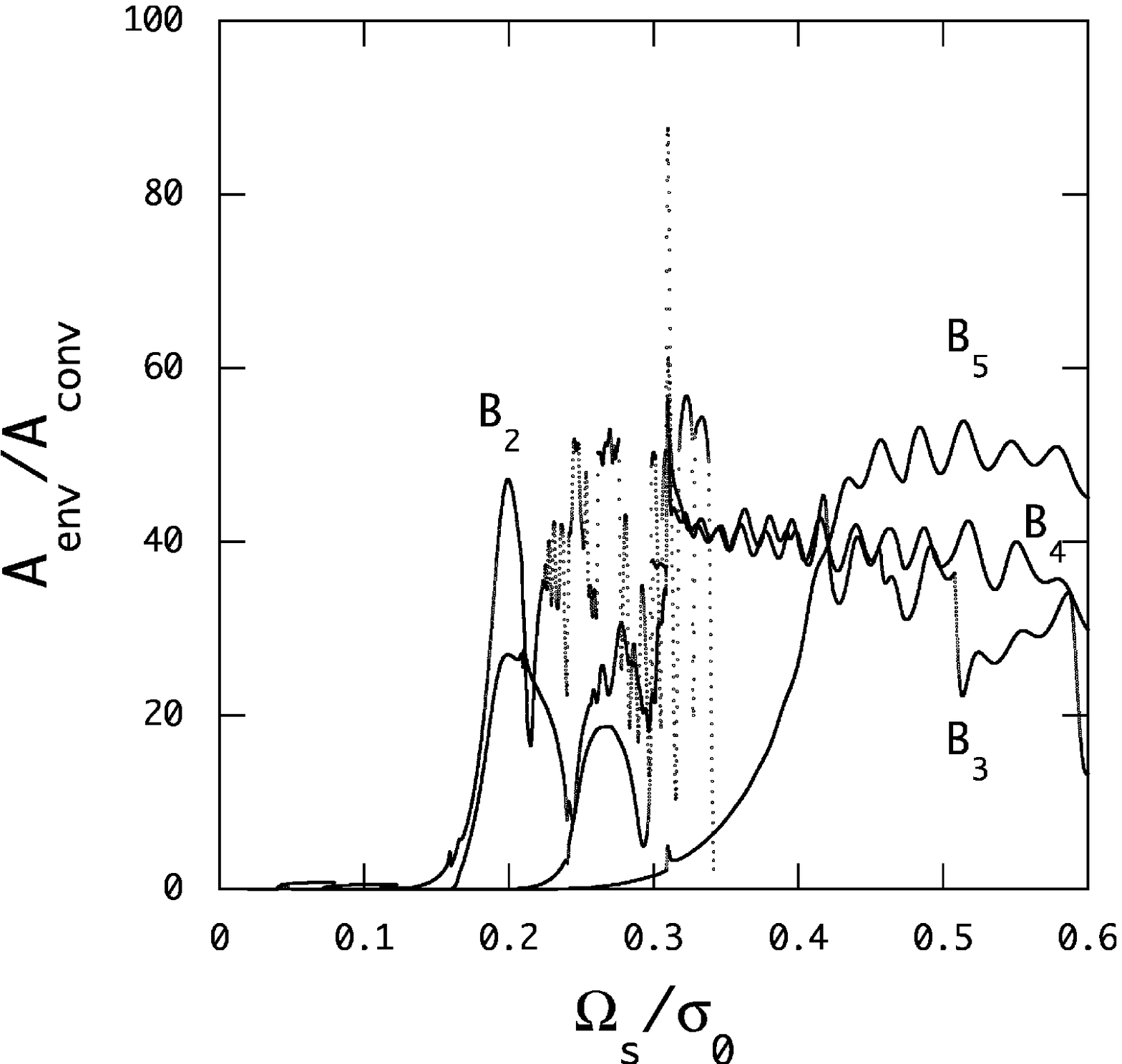}}
\resizebox{0.66\columnwidth}{!}{
\includegraphics{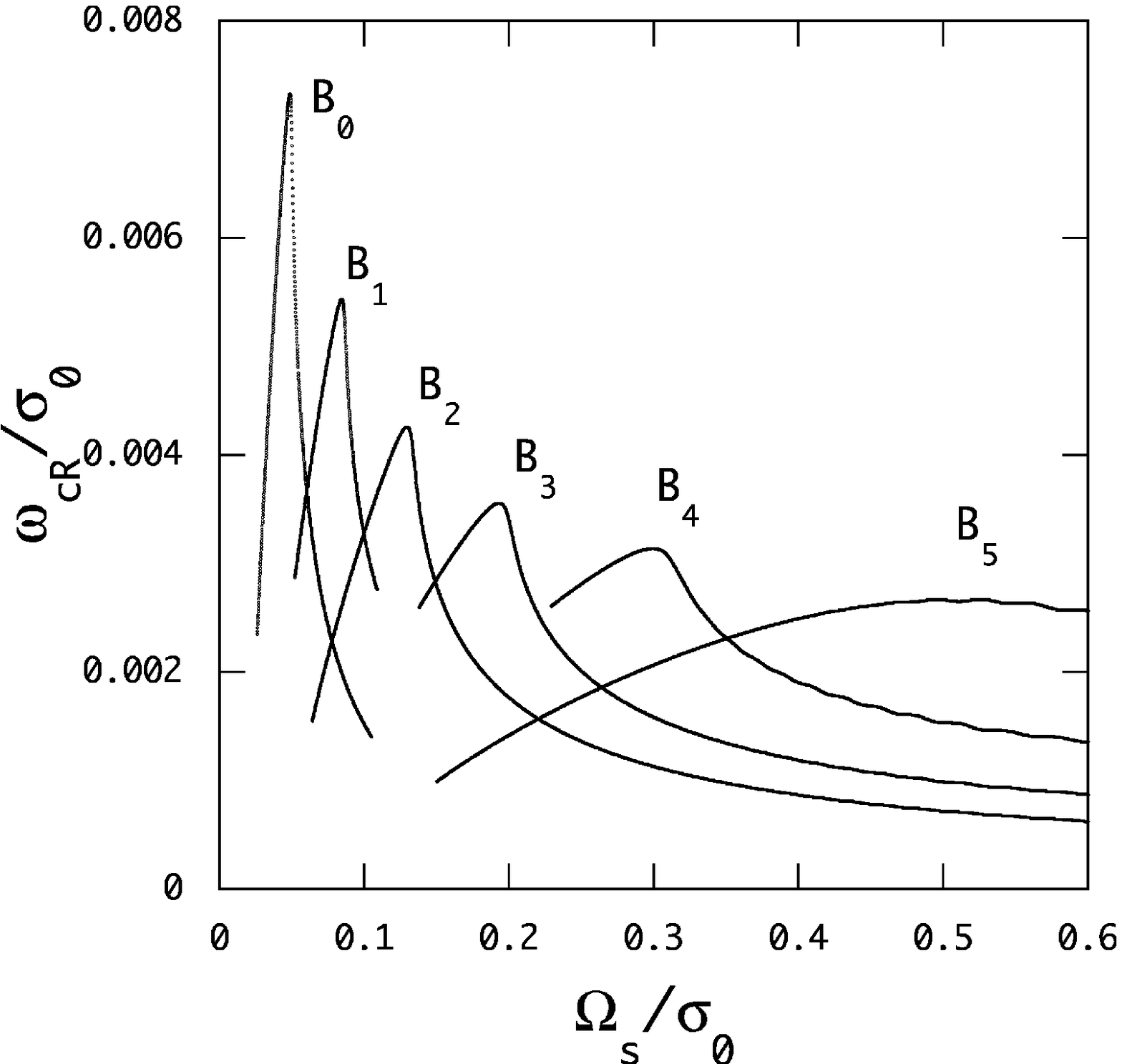}}
\resizebox{0.66\columnwidth}{!}{
\includegraphics{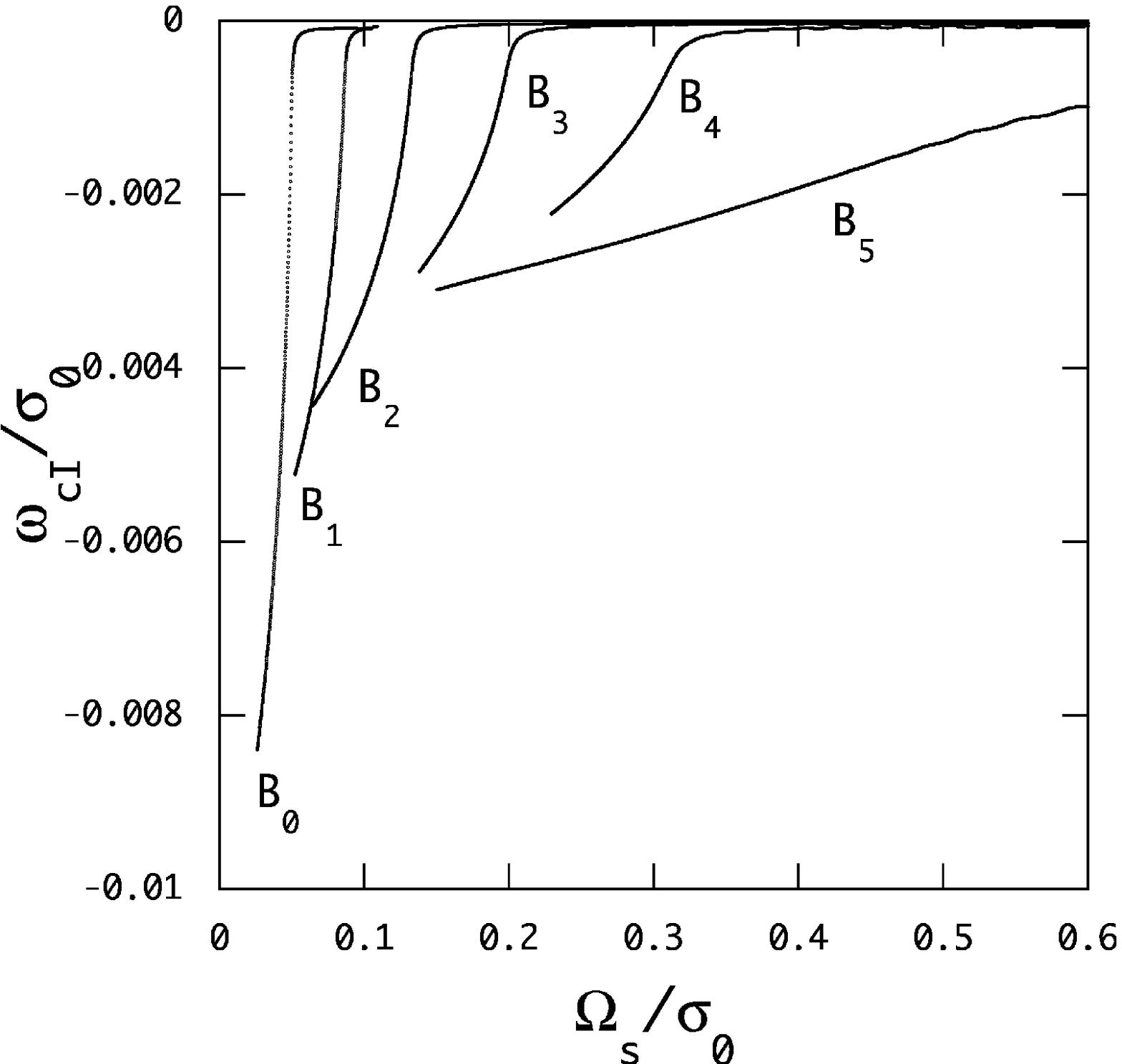}}
\resizebox{0.66\columnwidth}{!}{
\includegraphics{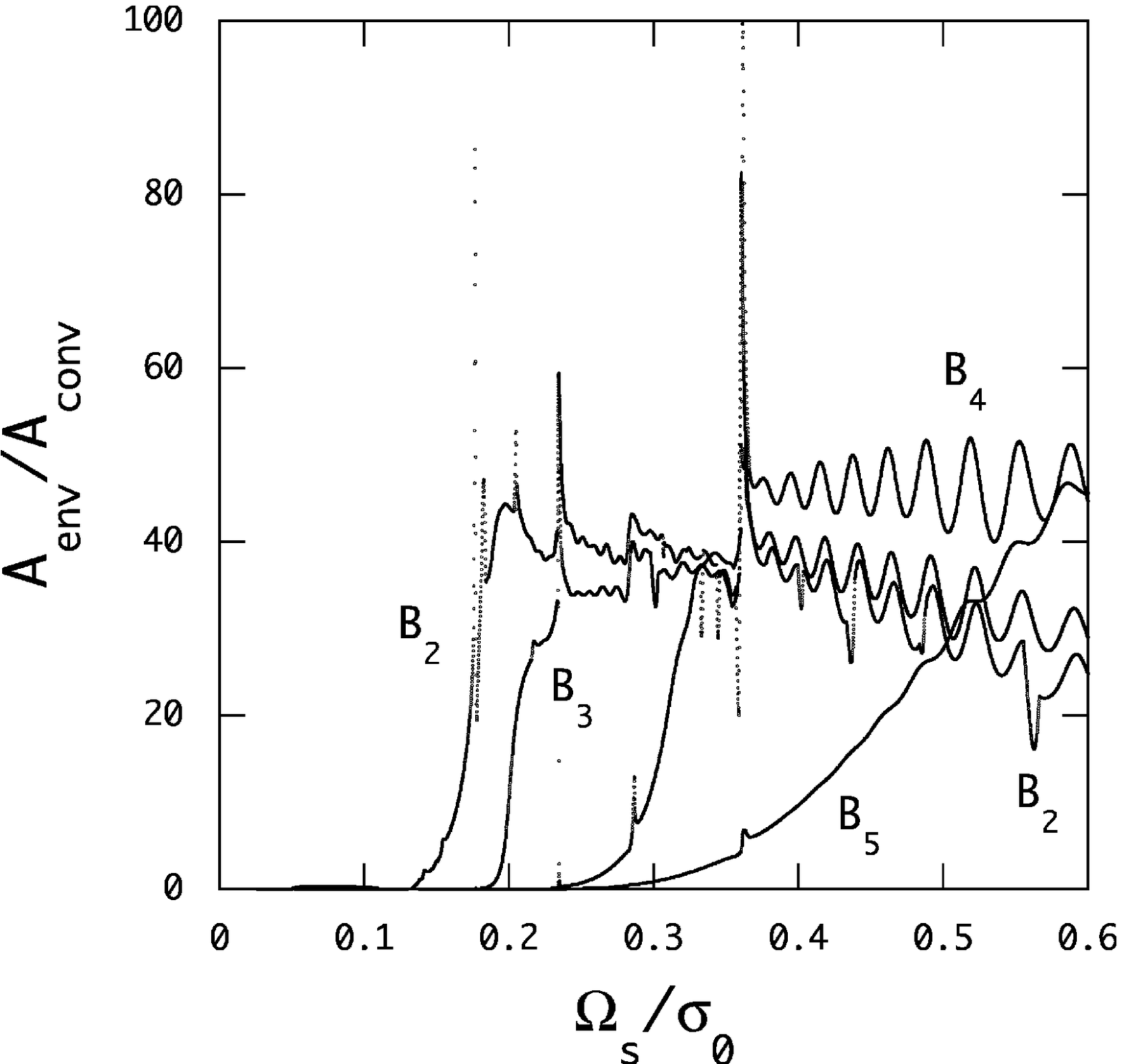}}
\caption{Same as Fig.\ref{fig:omega_m2md1b1p1} but for the $20M_\odot$ ZAMS model.}
\label{fig:omega_m20md1}
\end{figure*}

Fig. \ref{fig:sigmar} shows that the frequencies $\sigma_{\rm R}$ of 
the OsC modes and of the opacity driven $g$-modes are well separated 
from each other for slowly rotating SPB stars and that
the frequency separation becomes smaller as $\overline\Omega_s$ increases.
We find that mode crossings between OsC modes and opacity driven $g$-modes take place, for example,
for $\overline\Omega_s\gtrsim 0.5$ for the ZAMS model and that
such mode crossings in rapidly rotating SPB stars become more likely as $|m|$ and $b$ increase.
As an example, Fig.\ref{fig:osc-gmode} depicts mode crossings between the $B_2$-mode and $g$-modes for $m=-1$ and $b=1.2$.
Because of the mode crossings, $\overline\omega_{c{\rm R}}(\overline\Omega_s)$ of the $B_2$-mode describes
a zig-zag curve as a function of $\overline\Omega_s$.
At $\overline\Omega_s=0.6$, for example, the $B_2$-mode stands between
$g_{15}$- and $g_{16}$-modes, which are both unstable, and hence the period spacings of the low frequency modes
including $g$-modes 
will be different depending on whether or not we count the $B_2$-mode as an observable low frequency mode.

Using the periods of opacity driven $g$-modes and OsC modes we discuss how the relations between the period $P$ and
period spacing $\Delta P$ look like for the low frequency modes in
rapidly rotating SPB stars, particularly when
an OsC mode comes in between opacity driven $g$-modes.
In Fig.\ref{fig:pdp} we plot $\Delta P$ as a function of $P$ for the opacity driven $g$-modes
and the $B_2$ mode of the $4M_\odot$ ZAMS star at $\overline\Omega_s=0.6$ for $m=-1$ and $b=1.2$ 
where $\Delta P=P_{n+1}-P_n$ and $P=(P_{n+1}+P_n)/2$ with $P_n$ being the period of the low frequency modes.
From the ranges of the periods and the period spacings in the figure, we find that 
the relation $P-\Delta P$ may be located between the relations 
for KIC 9020774 and KIC 11971405 in Fig. 27 of \citet{PapicsTkachenkoVanReethetal2017},
who provided $P-\Delta P$ relations for $g$-modes observationally identified in several rotating SPB stars.
Although the observational $P-\Delta P$ relations
do not necessarily show deep dips in the plots of $\Delta P$ versus $P$,
the theoretical $P-\Delta P$ relation does
have such deep dips, which are produced by resonances of the envelope $g$-modes with an inertial mode in the core.
As a result of the resonance,
the radial component of the displacement vector, $\xi_r$, of a $g$-mode has comparable amplitudes both in the core and in the envelope,
an example of which is shown in Fig.\ref{fig:b2-inertial}, where the real parts of the expansion coefficient $xS_l$
are plotted versus $x=r/R$ for the $m=-1$ $g_8$-mode.
Similar dips are known to exist in the $P-\Delta P$ relations observationally obtained for $\gamma$ Dor stars 
(e.g., \citealt{LiVanReethBeddingetal20}) and they
are believed to be produced as a result of resonances between $g$-modes and inertial modes as discussed by 
\citet{Quazzanietal20} and \citet{Saioetal21}.
Although the resonances with core inertial modes produce rather prominent features in $P-\Delta P$ relations,
the OsC modes are likely to disturb only the longest period parts of the $P-\Delta P$ relations.
If we define the spin parameter $s_c\equiv 2\Omega_c/\omega_{c{\rm R}}$ evaluated in the core,
we obtain $s_c=1.04\times 10^1$ for the $g_8$-mode and $s_c=1.15\times 10^3$ for the $B_2$-mode,
and the resonances of opacity driven $g$-modes are more likely to occur with an inertial mode
than with an OsC mode.

\subsection{$20M_\odot$ models}

Fig.\ref{fig:omega_m20md1} plots $\overline\omega_c$ and $A_{\rm env}/A_{\rm core}$ of $m=-1$ and $m=-2$
OsC modes of the $20M_\odot$ ZAMS model for $b=1.1$.
The behavior of $\overline\omega_c$ is essentially the same as that obtained for the $2M_\odot$ and $4M_\odot$ models although there exists some minor differences.
For $b=1.1$, for example, strong rotational stabilization of the OsC modes with increasing $\overline\Omega_s$ occurs 
for the $B_0$- to $B_4$-modes for $m=-1$, although only the $B_0$- to $B_2$-modes are subject to 
such strong stabilization for the $2M_\odot$ model.
The difference may be due to the fact that the fractional radius $x_c$ of the convective core of the $20M_\odot$ model 
is larger by a factor $\sim2$ than that of the $2M_\odot$ model.
For the rotation law given by equation (\ref{eq:difrot}), the width of the fractional radius, $\Delta x$, over which
$\Omega(r)$ rapidly changes from $\overline\Omega_c$ to $\overline\Omega_s$ 
may be estimated as $\Delta x\approx 2/a=0.02$.
When the fractional wavelengths $\sim x_c/(n+1)$ are comparable to or less than $\Delta x$,
the effects of the differential rotation on OsC modes are significant and
strong rotational stabilization of the OsC modes with increasing $\overline\Omega_s$ cannot occur.
In other words, strong stabilization of OsC modes takes place when $x_c/(n+1)\lesssim \Delta x$,
which may be consistent to the numerical results in this paper.

It is interesting to note that 
since $\overline\omega_{c{\rm R}}$ of $B_n$-modes is much smaller than $|m\overline\Omega_s(b-1)|$ for rapid rotation
unless $b\approx 1$, 
their Doppler shifted frequencies in the envelope take almost the same value given by
$\overline\omega_{s{\rm R}}\approx-m\overline\Omega_s(b-1)$ and in the inertial frame by 
$\overline\sigma_{{\rm R}}\approx-m\overline\Omega_sb$.
This suggests that several OsC $B_n$-modes with different radial orders $n$ can be in resonance with
a low frequency envelope $g$-mode having the frequency $\overline\omega_g\approx-m\overline\Omega_s(b-1)$
to obtain large amplitudes at the surface.
If this multiple excitation of $B_n$-modes happens at a given $\overline\Omega_s$,
we would observe a fine structure of frequency around $-m\overline\Omega_s b$ in the frequency spectrum
and a frequency separation in the fine structure may be given by 
$|\overline\omega_{c{\rm R}}^{n_2}-\overline\omega_{c{\rm R}}^{n_1}|$ where $n_1$ and $n_2$ are the radial orders of
OsC $B_n$-modes in resonance with a $g$-mode.
For example, for $m=-1$ and $b=1.1$ $B_n$-modes at $\overline\Omega_s=0.5$
we find at least three $B_n$-modes, $B_3$-, $B_4$-, and $B_5$-modes, are in resonance with the $g_{20}$-mode
to have large amplitudes at the surface.
The frequency separation between $B_3$- and $B_4$-modes, for example, is given by 
$|\overline\omega_{c{\rm R}}^{4}-\overline\omega_{c{\rm R}}^{3}|\approx 1.8\times10^{-4}$, which makes a fine frequency structure around $\overline\sigma\approx 0.55$.

\section{Angular momentum transport by OsC modes}

\begin{figure}
\resizebox{0.9\columnwidth}{!}{
\includegraphics{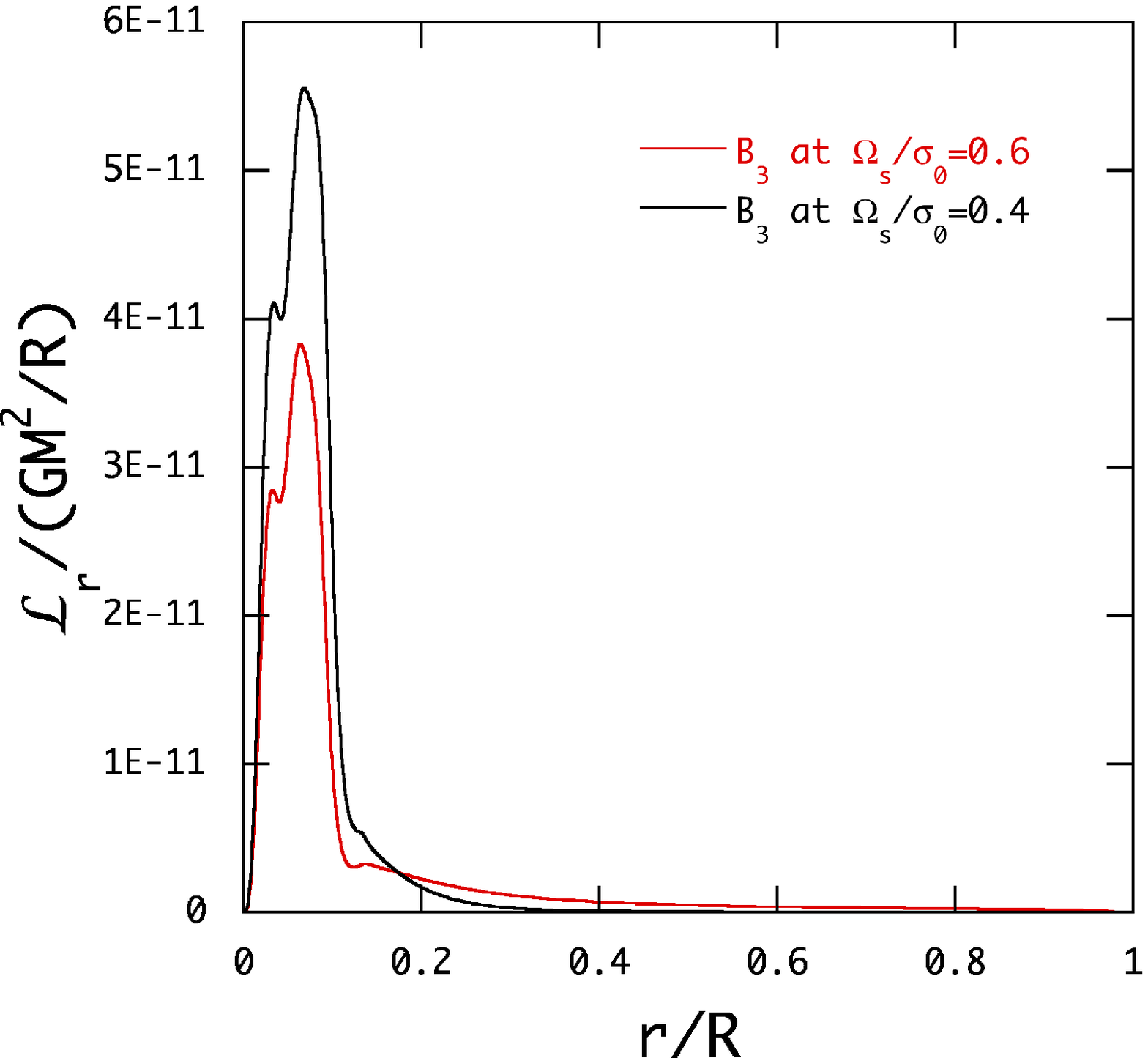}}
\resizebox{0.9\columnwidth}{!}{
\includegraphics{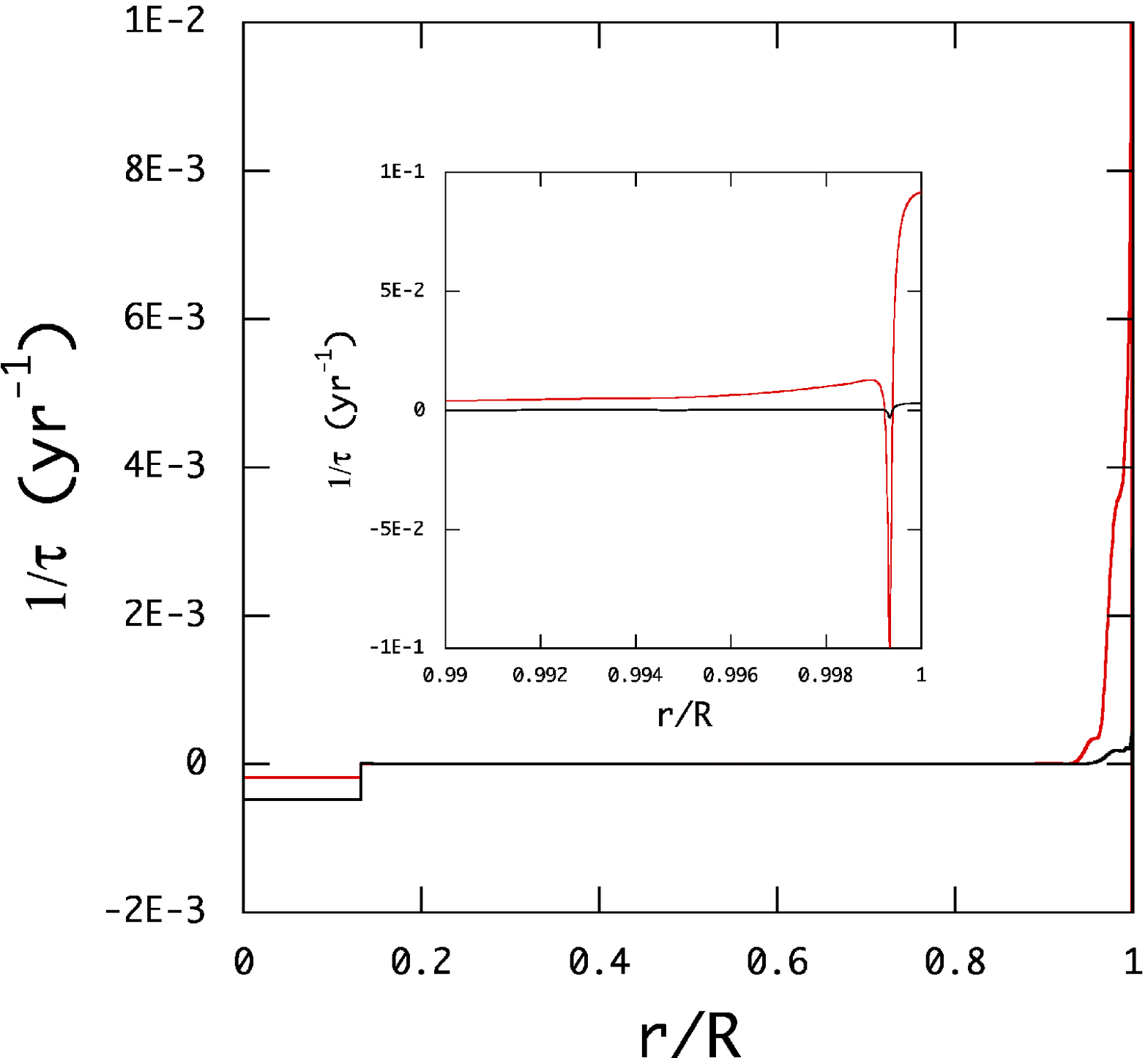}}
\caption{${\cal L}_r/(GM^2/R)$ and $1/\tau$ versus $r/R$ for the $m=-1$ $B_3$-modes of the $2M_\odot$ ZAMS model at
$\overline\Omega_s=0.4$ and 0.6 for $b=1.1$, where the oscillation amplitude is given by (\ref{eq:ampnorm}).
}
\label{fig:Lr_m2md1b1p1mm1B3_0406}
\end{figure}

Since OsC modes that excite envelope $g$-modes have amplitudes both in the core and in the envelope,
it is worth examining angular momentum transport by the OsC modes.
In the Cowling approximation (\citealt{Cowling41}), angular momentum transport by low frequency oscillations in rotating stars
may be described by (e.g., \citealt{Pantillonetal07}; \citealt{Mathis09}; \citealt{Lee13}; see also \citealt{BelkacemMarquesGoupiletal2015})
\be
\rho{d{<\ell>}\over dt}=-{1\over 4\pi r^2}{\partial\over\partial r}{\cal L}_r,
\ee
where $\ell=r^2\sin^2\theta \Omega$ denotes the specific angular momentum around the rotation axis, and 
\be
{\cal L}_r\equiv 4\pi r^2\left<\rho r\sin\theta v'_r\left(v'_\phi+2\Omega \cos\theta\xi_\theta\right)\right>,
\ee
and $\left<{f}\right>=\int_0^\pi\int_0^{2\pi} f\sin\theta d\theta d\phi$.
Note that to evaluate ${{\cal L}_r}$, we use $\pmb{v}'=\rmi\omega\pmb{\xi}-r\sin\theta\left(\pmb{\xi}\cdot\nabla\Omega\right)\pmb{e}_\phi$.
We regard ${\cal L}_r$ as the angular momentum luminosity transported by the waves.
If ${\cal L}_r$ increases with increasing $r$, we consider that the waves extract angular momentum from rotating fluids, 
decelerating the rotation.
On the other hand, if ${\cal L}_r$ decreases, the waves deposit angular momentum to the fluids, accelerating
the rotation.

It is also useful to calculate the time scale $\tau$ of angular momentum changes expected in the interior.
In the envelope, we may define $\tau$ by
\be
{1\over\tau}={1\over \ell}{d{<\ell>}\over dt},
\ee
and for the convective core, we calculate the averaged time scale $\tau$ defined by
\be
{1\over\tau}=-{{\cal L}_{r}(r_c)\over J_c},
\label{eq:taucore}
\ee
where $J_c=\int_0^{r_c}\ell\rho dV$ and $-{\cal L}_{r}(r_c)=\int_0^{r_c}dV\rho{d{<\ell>}/ dt}$.
Positive (negative) $\tau$ indicates acceleration (deceleration) of rotation of the fluids.
To estimate ${\cal L}_r$ and $\tau$, we normalize the amplitudes of OsC modes by assuming
\be
\int_0^R\omega_{\rm R}^2\pmb{\xi}\cdot\pmb{\xi}^*\rho dV=\int_0^{r_c} v_{\rm turb}^2 \rho dV,
\label{eq:ampnorm}
\ee 
where $v_{\rm turb}$ is the turbulent velocity in the convective core and
is computed using the mixing length theory of convection for non-rotating stars.
Equation (\ref{eq:ampnorm}) may suggest that the kinetic energy of the turbulent fluids in the core
is redistributed over the entire interior by the OsC modes when the modes have amplitudes both in the core and in the envelope.

\begin{figure}
\resizebox{0.9\columnwidth}{!}{
\includegraphics{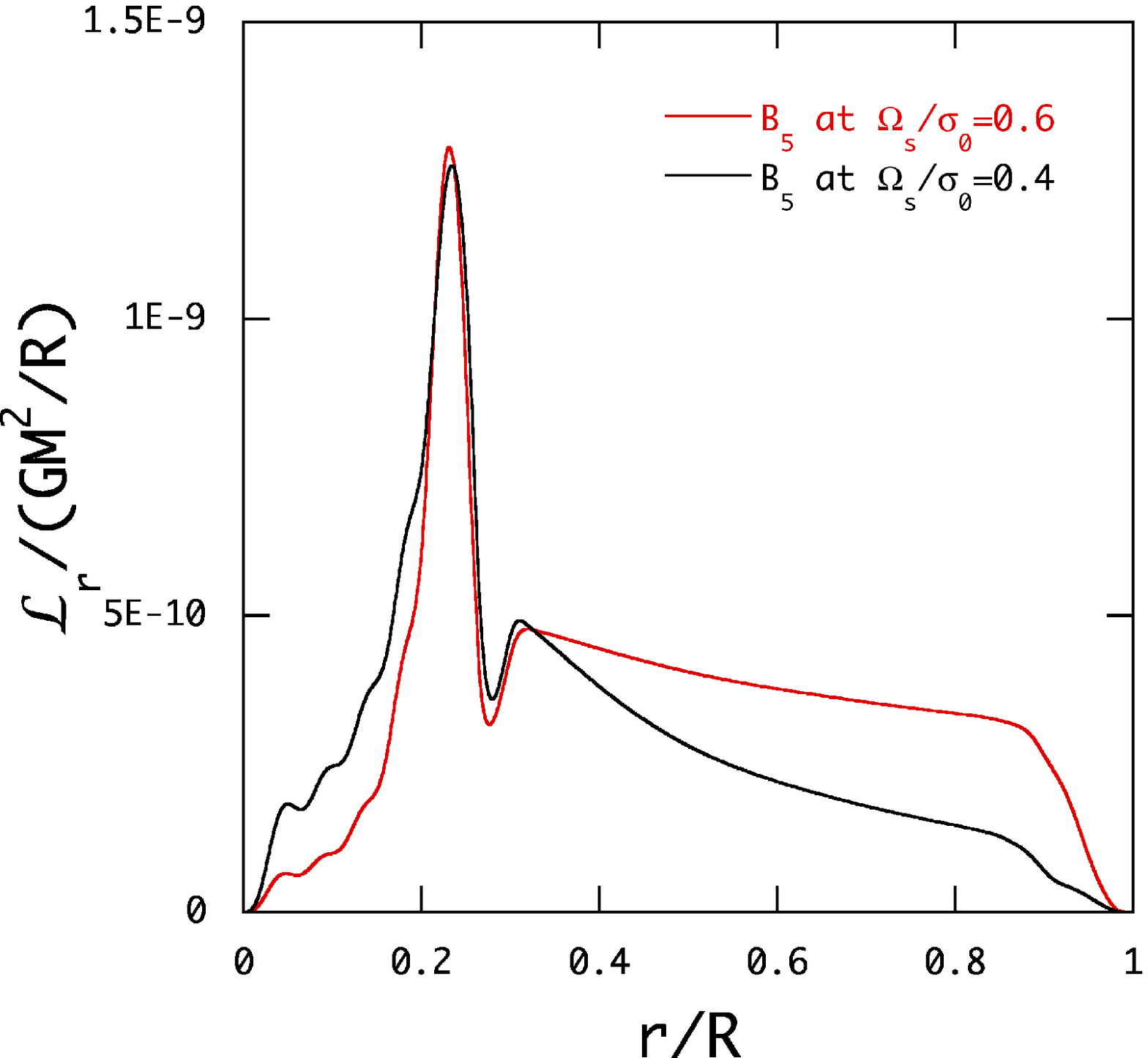}}
\resizebox{0.9\columnwidth}{!}{
\includegraphics{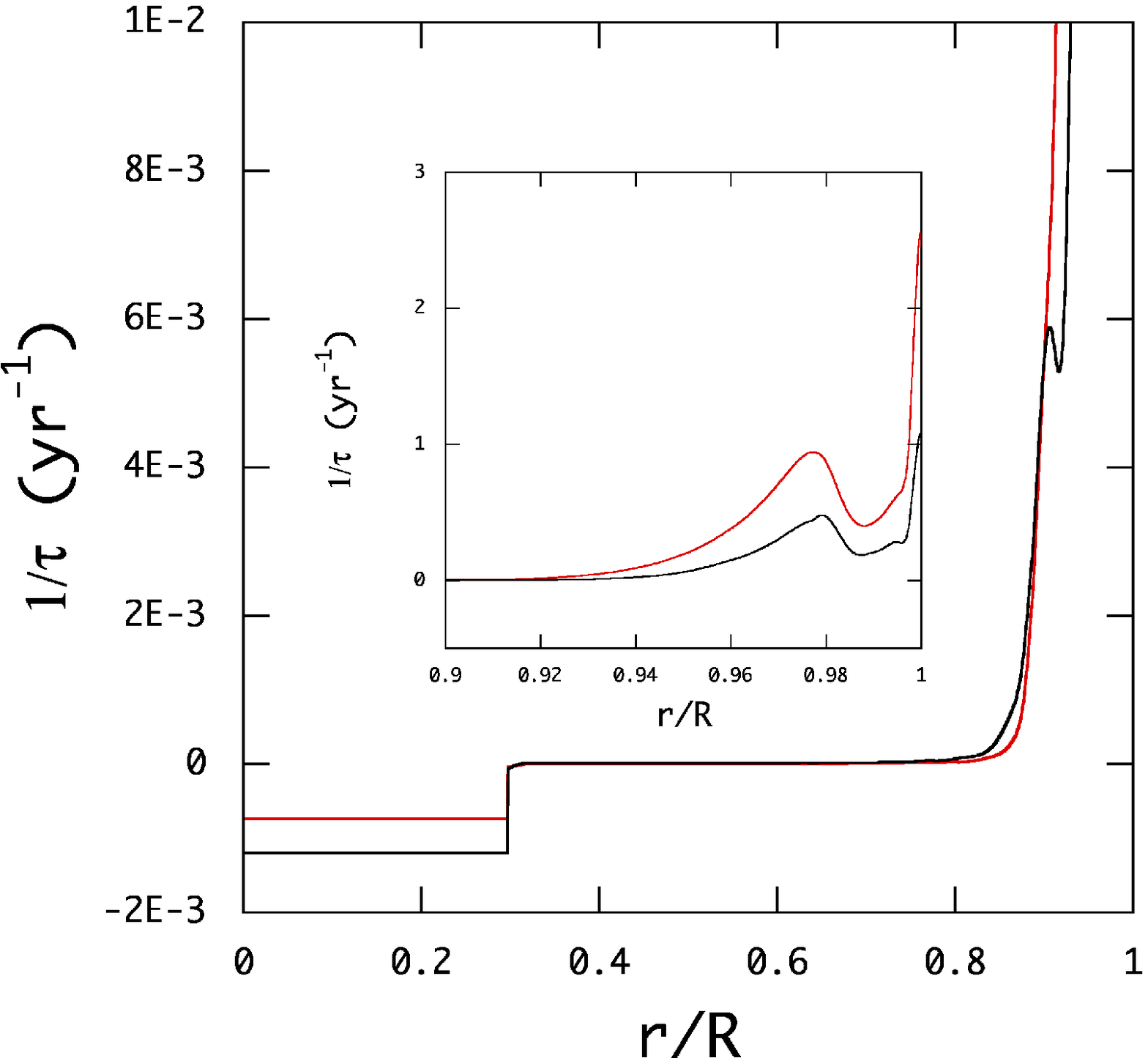}}
\caption{Same as Fig. \ref{fig:Lr_m2md1b1p1mm1B3_0406} but for the $B_5$ modes 
of the $20M_\odot$ ZAMS model at $\overline\Omega_s=0.4$ and 0.6.
}
\label{fig:Lr_m20md1b1p1mm1B5_0406}
\end{figure}

In Fig. \ref{fig:Lr_m2md1b1p1mm1B3_0406}, ${\cal L}_r/(GM^2/R)$ and $1/\tau$ of the $m=-1$ $B_3$-modes 
at two different $\overline\Omega_s$s are plotted versus $r/R$ for $b=1.1$ for the $2M_\odot$ ZAMS model.
As $r/R$ increases from the centre, ${\cal L}_r$ increases from zero to the maximum within the core
and then decreases towards the core boundary at $r_c/R=0.132$.
Positive ${\cal L}_{r}(r_c)$ at the boundary indicates that 
the rotation speed of the core is decelerated by exciting the prograde OsC modes as suggested by equation (\ref{eq:taucore}).
In the envelope, on the other hand, ${\cal L}_r$ gradually decreases as $r/R$ increases from the core boundary 
to $r/R\sim 0.9$, from which it decreases steeply towards the surface.
This suggests that angular momentum deposition takes place in the envelope, particularly in the layers close to the stellar surface. 
From the plot of $1/\tau$ against $r/R$ we find that the time scale $\tau$ in the envelope is generally positive and 
it can be very small in the surface layers where the density is low and the dissipations are large.
This may be suggested by the inset, which is a magnification to show $1/\tau$ in the region close to the surface.
On the other hand, we have negative $\tau$ in the convective core and $|\tau|$ can be small when
a large amount of angular momentum extraction by OsC modes occurs.
We note that, for the amplitude normalization given by equation (\ref{eq:ampnorm}), the time scale $|\tau|$ in the core can be 
longer for the OsC modes that excite envelope $g$-modes than for the OsC modes that tend to be confined in the core.
From Fig. \ref{fig:Lr_m2md1b1p1mm1B3_0406} we consider that the OsC modes that excites $g$-modes can transfer
angular momentum from the convective core to the envelope and to the surface of the stars.

Fig. \ref{fig:Lr_m20md1b1p1mm1B5_0406} plots ${\cal L}_r/(GM^2/R)$ and $1/\tau$ of the $m=-1$ $B_5$-modes
against $r/R$ for the $20M_\odot$ ZAMS model.
Their behavior as a function of $r/R$ is similar to that found for the $2M_\odot$ model
and $|\tau|$ in the core is much longer than that at the stellar surface.
We also note that $|\tau|$ in the core is by a factor $\sim 3$ shorter than that for the $2M_\odot$ stars
for the normalization (\ref{eq:ampnorm}).

\section{discussions}

Our numerical investigations in this paper have suggested that  
low radial order $B_n$-modes in weakly differentially rotating stars are strongly stabilized by rotation 
to obtain vanishingly small $|\overline\omega_{c{\rm I}}|$ as $\overline\Omega_s$ increases, and that
as the radial oder $n$ increases rotational stabilization of the $B_n$-modes become weaker
so that the magnitudes of $\overline\omega_{c{\rm I}}$ only gradually decrease with increasing $\overline\Omega_s$.
For the low radial order $B_n$-modes, 
it is important to note that even if they are stabilized to have vanishingly
small $|\overline\omega_{c{\rm I}}|$, they remain unstable with small but finite $|\overline\omega_{c{\rm I}}|\not=0$
and excite $g$-modes in the envelope.
\citet{LeeSaio89,LeeSaio90} suggested that this destabilization of $g$-modes 
occurs when oscillatory convective modes with negative energy of oscillation 
are in resonances with envelope $g$-modes with positive energy of oscillation.
In other words, because of the resonances
OsC modes cannot be completely stabilized by rotation, that is, they cannot be an
oscillatory convective mode having a pure real frequency $\overline\omega_{c}$.

\begin{figure}
\resizebox{0.9\columnwidth}{!}{
\includegraphics{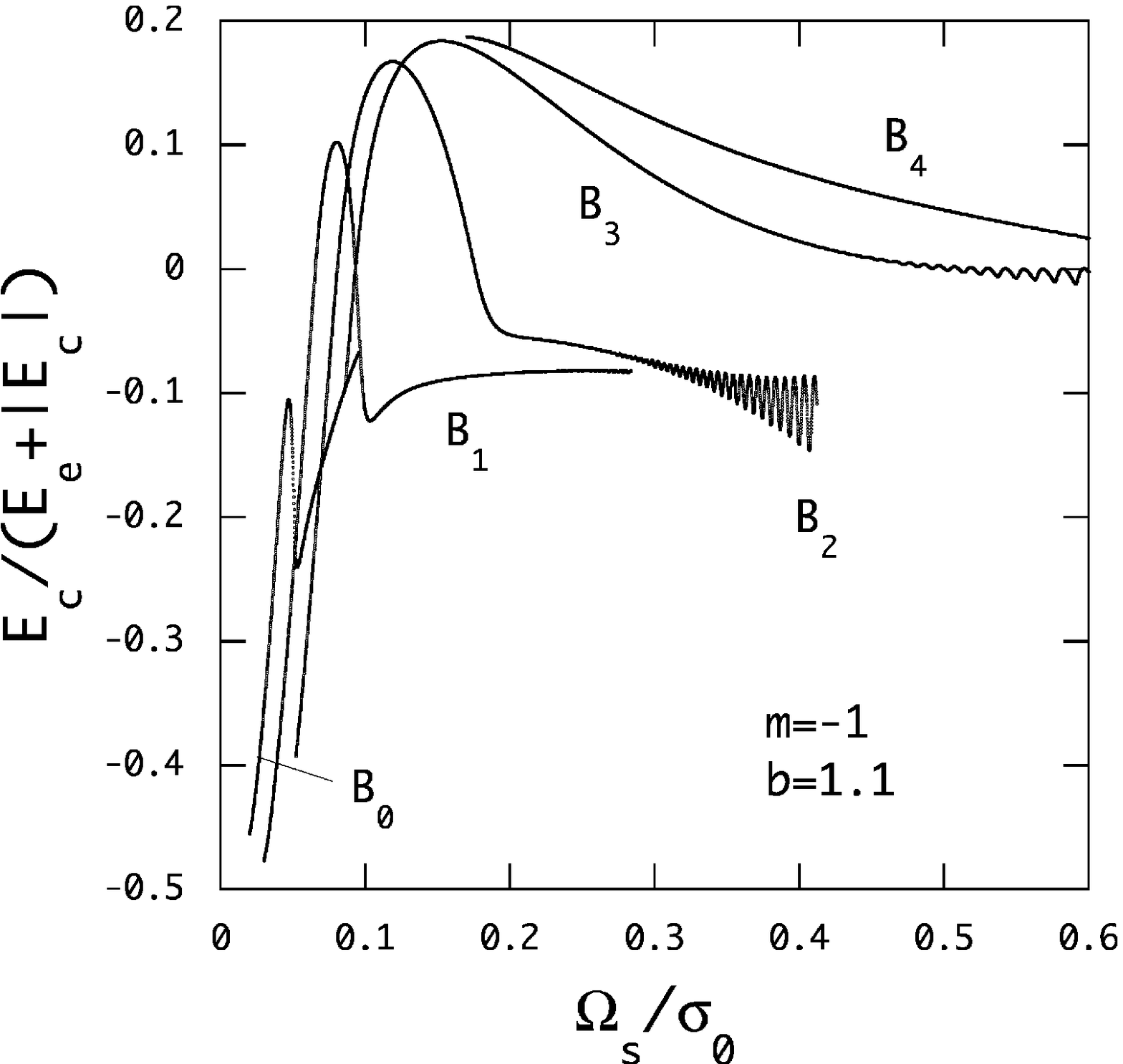}}
\resizebox{0.9\columnwidth}{!}{
\includegraphics{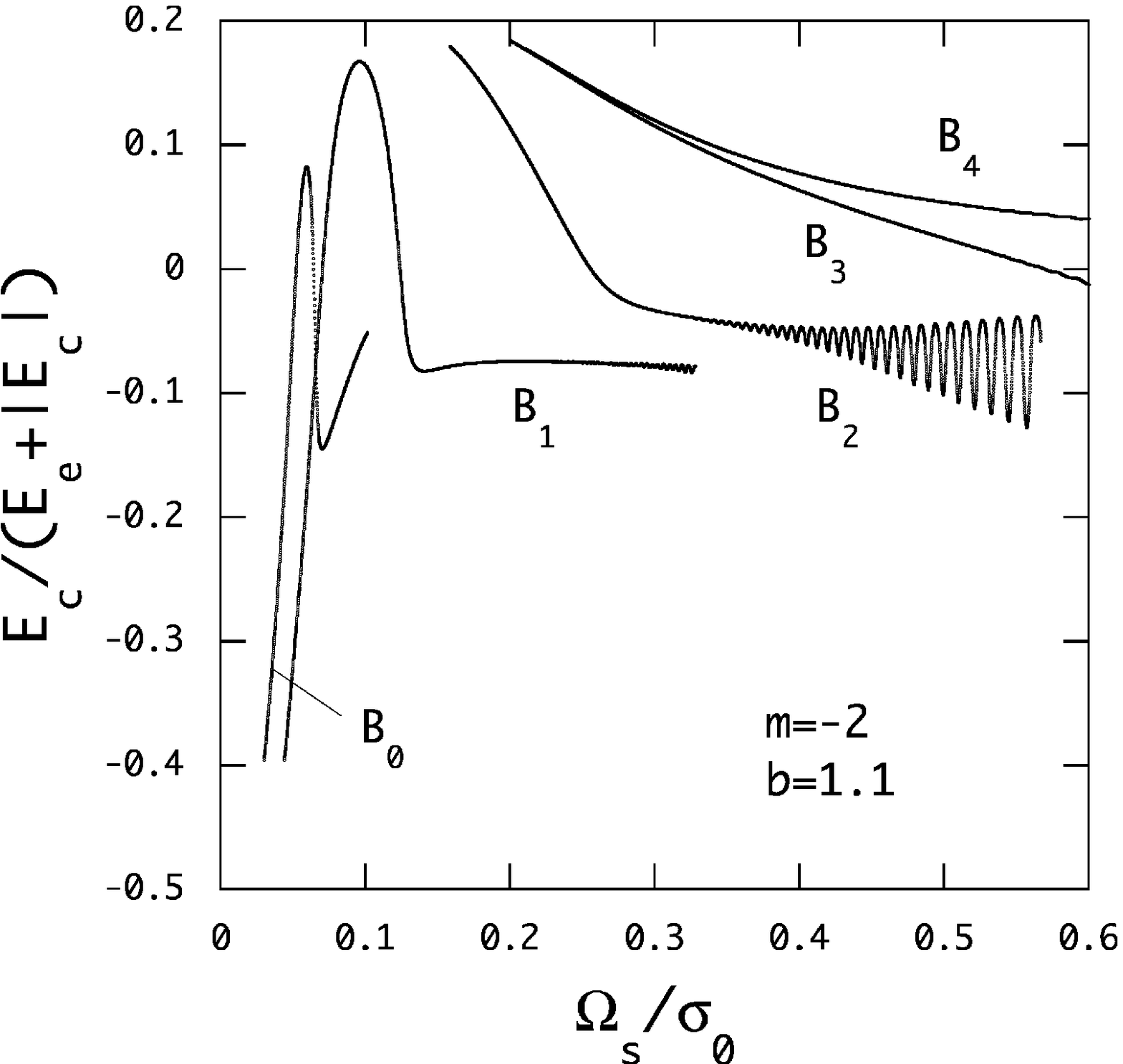}}
\caption{$E_c/(E_e+|E_c|)$ versus $\Omega_s/\sigma_0$ for $m=-1$ and $m=-2$ OsC modes of
the $2M_\odot$ ZAMS model for $b=1.1$.
}
\label{fig:Ec_m2md1b1p1}
\end{figure}

\begin{figure}
\resizebox{0.9\columnwidth}{!}{
\includegraphics{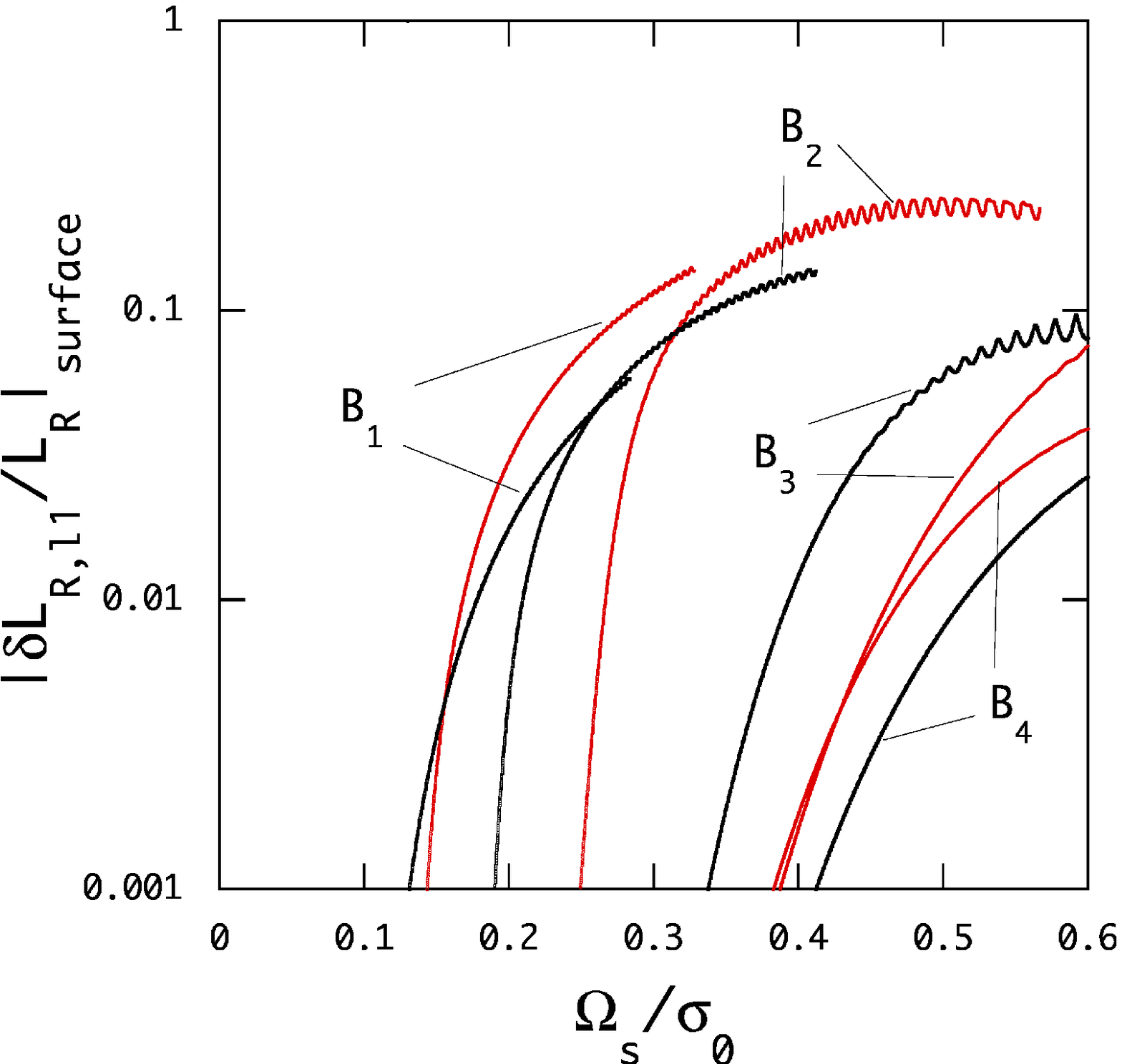}}
\resizebox{0.9\columnwidth}{!}{
\includegraphics{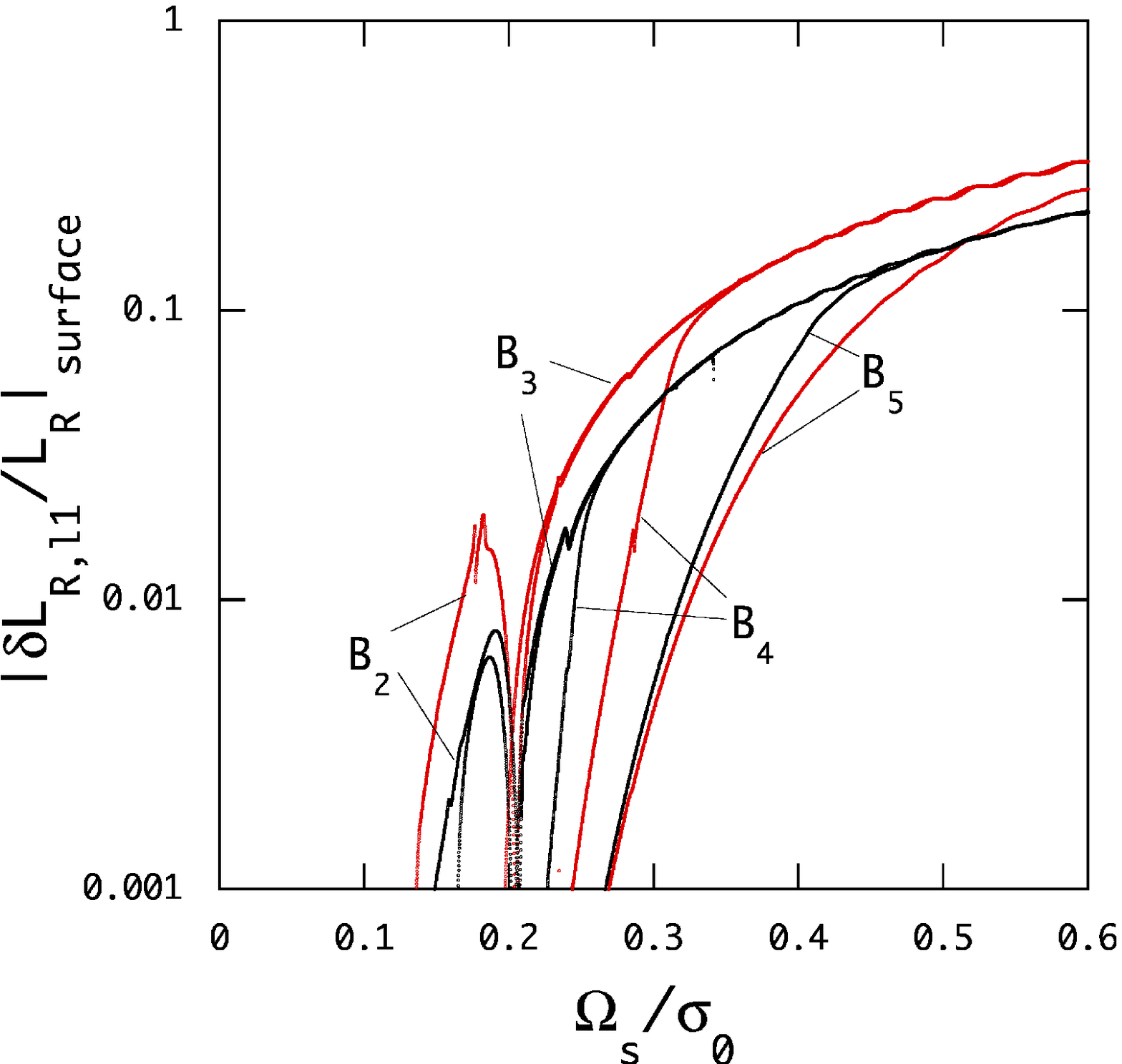}}
\caption{$(\delta L_{r,l_1}/L_r)_{\rm surface}$ versus $\overline\Omega_s$ for $m=-1$ (black lines) and $m=-2$ (red lines)
OsC modes of the $2M_\odot$ ZAMS model (upper panel) and of the $20M_\odot$ ZAMS model (lower panel) for $b=1.1$ where
the oscillation amplitude is normalized by equation (\ref{eq:ampnorm}).}
\label{fig:deltaLr}
\end{figure}

To probe the suggestion given above, we compute the oscillation energy of OsC modes.
We define the specific energy $e_W$ of oscillation as the sum of the kinetic energy $e_K$ and
the potential energy $e_P$ of oscillation, $e_W=e_K+e_P$, where in the Cowling approximation
\be
e_K=\overline{\pmb{v}^{\prime 2}}/2, \quad e_P=\left[\overline{\left({p^\prime/ \rho c}\right)^2}+N^2\overline{\xi_r^2}\right]/2,
\label{eq:ew}
\ee
and $N^2=-gA$ with $N$ being the Brund-V\"ais\"al\"a frequency,
$c=\sqrt{\Gamma_1p/\rho}$ with $\Gamma_1=(\partial\ln p/\partial\ln\rho)_{ad}$
is the adiabatic sound velocity, $g=GM_r/r^2$ with $M_r=\int_0^r4\pi r^2\rho dr$ and $G$ the gravitational constant, and
\be
rA={d\ln \rho\over d\ln r}-{1\over\Gamma_1}{d\ln p\over d\ln r}.
\ee
Note that $\overline{f}$ indicates the time average of the quantity $f$.
In non-rotating stars, we usually have the equipartition of energy as given by $\int e_KdV=\int e_P dV$, but
this equipartition of energy is not always satisfied for rotating stars.
In Fig.\ref{fig:Ec_m2md1b1p1}
we plot the ratio $\eta\equiv E_c/(E_e+|E_c|)$ as a function of $\overline\Omega_s$ for OsC modes of the $2M_\odot$ ZAMS model for $b=1.1$ where $E_c$ and $E_e$ are defined as
\be
E_c=\int_0^{r_c}\rho e_WdV, \quad E_e=\int_{r_c}^R\rho e_W dV.
\ee
As $\overline{\Omega}_{s}\rightarrow 0$, the modes tend towards pure convective modes confined in the core, whose
$\overline\omega_c$ is pure imaginary.
This takes place when $\int_0^Re_PdV\approx \int_0^{r_c}e_PdV<0$
because $N^2<0$ and $N^2\overline{\xi_r^2}$ dominates $\overline{\left({p^\prime/ \rho c}\right)^2}$ in the convective core.
As $\overline\Omega_s$ increases from $\overline\Omega_s\sim0$, 
the convective modes are stabilized to be overstable and $E_c$ of the modes
changes its sign from being negative to being positive, although
$E_c$ of the $m=-1$ $B_0$-mode is an exception to this rule.
As $\overline\Omega_s$ further increases, we note that low radial order $B_n$-modes are likely to suffer strong 
rotational stabilization.
As the low radial order $B_n$-modes get close to a state of complete stabilization with $\overline\omega_{c{\rm I}}\sim 0$,
their oscillation energy $E_c$ in the core changes its sign from being positive to being negative as shown by
Fig.\ref{fig:Ec_m2md1b1p1}.
This may confirm the interpretation that excitation of envelope $g$-modes by low radial order $B_n$-modes occurs as a result of resonant couplings between oscillatory convective modes with negative energy
and envelope $g$-modes with positive energy of oscillation (\citealt{LeeSaio90}).

As the radial order $n$ increases, the $B_n$-modes in differentially rotating stars
tend to be only weakly stabilized by rotation and do not always
obtain vanishingly small $|\overline\omega_{c{\rm I}}|$.
In this case, the oscillation energy $E_c$ is likely to stay positive even for rapid rotation.
However, as suggested by Figs. \ref{fig:omega_m2md1b1p1} and \ref{fig:omega_m2md1b1p2}, the modes still can 
excite envelope $g$-modes for rapid rotation speeds, e.g., for $\overline\Omega_s\gtrsim 0.5$.
This can be understood by using the quantity $\Phi_1^R$ defined by
\be
\Phi_1^R=-{\omega_{s{\rm I}}\over\omega_{s{\rm R}}}{\sqrt{\lambda_{km}}\over\omega_{s{\rm R}}}\int_{r_c}^R N {dr\over r}
\sim -{\omega_{s{\rm I}}\over\omega_{s{\rm R}}} \pi n_e,
\ee
where $\lambda_{km}$ is the eigenvalue of Laplace's tidal equation (e.g., \citealt{LeeSaio97}) and 
for prograde sectoral $g$-modes in the envelope we can assume $\sqrt{\lambda_{km}}\sim |m|$ (e.g., \citealt{Townsend03}),
and $n_e$ is the number of radial nodes of the $g$-modes in the envelope (\citealt{LeeSaio20}).
Note that $\omega_{s{\rm I}}=\omega_{c{\rm I}}$.
\citet{LeeSaio20} used $|\Phi_1^R|\lesssim 1$ as the condition that effective excitation of envelope $g$-modes by OsC modes can take place.
This condition may be satisfied when $|\omega_{s{\rm I}}/\omega_{s{\rm R}}|\ll 1$ and/or the number of nodes, $n_e$, is not extremely large.
As $\overline\Omega_s$ increases, the Doppler shifted frequency $\overline\omega_{s{\rm R}}$ of the $B_n$-modes 
is approximately given by $\overline\omega_{s{\rm R}}\approx -m\overline\Omega_s(b-1)$ in the envelope.
For example, for $b=1.1$ we have $\overline\omega_{s{\rm R}}\gtrsim 0.05|m|$ for $\overline\Omega_s\gtrsim 0.5$.
Even if the ratio $|\overline\omega_{s{\rm I}}/\overline\omega_{s{\rm R}}|$ is not necessarily very small, 
we may have $|\Phi_1^R|\lesssim 1$ for $n_e\sim 10$ and hence even the $B_n$-modes whose $|\overline\omega_{s{\rm I}}|$ is 
not necessarily vanishingly small can excite envelope $g$-modes in rapidly rotating stars.

Using the amplitude normalization given by equation (\ref{eq:ampnorm}),
we compute the relative luminosity variation $\delta L_r/L_r$ at the stellar surface.
An example of such computations is shown in Fig. \ref{fig:deltaLr} for the OsC modes of $m=-1$ (black lines) 
and $m=-2$ (red lines)
for the $2M_\odot$ (upper panel) and $20M_\odot$ (lower panel) ZAMS models where $b=1.1$ is assumed.
In general, $|\delta L_r/L_r|$ increases as $\overline\Omega_s$ increases and is saturated to be
of order of $\sim 0.1$ for $\overline\Omega_s\gtrsim 0.5$.
Observationally, the amplitudes of the variations are of order of $\sim 10^{-3}$ (e.g., \citealt{Balona16,Balonaetal19}), 
suggesting $|\delta L_r/L_r|\lesssim 10^{-3}$, and hence
the normalization (\ref{eq:ampnorm}) leads to an overestimation of the oscillation amplitudes by a factor  
$f\gtrsim 10^2$.
If we let ${\cal L}_r^0$ and $\tau^0$ denote the evaluations by equation (\ref{eq:ampnorm}), 
we have ${\cal L}_r={\cal L}_r^0/f^2$ and $\tau= \tau^0\times f^{2}$.
For example, the time scale $\tau^0$ in the core of the $2M_\odot$ ZAMS star is of order of $10^4$ year
(see Fig.\ref{fig:Lr_m2md1b1p1mm1B3_0406}), and if we use $f\sim 10^2$ we obtain $\tau\sim 10^8$ year, which is
much less than the lifespan of $2M_\odot$ main sequence stars (e.g., \citealt{Kippenhahn2012}).
This may suggest that angular momentum transport by the OsC modes can be a viable mechanism for
extracting excess angular momentum from the core of the main sequence stars.

Hot main sequence stars may suffer mass loss due to optically thin winds from the stellar surface 
(e.g., \citealt{KrtickaKubat2010,KrtickaKubat2014}).
The properties of stellar pulsations of mass losing stars with stellar winds are not necessarily well understood.
It is difficult to properly treat pulsations of optically thin winds moving with supersonic velocities.
However, the effects of stellar winds on $g$-mode excitation by OsC modes are probably insignificant since the mode excitation
takes place in the deep interior of the envelope and is unlikely to be affected by the winds from the surface.
Microscopic diffusion processes in the envelope of stars with stellar winds and surface magnetic fields
have been assumed to explain the existence of chemically peculiar stars (e.g., \citealt{Michaudetal1983})
and of stars with helium inhomogeneities (\citealt{Vauclairetal1991,LeoneLanzafame1997}).
So long as these chemical peculiarities occur in a thin surface layer occupying a tiny fraction of the stellar mass,
however, their effects on the $g$-mode excitation by OsC modes will also be insignificant, although angular momentum
transport near the stellar surface could be somewhat affected since thermal properties as represented by opacity
in the surface layer would be modified by chemical peculiarities, particularly, by helium stratification.


\section{Conclusions}

We have computed low $|m|$ OsC ($B_n$-) modes of $2M_\odot$, $4M_\odot$ and $20M_\odot$ main sequence stars assuming that the core rotates slightly faster than the envelope.
We find that the OsC modes in rapidly rotating stars can resonantly excite prograde sectoral 
$g$-modes in the envelope, which will be observed as rotational modulations in early type stars.
We find that low radial order $B_n$-modes in differentially rotating stars are likely subject to strong
stabilization by rotation, but as the radial order $n$ increases, the stabilizing effect of rotation on the $B_n$-modes
becomes weak.
We find that the general properties of the OsC modes do not strongly depend on the mass of the stars.

To compute non-adiabatic OsC modes in this paper, we have employed the prescription $(\nabla\cdot\pmb{F}_C)^\prime=0$ 
for the approximation of frozen-in convection in pulsating stars.
We have compared the results for the $2M_\odot$ models to those obtained by \citet{LeeSaio20}, who
used $\delta(\nabla\cdot\pmb{F}_C)=0$ to compute the OsC modes.
We find that the general properties of the OsC modes obtained by applying the two different prescriptions for the convective energy flux are roughly the same, except that for $(\nabla\cdot\pmb{F}_C)^\prime=0$, no OsC modes
behave like inertial modes that satisfy the relation $\overline\omega_{c}\propto\overline\Omega_c$, when they
tend toward complete stabilization with increasing $\overline\Omega_c\approx b\overline\Omega_s$.

$4M_\odot$ main sequence stars correspond to SPB variables, in which many low frequency $g$-modes are
excited by the iron opacity bump mechanism and OsC modes are expected to coexist with such opacity driven $g$-modes.
We have compared the frequency of the OsC modes to that of prograde sectoral $g$-modes driven by
the opacity mechanism.
The frequency of the OsC modes in the inertial frame 
is in general smaller than that of the opacity driven $g$-modes and
the OsC modes and the $g$-modes at a given $\overline\Omega_s$
are well separated when $\overline\Omega_s$ is not large.
In this case, the OsC modes will be observed as rotational modulations.
As $\overline\Omega_s$ increases, however, 
the OsC modes will come close to or stand among the $g$-modes.
If the OsC modes and $g$-modes are not well separated, 
some complexities due to the OsC modes will arise in the analyses of 
low frequency modes using $P-\Delta P$ relations.
For $\overline\Omega_c/\overline\Omega_s\approx 1.2$, for example, we find that the period spacings of opacity driven $g$-modes of the $4M_\odot$ ZAMS model are disturbed by low $|m|$ OsC modes
for $\overline\Omega_s\gtrsim 0.5$.
For rapidly rotating SPB stars, we also find that opacity driven $g$-modes can resonate with an inertial mode in the core and
the periods and period spacing relations of the $g$-modes are disturbed to describe deep dips in the $P-\Delta P$ plots.
See \citet{Quazzanietal20} and \citet{Saioetal21} for similar
phenomena in $\gamma$ Dor stars.
Fittings of theoretical $P-\Delta P$ relations to observational ones (e.g., \citealt{PapicsTkachenkoVanReethetal2017})
will provide us with important information concerning the interior structure of SPB stars.

To see the mass dependence of resonant $g$-mode excitation by OsC modes, we compute OsC modes in 
$20M_\odot$ main sequence stars and find that the OsC modes excite envelope $g$-modes when the core rotates
slightly faster than the envelope, as found for $2M_\odot$ and $4M_\odot$ main sequence stars.
We confirm no strong mass dependence of resonant $g$-mode excitation by OsC modes. 
Of course, there exist some minor differences that depend on the mass of stars.
For example,
the density in the envelope of $20M_\odot$ main sequence stars is lower than that of $2M_\odot$ stars
if plotted as a function of the fractional radius
and the magnitudes of the growth rate $\eta\equiv -\omega_{s{\rm I}}/\omega_{s{\rm R}}$ for
envelope $g$-modes are much larger for the former than for the latter.
This may explain why resonant fluctuations of $\overline\omega_c$ and $A_{\rm env}/A_{\rm core}$
as a function of $\overline\Omega_s$ for rapid rotation are much smoother for the $20M_\odot$ stars than
for the $2M_\odot$ stars.

Calculating ${\cal L}_r$ and $1/\tau$ for OsC modes in rotating main sequence stars, 
we have shown that the OsC modes in resonance with envelope $g$-modes can transport angular momentum 
from the core to the envelope of the stars.
We have also suggested that if the angular momentum transport from the core to the envelope occurs efficiently
in weakly differentially rotating main sequence stars,
the rotation rate of the core is kept low, which helps the stars rotate uniformly during 
their main sequence evolution.

\begin{appendix}

\section{two prescriptions for frozen-in convection}

Energy in the stellar interior is transported by radiation and/or convection.
Although energy transport by radiation occurs without fluid motions,
convective energy transport is accompanied by fluid motions which are turbulent in the stellar interior.
In pulsating stars we have to calculate perturbations of both radiative energy flux and convective energy flux.
Since we usually employ the diffusion approximation for the radiative energy transfer in the interior, the perturbed
radiative energy flux may be governed by the temperature perturbation.
However, for the convective energy transfer, 
we have to consider the perturbations of turbulent fluid motion, to which statistical description
should be applied.
Although this is not necessarily an easy problem to solve, 
there have been several attempts to describe the perturbations of turbulent fluids
in pulsating stars 
(e.g., \citealt{Unno67,Gabrieletal75,Gough77,Xiong77,Grigahceneetal05}).

\begin{table*}
\begin{center}
\caption{Complex eigenfrequency $\overline\omega=\overline\omega_{\rm R}+\rmi\overline\omega_{\rm I}$
of low $l$ core modes in the $4M_\odot$ ZAMS model with $X=0.7$ and $Z=0.02$ for $\Omega_s=0$, where $\overline\omega=\omega/\sqrt{GM/R^3}$ and
$n_c$ is the number radial nodes of $S_{l_1}$ in the core. 
The notation $a(b)$ implies a number given by $a\times10^b$.}
\begin{tabular}{@{}ccccccc}
\hline
    & \hspace*{2.2cm}$~~l=1$ && \hspace*{2.2cm}$~~l=2$ && \hspace*{2.2cm}$~~l=3$  \\
\hline
 $n_c$   & $\overline\omega_{\rm R}$ & $\overline\omega_{\rm I}$ & $\overline\omega_{\rm R}$ & $\overline\omega_{\rm I}$ & $\overline\omega_{\rm R}$ & $\overline\omega_{\rm I}$ \\
\hline
$0$  & $2.56(-3)$ & $-1.48(-3)$ & $2.91(-3)$ & $-1.68(-3)$ & $3.06(-3)$ & $-1.77(-3)$  \\
$1$  & $1.79(-3)$ & $-1.03(-3)$  & $2.23(-3)$ & $-1.29(-3)$  & $2.49(-3)$ & $-1.44(-3)$  \\
$2$  & $1.42(-3)$ & $-8.17(-4)$  & $1.84(-3)$ & $-1.06(-3)$  & $2.12(-3)$ & $-1.22(-3)$  \\
\hline
\end{tabular}
\medskip
\end{center}
\end{table*}

Since it is difficult to properly treat the coupling between convective fluid motions and pulsations in rotating stars
(see \citealt{BelkacemMarquesGoupiletal2015}),
we usually employ a simplifying assumption called frozen-in convection to compute non-adiabatic pulsations of the stars. 
We may perturb the entropy equation to obtain
\be
\rmi\omega\rho T c_p{\delta s\over c_p}
=\delta (\rho\epsilon)-\delta\left(\nabla\cdot\pmb{F}\right)
=(\rho\epsilon)^\prime-\left(\nabla\cdot\pmb{F}\right)^\prime
\label{eq:enteq0}
\ee
where $\pmb{F}=\pmb{F}_{\rm R}+\pmb{F}_{\rm C}$, and $\pmb{F}_{\rm R}$ and $\pmb{F}_{\rm C}$ are
the radiative energy flux and the convective energy flux, respectively, and other symbols have their usual meanings.
Note that in equilibrium we have $\nabla\cdot\pmb{F}=\rho\epsilon$.
Approximation called frozen-in convection is given in various ways.
For example,
we may assume $\delta L_{\rm C}=\delta F_{{\rm C},H}=0$
or $L^\prime_{\rm C}=F_{{\rm C},H}^\prime=0$ (e.g., \citealt{GlatzelMehren96})
for non-radial oscillations of non-rotating stars where
$L_C=4\pi r^2F_{\rm C}$, and $F_{\rm C}$ and $F_{{\rm C},H}$ are
the convective energy flux in radial and horizontal directions, respectively.
As discussed in \citet{Unnoetal1989}, we may also assume $\delta(\nabla\cdot\pmb{F}_{\rm C})=0$ or
$\delta (\rho^{-1}\nabla\cdot\pmb{F}_{\rm C})=0$, and
there can be other prescriptions to give the approximation of frozen-in convection.
For rotating stars, for example, \citet{LeeSaio93,LeeSaio20} assumed $\delta(\nabla\cdot\pmb{F}_{\rm C})=0$, while \citet{LeeBaraffe95}
used $(\nabla\cdot\pmb{F}_{\rm C})^\prime=0$.
We can usually expect that the different prescriptions for frozen-in convection do not lead to
significantly contradicting results for the stability of $g$-modes and $p$-modes in general.
But, this is not always the case for low frequency modes in the convective core of the stars.


Let us discuss 
low frequency non-adiabatic modes in 
the convective core of a non-rotating star.
In the Cowing approximation, non-adiabatic oscillations in the star are governed by (see, e.g., \citealt{Unnoetal1989})
\be
r{dy_1\over dr}=\left({V\over\Gamma_1}-3\right)y_1+\left({l(l+1)\over c_1\overline\omega^2}-{V\over\Gamma_1}\right)y_2+\alpha_T{\delta s\over c_p},
\label{eq:dy1}
\ee
\be
r{dy_2\over dr}=\left(c_1\overline\omega^2+rA\right)y_1+\left(1-U-rA\right)y_2+\alpha_T{\delta s\over c_p},
\label{eq:dy2}
\ee
where 
\be
y_1={\xi_r\over r}, \quad y_2={p'\over pV}.
\ee

If we assume $\delta(\nabla\cdot\pmb{F}_{\rm C})=0$,
we obtain from equation (\ref{eq:enteq0})
\be
\rmi \omega\rho T c_p{\delta s\over c_p} 
=\delta (\rho\epsilon)-\delta\left(\nabla\cdot\pmb{F}_{\rm R}\right).
\label{eq:deltas}
\ee
Considering most of the energy generated in the convective core is transported by convection,
we may ignore the term $\delta(\nabla\cdot\pmb{F}_{\rm R})$ in the convective core.
Substituting $\delta s/c_p$ given by equation (\ref{eq:deltas}) with $\delta(\nabla\cdot\pmb{F}_{\rm R})=0$
into equations (\ref{eq:dy1}) and (\ref{eq:dy2}) and using $\delta p/p=V(y_2-y_1)$, 
we obtain a set of differential equations with complex coefficients
for the dependent variables $y_1$ and $y_2$:
\be
r{dy_1\over dr}=\left({V\over\Gamma_1}-3+\rmi\gamma\right)y_1+\left({l(l+1)\over c_1\overline\omega^2}-{V\over\Gamma_1}-\rmi\gamma\right)y_2,
\label{eq:dy1c}
\ee
\be
r{dy_2\over dr}=\left(c_1\overline\omega^2+rA+\rmi\gamma\right)y_1+\left(1-U-rA-\rmi\gamma\right)y_2,
\label{eq:dy2c}
\ee
where
\be
\gamma=\alpha_T{c_3\over c_2\overline\omega}\left(\epsilon_{ad}+{1\over\Gamma_1}\right)V.
\ee

On the other hand, if we assume $(\nabla\cdot\pmb{F}_{\rm C})^\prime=0$, we obtain from equation (\ref{eq:enteq0})
\begin{align}
\rmi \omega\rho T c_p{\delta s\over c_p} 
=&(\rho\epsilon)^\prime-\left(\nabla\cdot\pmb{F}_{\rm R}\right)^\prime\nonumber\\
=&\delta(\rho\epsilon)-y_1{d\over d\ln r}\nabla\cdot\pmb{F}_{\rm C},
\end{align}
where we have neglected the term $\delta(\nabla\cdot\pmb{F}_{\rm R})$.
If we approximate $\rho\epsilon\approx\nabla\cdot\pmb{F}_{\rm C}$ in the convective core,
we obtain
\be
\rmi \overline\omega c_2{\delta s\over c_p} =c_3\left(\epsilon_{ad}+{1\over\Gamma_1}\right)Vy_2,
\ee
with which we obtain
\be
r{dy_1\over dr}=\left({V\over\Gamma_1}-3\right)y_1+\left({l(l+1)\over c_1\overline\omega^2}-{V\over\Gamma_1}-\rmi\gamma\right)y_2,
\label{eq:dy1d}
\ee
\be
r{dy_2\over dr}=\left(c_1\overline\omega^2+rA\right)y_1+\left(1-U-rA-\rmi\gamma\right)y_2.
\label{eq:dy2d}
\ee
Although the differences between the set of equations (\ref{eq:dy1c}) and (\ref{eq:dy2c}) and that of equations
(\ref{eq:dy1d}) and (\ref{eq:dy2d}) seem insignificant, 
we find the differences have important consequences for the modal property of
low frequency modes in the convective core.

To see this, we may employ a local analysis.
For the case of $\delta(\nabla\cdot\pmb{F}_{\rm C})=0$, substituting 
\be
y_1\propto \exp\left(\rmi k_r r+\rmi \omega t\right), \quad y_2\propto \exp\left(\rmi k_r r+\rmi \omega t\right),
\ee
into equations (\ref{eq:dy1c}) and (\ref{eq:dy2c}), we obtain for the radial wavenumber $k_r$ (e.g., \citealt{Unnoetal1989})
\be
-r^2k_r^2=\left(c_1\overline\omega^2+rA+\rmi\gamma\right)\left({l(l+1)\over c_1\overline\omega^2}-{V\over\Gamma_1}-\rmi\gamma\right),
\ee
which reduces to
\be
-r^2k_r^2
=l(l+1)\rmi\alpha_T{c_3\over c_1c_2}\left(\epsilon_{ad}V+{V\over\Gamma_1}\right){1\over\overline\omega^3},
\ee
for $|\overline\omega|\ll1$ and $rA=0$ in the convective core.
The factor $1/\overline\omega^3$ is the large parameter in our local analysis and is rewritten using the growth rate 
$\eta=-\omega_{\rm I}/\omega_{\rm R}$
as
\be
{1\over\overline\omega^3}={\overline\omega_{\rm R}^3\over|\overline\omega|^6}\left[1-3\eta^2+\rmi\eta\left(3-\eta^2\right)\right].
\ee
To make $k_r^2$ be real and positive, we assume $\eta=1/\sqrt{3}\approx 0.577$ to obtain
\be
r^2k_r^2={8\over3\sqrt{3}}
l(l+1)\alpha_T{c_3\over c_1c_2}\left(\epsilon_{ad}V+{V\over\Gamma_1}\right)
{\overline\omega_{\rm R}^3\over|\overline\omega|^6},
\label{eq:kr}
\ee
which suggests the existence of low frequency modes propagative in the core.
The low frequency modes are called core modes in this paper.
The dispersion relation (\ref{eq:kr}) 
indicates that the existence of the core modes is closely related to
nuclear energy generation $c_3\epsilon_{ad}$.
Using the method of calculation by \citet{LeeSaio93} in which $\delta(\nabla\cdot\pmb{F}_{\rm C})=0$ is assumed, 
we compute low frequency core modes in the $4M_\odot$ ZAMS model assuming $\nabla-\nabla_{ad}=0$, 
and the results are summarized in the table A1.
As found from the table, the growth rates are nearly equal to $1/\sqrt{3}\approx 0.577$.
Since the core modes exist even for $rA=0$, they cannot be convective modes.
Note also that the core modes have no adiabatic counterparts.

If we assume $(\nabla\cdot\pmb{F}_{\rm C})^\prime=0$, on the other hand, we obtain for the wavenumber $k_r$
\be
-r^2k_r^2=\left(c_1\overline\omega^2+rA\right)\left({l(l+1)\over c_1\overline\omega^2}-{V\over\Gamma_1}-\rmi\gamma\right),
\ee
which reduces to
\be
r^2k_r^2=-l(l+1)
\ee
for $|\overline\omega|\ll 1$ and for $rA=0$ in the core.
This dispersion relation suggests that for $(\nabla\cdot\pmb{F}_{\rm C})^\prime=0$,
there appears no low frequency modes propagative in the core if $rA=0$.

\section{Oscillation equations for differentially rotating stars}

In this appendix, we present the set of linear ordinary differential equations for non-adiabatic oscillations
of differentially rotating stars in the Cowling approximation, 
obtained by assuming $(\nabla\cdot\pmb{F}_{\rm C})^\prime=0$ 
for the convective energy flux $\pmb{F}_{\rm C}$.
Using the dependent variables $\pmb{y}_1$ to $\pmb{y}_4$ defined as
\begin{align}
\pmb{y}_1=\left(S_{l_j}\right), \quad \pmb{y}_2=\left({p'_{l_j}\over\rho g r}\right), \quad
\pmb{y}_3=\left({L^\prime_{{\rm R},l_j}\over L_{\rm R}}\right), \quad \pmb{y}_4=\left({\delta s_{l_j}\over c_p}\right), 
\end{align}
where $g=GM_r/r^2$ with $M_r=\int_0^r4\pi r^2\rho dr$, $c_p$ is the specific heat at constant pressure, 
$L_{\rm R}$ is the radiative luminosity, and $L_{\rm R}^\prime$ is its Euler perturbation, 
the set of linear differential equations may be given by
\begin{align}
r{d \pmb{y}_1\over d r}=&\left[\left({V\over\Gamma_1}-3\right)\pmbmt{1}+\nu\pmbmt{WO}+{1\over\omega^2}{\partial\Omega^2\over\partial\ln r}\pmbmt{R}_1\right]\pmb{y}_1\nonumber\\
&+\left({\pmbmt{W}\over c_1\bar\omega^2}-{V\over\Gamma_1}\pmbmt{1}\right)\pmb{y}_2
+\alpha_T\pmb{y}_4,
\end{align}
\begin{align}
r{d\pmb{y}_2\over d r}=&\left[\left(c_1\bar\omega^2+rA\right)\pmbmt{1}-4c_1\bar\Omega^2\pmbmt{G}-c_1{\partial\bar\Omega^2\over\partial\ln r}\pmbmt{R}_2
\right]\pmb{y}_1\nonumber\\
&+\left[\left(1-U-rA\right)\pmbmt{1}-\nu\pmbmt{O}^T\pmbmt{W}\right]\pmb{y}_2+\alpha_T\pmb{y}_4,
\end{align}
\begin{align}
r{d\pmb{y}_3 \over d r}=&
\left[-c_3\left(1-{\epsilon_T\over\alpha_T}\right)rA\pmbmt{1}-\left(1-{\nabla_{ad}\over\nabla}\right)\pmbmt{\Lambda}_0\right]\pmb{y}_1\nonumber\\
&+\left[c_3\left(\epsilon_{ad}V+{V\over\Gamma_1}\right)\pmbmt{1}-{\nabla_{ad}\over\nabla}\pmbmt{\Lambda}_0\right]\pmb{y}_2
-\hat c_3\pmb{y}_3\nonumber\\
&
+\left[\left(c_3(-\alpha_T+\epsilon_T)-\rmi\overline\omega c_2\right)\pmbmt{1}-{\pmbmt{\Lambda}_0\over\nabla V}
\right]\pmb{y}_4
\end{align}
\begin{align}
{1\over\nabla V}r{d\pmb{y}_4\over d r}=&\left(D_1+1-\hat c_3\right)\pmb{y}_1
-D_1\pmb{y}_2\nonumber\\
&-\pmb{y}_3+\left(4-\kappa_T+\alpha_T\right)\pmb{y}_4\nonumber\\
&-\left(1-{\nabla_{ad}\over\nabla}\right){d\pmb{y}_1\over d\ln r}-{\nabla_{ad}\over\nabla}{d\pmb{y}_2\over d\ln r},
\end{align}
where $\nu=2\Omega/\omega$ and $\omega=\sigma+m\Omega$ with $\sigma$ being the frequency in the inertial frame, and
\be
U={d\ln M_r\over d\ln r}, \quad V=-{d\ln p\over d\ln r}, 
\quad 
\ee
\be
rA={d\ln\rho\over d\ln r}-{1\over\Gamma_1}{d\ln p\over d\ln r}, \quad 
\Gamma_1=\left({\partial\ln p\over\partial\ln\rho}\right)_{ad}, 
\ee
\be
c_1={(r/R)^3\over M_r/M}, \quad c_2={4\pi r^3\rho Tc_p\over L_{\rm R}}\sqrt{GM\over R^3}, 
\ee
\be
c_3={4\pi r^3\rho\epsilon_N\over L_{\rm R}}, \quad \hat c_3={d\ln L_{\rm R}\over d\ln r},
\ee
\be
\nabla={d\ln T\over d\ln p}, \quad \nabla_{ad}=\left({\partial\ln T\over \partial \ln p}\right)_{S}, 
\ee
\be
\epsilon_{ad}=\left({\partial\ln\epsilon\over\partial\ln p}\right)_S, 
\quad \epsilon_T=\left({\partial\ln\epsilon\over\partial\ln T}\right)_p, 
\ee
\be
\kappa_{ad}=\left({\partial\ln\kappa\over\partial\ln p}\right)_{S}, \quad 
\kappa_T=\left({\partial\ln\kappa\over\partial\ln T}\right)_p,
\ee
\begin{align}
&D_1={\nabla_{ad}\over\nabla}\left({d\ln\nabla_{ad}\over d\ln r}+rA-{V\over\Gamma_1}+U+V-1\right)\nonumber\\
&~~~~ -\left(4\nabla_{ad}-\kappa_{ad}-{1\over\Gamma_1}\right)V, \quad 
\alpha_T=-\left({\partial\ln\rho\over\partial\ln T}\right)_p,
\end{align}
and $\pmbmt{1}$ is the unit matrix and other matrices are defined as
\be
\pmbmt{R}_1=\pmbmt{WH}, 
\ee
\be
\pmbmt{R}_2=\nu\left(\pmbmt{O}^T\pmbmt{WH}+m\pmbmt{C}_0\pmbmt{L}_1^{-1}\pmbmt{\Lambda}_1^{-1}\pmbmt{Q}_1\right)+\pmbmt{1}-\pmbmt{Q}_0\pmbmt{Q}_1,
\ee
\be
\pmbmt{G}=\pmbmt{O}^T\pmbmt{WO}-\pmbmt{C}_0\pmbmt{L}_1^{-1}\pmbmt{K},
\ee
\be
\pmbmt{H}=m\pmbmt{M}_1\pmbmt{L}_1^{-1}\pmbmt{\Lambda}_1^{-1}\pmbmt{Q}_1-\pmbmt{\Lambda}_0^{-1}\left(
\pmbmt{Q}_0\pmbmt{C}_1+3\pmbmt{Q}_0\pmbmt{Q}_1-\pmbmt{1}\right),
\ee
and the definition of the matrices $\pmbmt{C}_0$, $\pmbmt{C}_1$, $\pmbmt{K}$, $\pmbmt{L}_0$, 
$\pmbmt{L}_1$, $\pmbmt{M}_0$, $\pmbmt{M}_1$, $\pmbmt{O}$, $\pmbmt{Q}_0$, $\pmbmt{Q}_1$, $\pmbmt{W}$,
$\pmbmt{\Lambda}_0$, and $\pmbmt{\Lambda}_1$
is given in \citet{LeeSaio90}.
If we define the vector $\pmb{h}$ and $\pmb{t}$ as
\be
\pmb{h}=\left(H_{l_j}\right), \quad \pmb{t}=\left(T_{l'_j}\right),
\ee
we can write the relations between ($\pmb{y}_1$,$\pmb{y}_2$) and ($\pmb{h}$,$\pmb{t}$) as
\be
\pmbmt{\Lambda}_0\pmb{h}=\left(\nu\pmbmt{WO}+{1\over\omega^2}{\partial\Omega^2\over\partial\ln r}\pmbmt{WH}\right)\pmb{y}_1+{\pmbmt{W}\over c_1\bar\omega^2}\pmb{y}_2,
\ee
\begin{align}
&2mc_1\bar\omega\bar\Omega\pmb{h}+ 2c_1\bar\omega\bar\Omega\pmbmt{C}_0\rmi\pmb{t}=\nonumber\\
&~~4c_1\bar\Omega^2\left[\pmbmt{G}+{1\over\omega}{\partial\Omega\over\partial\ln r}\left(\pmbmt{O}^T\pmbmt{WH}
+m\pmbmt{C}_0\pmbmt{L}_1^{-1}\pmbmt{\Lambda}_1^{-1}\pmbmt{Q}_1\right)\right]\pmb{y}_1
\nonumber\\
&~~+\nu\pmbmt{O}^T\pmbmt{W}\pmb{y}_2.
\end{align}
Applying appropriate boundary conditions at the centre and the surface of the stars, we solve the set of
linear ordinary differential equations as an eigenvalue problem for the inertial frame frequency $\sigma$.

\end{appendix}

\bigskip
\bigskip
\noindent
Data Availability: The data underlying this article will be shared on reasonable request to the corresponding author.

\bibliographystyle{mnras}
\bibliography{myref}

\end{document}